\documentclass{article}

\usepackage{arxiv}

\usepackage[english]{babel} 

\usepackage{amsmath}
\usepackage{amssymb}
\usepackage{amsfonts}
\usepackage{amsthm}
\usepackage{mathrsfs}
\usepackage{mathtools}
\usepackage{bm}
\usepackage{commath} % \dif
\usepackage{hyperref}  
\usepackage[english]{fancyref}
\usepackage{dsfont} % indicatrice
\usepackage{ctable}
\usepackage{colortbl}
\usepackage{multirow}
\usepackage{enumitem}
\usepackage{caption}
\usepackage{subfig}
\usepackage[ruled]{algorithm2e}
\usepackage{varioref} % loadbefore cleveref : https://tex.stackexchange.com/questions/341268/cleveref-package-and-conditional-page-references
\usepackage[noabbrev]{cleveref}
\usepackage[backend=bibtex, 
style=authoryear,
hyperref = true, 
backref=true,
doi=false, isbn=false, url=false, eprint=false, dashed=false,
maxcitenames=2]{biblatex}

\newtheorem{proposition}{Proposition}
\newtheorem{lemma}{Lemma}
\definecolor{SchoolColor}{HTML}{003262} %berkeley blue
\hypersetup{
	colorlinks,
	citecolor=SchoolColor,
	filecolor=black,
	linkcolor=black,
	urlcolor=SchoolColor,
}
\newcommand{\algorithmstyle}[1]{\renewcommand{\algocf@style}{#1}}

\newcommand{\llhood}{L}
\DeclareMathOperator{\Umethod}{F-procedure}
\DeclareMathOperator{\temp}{temp}
\DeclareMathOperator{\aitken}{aitken}
\newcommand{\p}{p} % proba
\newcommand{\q}{q} %{\mathcal{R}} % proba

\DeclareMathOperator{\ICLbic}{ICL_{\textit{BIC}}}

\DeclarePairedDelimiterX{\KLx}[2]{(}{)}{%
	#1\,\delimsize\|\,#2%
}
\DeclareMathOperator{\KLy}{KL}
\newcommand{\KL}{\KLy \KLx}
\DeclareMathOperator{\entropy}{H}

\newcommand{\pkg}[1]{{\fontseries{b}\selectfont #1}}
\newcommand{\bZ}{\bm{Z}}

\newcommand{\R}{\mathbb{R}}

\bmdefine{\bG}{G}
\bmdefine{\bT}{T}
\bmdefine{\bJ}{J}

\bmdefine{\bxi}{\xi}

\bmdefine{\Obs}{\bm{Y}}
\bmdefine{\obs}{\bm{y}}

\bmdefine{\Clust}{\bm{Z}}
\bmdefine{\clust}{\bm{z}}
\newcommand{\rawclust}{z}
\bmdefine{\Scores}{\bm{X}}
\bmdefine{\scores}{\bm{x}}

\bmdefine{\Latent}{\bm{\eta}}
\bmdefine{\latent}{\bm{\eta}}

\bmdefine{\param}{\bm{\theta}}

\bmdefine{\globalparam}{\vartheta}

\bmdefine{\bU}{\bm{U}}
\bmdefine{\bu}{\bm{u}}

\bmdefine{\bW}{W}
\bmdefine{\bC}{C}
\bmdefine{\bR}{R}
\bmdefine{\bM}{M}
\bmdefine{\bL}{L}
\bmdefine{\bP}{P}
\bmdefine{\bI}{I}
\bmdefine{\bH}{H}
\bmdefine{\bH}{H}
\bmdefine{\bA}{A}
\bmdefine{\bB}{B}
\bmdefine{\bJ}{J}
\bmdefine{\bV}{V}
\bmdefine{\bP}{P}
\bmdefine{\ba}{a}
\bmdefine{\bDelta}{\Delta}

\bmdefine{\bBeta}{\beta}

\newcommand{\alphaLDA}{\delta}
\bmdefine{\balphaLDA}{\alphaLDA}

\bmdefine{\bnu}{\nu}
\DeclareMathOperator{\BDLM}{BDLM}
\DeclareMathOperator{\Fisher}{F}
\newcommand{\bvarcovar}{\tilde{\mathbf{M}}} % variat covar
\newcommand{\bvarmean}{\tilde{\bmu}} % variat mean
\newcommand{\varn}{\tilde{n}} % variational cluster counts

\newcommand{\btildeC}{\tilde{\bC}}
\newcommand{\btildeS}{\tilde{\bS}}

\newcommand{\bhatC}{\hat{\bC}}
\bmdefine{\bLambda}{\Lambda}
\DeclareMathOperator{\vect}{vect}
\bmdefine{\bv}{v}

\newcommand{\Model}{\mathcal{M}}
\newcommand{\dimparam}{\gamma}

\bmdefine{\ClustLDA}{T}
\bmdefine{\clustLDA}{t}

\bmdefine{\bPhi}{\phi}
\bmdefine{\bGamma}{\gamma}

\bmdefine{\bS}{\bm{S}}
\bmdefine{\bSigma}{\bm{\Sigma}}
\bmdefine{\bmu}{\bm{\mu}}
\newcommand{\nb}{n}
\renewcommand{\dim}{p}
\newcommand{\latentdim}{d}
\newcommand{\K}{K}
\bmdefine{\bPi}{\bm{\pi}}
\bmdefine{\bw}{\bm{w}}

\bmdefine{\bpatch}{t}

\bmdefine{\bmean}{\bm{m}}
\bmdefine{\balpha}{\bm{\alpha}}
\bmdefine{\bD}{\bm{D}}
\bmdefine{\bPsi}{\bm{\Psi}}

\DeclareMathOperator{\cst}{const\,}
\DeclareMathOperator{\Expectation}{\mathbb{E}}
\DeclareMathOperator{\Tr}{Tr}
\DeclareMathOperator{\diag}{diag}

\bmdefine{\Clustrow}{\bm{Z}_{r}}
\bmdefine{\clustrow}{\bm{z}}

\bmdefine{\Clustcol}{\bm{Z}_{c}}
\bmdefine{\clustcol}{\bm{z}}

\bmdefine{\bPirow}{\bm{\pi}_{r}}
\bmdefine{\bPicol}{\bm{\pi}_{c}}

\DeclareMathOperator{\Gaussian}{\mathcal{N}}
\DeclareMathOperator{\Mult}{\mathcal{M}}

\DeclareMathOperator{\Dir}{\mathcal{D}} % Dirichlet density
\DeclareMathOperator{\J}{\mathcal{J}}
\bmdefine{\bTau}{\bm{\tau}}

\newcolumntype{M}[1]{>{\centering\arraybackslash}m{#1}}
\newcolumntype{R}[1]{>{\raggedright\arraybackslash}m{#1}}
\newcolumntype{C}[1]{>{\centering\arraybackslash}m{#1}}

\DeclareMathOperator*{\argmax}{arg\,max}
\DeclareMathOperator*{\argmin}{arg\,min}

\graphicspath{{./figures}{./images/}{./images/sigma20/}{./images/sigma30/}{./images/OtherBenchmarks/NLB/}{./images/OtherBenchmarks/S-PLE/}{./images/OtherBenchmarks/HDMI/}}

\title{A Bayesian Fisher-EM algorithm for discriminative Gaussian subspace clustering}

\author{Nicolas Jouvin \\
	Universit\'{e} Paris 1 Panth\'{e}on-Sorbonne, SAMM, France\\
	FP2M, CNRS FR 2036, Paris, France \\
	\And 
	Charles Bouveyron \\
	Universit\'{e} C\^{o}te d’Azur, Inria, CNRS, Laboratoire J.A. Dieudonné\\
	Maasai research team, Nice, France
	\And 
	Pierre Latouche\\
	Universit\'{e} de Paris, MAP5, CNRS,  \\
	FP2M, CNRS FR 2036, Paris, France \\ 
}

\begin{document}
	
	\maketitle
	
\begin{abstract}
	High-dimensional data clustering has become and remains a challenging task for modern statistics and machine learning, with a wide range of applications. We consider in this work the powerful discriminative latent mixture model, and we extend it to the Bayesian framework. Modeling data as a mixture of Gaussians in a low-dimensional discriminative subspace, a Gaussian prior distribution is introduced over the latent group means and a family of twelve submodels are derived considering different covariance structures. Model inference is done with a variational EM algorithm, while the discriminative subspace is estimated via a Fisher-step maximizing an unsupervised Fisher criterion. An empirical Bayes procedure is proposed for the estimation of the prior hyper-parameters, and an integrated classification likelihood criterion is derived for selecting both the number of clusters and the submodel. The performances of the resulting Bayesian Fisher-EM algorithm are investigated in two thorough simulated scenarios, regarding both dimensionality as well as noise and assessing its superiority with respect to state-of-the-art Gaussian subspace clustering models. In addition to standard real data benchmarks, an application to single image denoising is proposed, displaying relevant results. This work comes with a reference implementation for the \pkg{R} software in the \pkg{FisherEM} package\footnote{Available on CRAN, see \url{https://github.com/nicolasJouvin/FisherEM} for additional information.} accompanying the paper.
	\keywords{Mixture model \and High dimensionality \and Dimensionality reduction \and Linear Discriminant Analysis}
	% \PACS{PACS code1 \and PACS code2 \and more}
	% \subclass{MSC code1 \and MSC code2 \and more}
\end{abstract}

\section{Introduction}
\label{bfem:sec:Intro}

Clustering has become an important part of contemporary statistics and machine learning,
%	Consisting in the unsupervised task of grouping $\nb$ observations into $\K$ groups of similar objects, 
with applications ranging from DNA microarray analysis in biology \parencite{ghosh2002mixture} to text analysis \parencite{aggarwal2012survey} and image processing \parencite{coleman1979image, jegou2010aggregating}. However, the high dimensionality characterizing modern datasets causes geometrical and statistical issues, often subsumed under the term of \textit{curse of dimensionality} \parencite{bellman1957dynamic, giraud2014introduction}, and affecting traditional approaches. In this paper, we focus on model-based clustering with Gaussian mixture models, and we propose an extension to the Bayesian framework of the work of \textcite{bouveyron2012simultaneous} along with a novel algorithm for simultaneously clustering and visualizing high-dimensional continuous data.

\subsection{High-dimensional Gaussian clustering}
%	 Grounded on statistical theory, each observation $\obs_i$ is assigned to a multinomial random variable $\clust_{i}$ representing its cluster assignment. Then, observations belonging to the same cluster are supposed to be independent and indentically distributed. The partition $\Clust$ being unobserved, clustering is cast as an inference problem, seeking the partition and parameters that best fits the observed data according to the underlying statistical model. In the case of continuous observations in dimension $\dim$, 
Finite mixture models constitute one of the most popular approaches to model-based clustering \parencite{mclachlan2004finite}. Let us consider $\nb$ continuous observations $\{ \obs_i\}$ in dimension $\dim$, summarized in a data matrix $\Obs \in \R^{\nb \times \dim}$, that we want to cluster into $\K$ groups. 
%	Finite mixture models assign each observation to an unobserved multinomial variable representing its cluster assignment. Then, the partition $\Clust$ being represented by the collection of latent varaiable, clustering is cast as a statistical inference task seeking to estimate 
The Gaussian mixture model \parencite[GMM,][]{bouveyron2019model} posits the following distribution:
\begin{align}
\label{bfem:eq:ObservedLlhoodGMM}
\p(\Obs \mid \bPi, \bmean, \bS) = \prod_{i=1}^{\nb} \sum_{k=1}^{\K} \pi_k \Gaussian_{\dim}(\obs_{i} \mid \bmean_k, \bS_k),
\end{align}
where $\bPi$ denotes the mixture proportions and $(\bmean_{k}, \bS_k)$ respectively corresponds to the mean and covariance matrix of the $k$-th component. Clustering arises from this probabilistic formulation with the introduction of an unobserved random variable $\clust_i$ following a multinomial distribution with parameter $\bPi$, and characterizing the cluster assignment of observation $\obs_i$. The corresponding partition $\Clust = \{\clust_{i}\}$ is then considered as the set of discrete latent variables which are to be estimated, along with the model parameters, traditionally by an Expectation-Maximization algorithm \parencite[EM,][]{dempster1977maximum}. Here, the curse of dimensionality comes from the covariance matrices $\bS_k$, which involve a number of parameters growing with the squared of the dimension. In such context, the number of observations required to fit high-dimensional data may be very large and computationally impractical.

\begin{figure}[t]
	\centering
	\includegraphics[width = 0.98\linewidth]{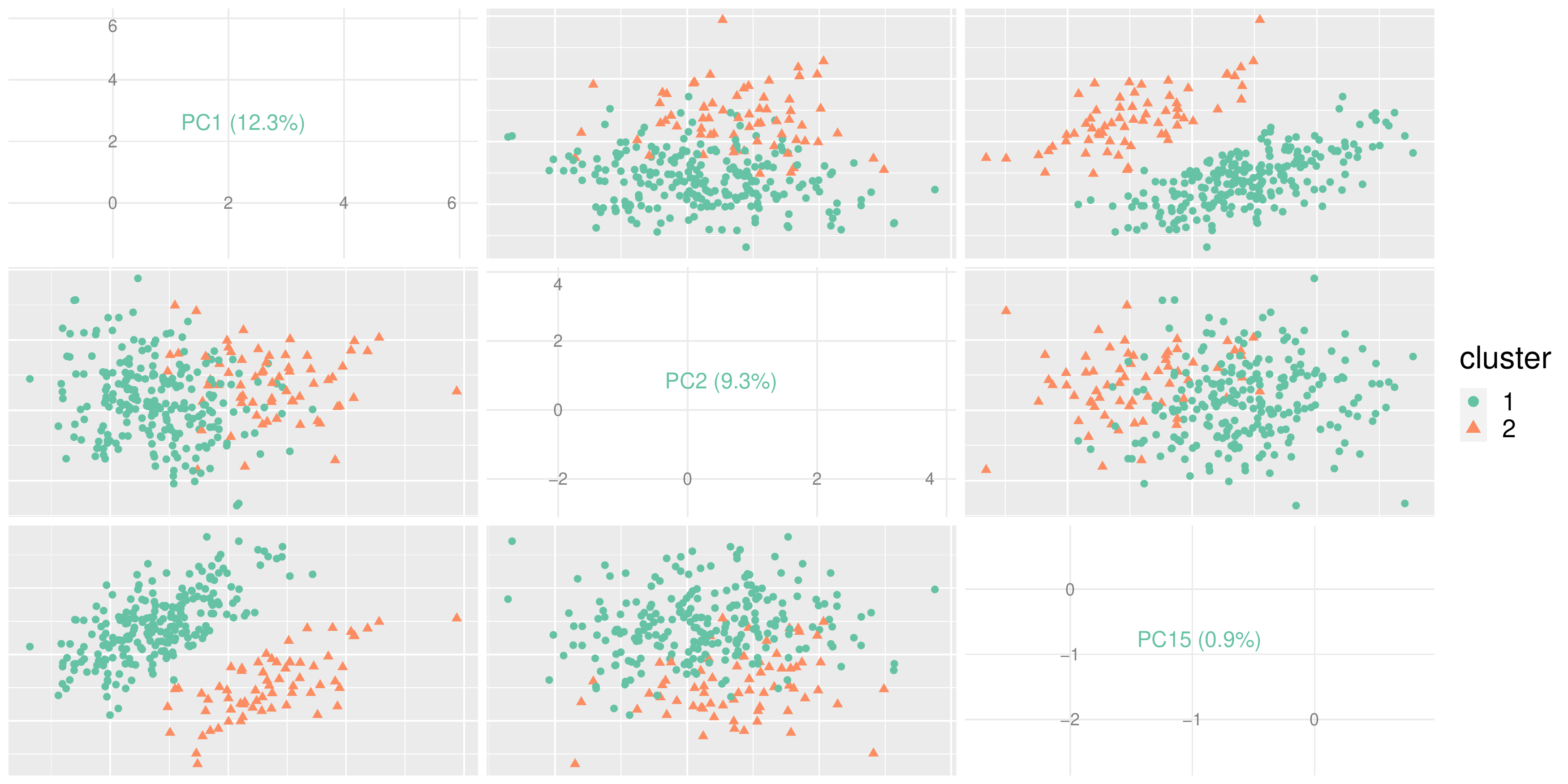}
	\caption{\textcite{chang1983using} data set with $\nb = 300$ observations projected on the 1st, 2nd and 15th principal components respectively. Colors and shapes indicate the true cluster membership. We see that the last principal component contains important discriminative information in terms of clustering, while the second principal component is not suited for the task.}
	\label{fig:chang1983}
\end{figure}

On the one hand, some approaches rely on unsupervised dimension reduction such as principal component analysis or factor analysis to project the data prior to model fitting \parencite{ghosh2002mixture}. However, such transformations do not take into account the clustering task at hand and might induce a loss of relevant discriminative information, in addition to losing the principled approach to model-based clustering. In order to demonstrate the importance of a relevant transformation in a clustering context, \textcite{chang1983using} exhibited a $2$-components simulation setting in dimension $\dim=15$ where the groups are best discriminated on the space defined by the first and the last components, as represented in \Cref{fig:chang1983}. Another example of this phenomenon is also given in \textcite[sec. 8.2]{mclachlan2004finite} in dimension $\dim = 5$. Model-free heuristics have also been proposed for subspace clustering, seeking for regions of high-density within the observed space. The CLIQUE algorithm \parencite{agrawal1998automatic} is a popular instance of such methods, and the building blocks for many others. We refer to \textcite{parsons2004subspace} for a comprehensive review on this subject.

On the other hand, a wealth of literature has focused on developing parsimonious models based on \Cref{bfem:eq:ObservedLlhoodGMM}, consisting of constrained covariance matrices $\bS_k$. \textcite{banfield1993model} and \textcite{celeux1995gaussian} proposed to perform a spectral decomposition of the latter, and derived a family of fourteen submodels by considering different constraints on the eigenvectors and eigenvalues \parencite[see \textit{e.g.}][]{scrucca2016mclust}. Although efficient and flexible, this method still requires the estimation of full-rank $\dim \times \dim$ matrices in the observation space for complex models, which can be impractical. Other types of restrictions were considered by \textcite{ghahramani1996algorithm} and \textcite{tipping1999mixtures}, with low-rank factorizations $\bS_k = \bU_k \bU_k^\top + \bPsi_k$ where $\bU_k$ is a $\dim \times \latentdim$ matrix. Based on a factor analysis formulation, these models have a geometric interpretation: integrating model-based clustering and probabilistic linear dimension reduction, they seek to cluster the data in $\K$ low-dimensional subspaces $\bU_k$ of dimension $\latentdim$. This formulation was further refined to impose a common subspace $\bU$ accross cluster, restricting the number of parameter and allowing a common vizualisation of the data points \parencite{yoshida2004mixed,baek2009mixtures,mcnicholas2008parsimonious,montanari2010heteroscedastic}. At the intersection of the spectral and geometric approaches, \textcite{bouveyron2007high} proposed a family of 28 submodels, separating between signal and noisy directions. These models are often refered to as Gaussian subspace clustering, and maximum likelihood inference is always preferred, usually via an EM algorithm. We refer the reader to \textcite{bouveyron2014model} for a thorough review of model-based high-dimensional clustering.

However, without further clustering information, the estimated latent subspace may be biased toward density estimation, preserving the variance of the observed data as much as possible, rather than clustering and explicit separation of the groups. These objectives are not always aligned as \Cref{fig:chang1983} suggests, and in order to circumvent this issue, several works introduced the notion of a discriminative subspace.

\subsection{Discriminative subspace: from classification to clustering}
In the supervised framework, where the partition $\Clust$ is observed, the tension between signal \textit{representation} and signal \textit{classification} is well known and analogous to the distinction between density estimation and clustering. \textcite[chap. 10]{fukunaga1990introduction} discusses this in detail, introducing the notion of class separability along with four different possible criteria to measure it. The idea is to find a linear subspace $\bU$ in which the group means are well separated while the within-class variance is small. The most popular criterion for such a task is a generalization of Fisher's Linear Discriminant Analysis \parencite[LDA,][chap. 4]{fisher1936use,duda2000pattern}:
\begin{align}
\label{eq:FisherCriterion}
\Fisher(\bU) = \Tr\left[(\bU^\top \bS_W \bU)^{-1} \bU^\top \bS_B \bU\right],
\end{align}
where $\bS_W = (1/\nb) \sum_k \sum_{i} \rawclust_{ik}  (\obs_{i} - \bmean_k) (\obs_{i} - \bmean_k)^\top$ is the within-class covariance matrix of the data, while $\bS_B = (1/\nb) \sum_k \nb_k(\bmean_{k} - \bar{\obs}) (\bmean_{k} - \bar{\obs})^\top$ is the between-class covariance, with $n_k = \sum_i \rawclust_{ik}$ and $\bmean_{k} = (1/ \nb_k) \sum_i \rawclust_{ik} \obs_i$. This criterion computes the trace of the ratio between the within-class and between-class covariance matrices in the latent space, and its maximization with respect to $\bU$ translates the goal of a discriminative subspace. Without orthogonality constraints, the maximization of \Cref{eq:FisherCriterion} happens to be equivalent to a generalized eigenvalue problem $\bS_B \bu = \lambda \bS_W \bu$ which can be solved efficiently, even when $\bS_W$ is singular \parencite{ye2005characterization}. Moreover, since the rank of the matrix $\bS_B$ is at most $\K-1$, there is only $\latentdim \leq \K-1$ dimensions of interest. 

In the unsupervised context, which is of interest in this paper, $\Clust$ is unknown, and the scatter matrices $\bS_W$ and $\bS_B$ cannot be formed. Still, building on these ideas, some works have been proposed to adapt the criterion. In the goal of feature selection for clustering, \textcite{dy2004feature} compared maximum likelihood approaches with the maximization criterion of \Cref{eq:FisherCriterion}, highlighting the interest of both in different contexts. Feature selection can be cast in the framework of dimension reduction where $\bU$ is forced to be a $(0,1)$-matrix with non-zero index in each columns indicating the subset of $\latentdim$ selected variables \parencite{nie2008trace}. For clustering applications, \textcite{de2006discriminative} proposed the discriminative cluster analysis, combining $k$-means and linear discriminant analysis. In a visualization approach, \textcite{scrucca2010dimension} proposed to project the data in a subspace minimizing the Fisher criterion, using the partition given by the model of \Cref{bfem:eq:ObservedLlhoodGMM}. It relies on a modified version of $\bS_B$ taking into account variability between within-class covariance, and demonstrates good visualization power. However, this method is post-inference and still requires to fit a GMM in the observation space which is prohibitive in real high-dimensional scenarios. Finally, \textcite{bouveyron2012simultaneous} proposed the discriminative latent mixture (DLM) model, which treats the estimation of the latent subspace as a separate problem from maximum likelihood estimation. The proposed model is closed to the mixture of common factor analyzers \parencite[MCFA,][]{baek2009mixtures}, although inference is different and done via the Fisher-EM algorithm which mixes the EM strategy with a specific Fisher-step. In the latter, the current posterior membership probabilities are used to compute the scatter matrices $\bS_W$ and $\bS_B$, and $\bU$ is supposed to be discriminant, maximizing \Cref{eq:FisherCriterion} with orthogonality constraints.

Albeit not directly related to the high-dimensional setting, the idea of incorporating clustering information is popular in the context of mixture modeling. It dates back to the CEM algorithm \parencite{celeux1992classification} aiming at maximizing the classification likelihood for inference. It is also present in the integrated classification likelihood \parencite[ICL,][]{biernacki2000assessing} criterion for selecting the number of clusters. This trade-off between clustering and density estimation also led to the proposition of differentiating between the notion of mixture component and cluster in Gaussian mixtures \parencite{Baudry2010}, the former being a combination of the latter. 

\subsection{Contribution and organization of the paper}

This paper introduces a new algorithm for the clustering of high-dimensional data with a constrained Gaussian mixture model. In \Cref{bfem:sec:BDLM}, we introduce a Bayesian formulation of the DLM model putting a prior distribution on the mean in the latent space, with a hyper-parameter $\lambda$ controlling the between-class variance. Following \textcite{bouveyron2012simultaneous}, we derive a family of submodels with constraints on the latent covariance matrices, and discuss its links to existing methods. Then, the posterior distribution now being intractable, \Cref{bfem:sec:Inference} introduces a variational extension of the Fisher-EM algorithm for simultaneous clustering and dimension reduction.
%We also investigate other generalization of Fisher criterion, such as the ratio of traces problem mentioned above, and discuss their pros and cons both theoretically and empirically. 
In \Cref{bfem:sec:NumericalExpe}, two carefully designed scenarios, controlling for dimension and noise, demonstrate the superiority of the corresponding Bayesian Fisher-EM over state-of-art model-based subspace clustering models. Traditional benchmarks of the litterature are also used to assess the performances on real data. Finally, \Cref{bfem:sec:application} proposes an application to the problem of image denoising, demonstrating the interest of the proposed methodology compared to state-of-the-art approaches and paving way for future improvements. 
%	\Cref{bfem:sec:NumericalExpe} assesses the performance and stability corresponding Bayesian Fisher-EM algorithm in several high-dimensional scenarios on both simulated and real data, while diving a detailed comparison with state-of-the-art model-based subspace clustering models. 
%Finally, \Cref{bfem:sec:RealData} proposes a real-data analysis on WWW data.

\section{The Bayesian discriminative latent mixture}
\label{bfem:sec:BDLM}

\subsection{Discriminative latent mixture}
\label{bfem:subsec:DLM}
\textcite{bouveyron2012simultaneous} proposed the following generative model, relying on the idea that $\K - 1$ properly chosen dimensions are sufficient to discriminate between $\K$ classes. It is based on a Factor-Analysis-like formulation where the latent scores $\scores_{i} \in \R^{\latentdim}$ are low-dimensional representation of the observations $\obs_{i}$, and are assumed to follow a Gaussian mixture model in the subspace:
\begin{equation}
\label{eq:DLM}
\begin{aligned}
\clust_i &\sim \Mult_{\K}(1, \bPi), \\
\scores_{i} \mid \{\rawclust_{ik} = 1\} &\sim \Gaussian_\latentdim(\bmu_{k}, \bSigma_{k}), \\
%		\epsilon_i \mid \{\rawclust_{ik} = 1\} & \sim \Gaussian_{\dim}(\bm{0}_\dim, \bPsi_k), \\
\obs_{i} & = \bU \scores_{i} + \bm{\epsilon}_i, & & \bm{\epsilon}_i \mid  \{\rawclust_{ik}=1\}  \sim \Gaussian_{\dim} (\bm{0}_{\dim}, \bPsi_k).
%\bm{\epsilon}_i \mid  \{\rawclust_{ik}=1\} & \sim \Gaussian_{\dim} (\bm{0}_{\dim}, \bPsi_k).
\end{aligned}
\tag{DLM}
\end{equation}
Here, the matrix $\bU$ is constrained to be column-orthonormal, $\bU^\top \bU = \bI_\latentdim$, and forms the basis of a low-dimensional subspace of dimension $\latentdim \leq \min(\K -1, \dim)$, which is called the \textit{discriminative subspace}. When the latent variables are integrated out, the marginal distribution is a constrained GMM:
\begin{align*}
\obs_i \sim \sum_{k=1}^{\K} \pi_k \Gaussian_{\dim}(\bU \bmu_k, \bU \bSigma_k \bU^\top + \bPsi_k), & & \bmean_{k} = \bU \bmu_k, & & \bS_{k} = \bU \bSigma_{k} \bU + \bPsi_k.
\end{align*}
This model relates to the MCFA model \parencite{baek2009mixtures}, except that the latent dimension is constrained to be at most $\K-1$, and $\bU$ is considered to be discriminative in the sense of Fisher's criterion. Moreover, the noise matrices $\bPsi_k$ are not constrained to be common across clusters nor diagonal anymore, but rather to be isotropic in the orthogonal of the subspace. Formally, let us define $\bD = [\bU, \bV]$ where $\bV \in \R^{\dim \times (\dim - \latentdim)}$ is the orthogonal complement of $\bU$ in $\R^{\dim}$. Then, $\bPsi_k$ is assumed to respect:
\begin{align*}
&\bV^\top \bPsi_k \bV = \beta_k \bI_{\dim - \latentdim},  \\
&\bU^\top \bPsi_k \bU = \bm{0}_{\latentdim \times \latentdim} \, \textrm{ and } \, \bV^\top \bPsi_k \bU = \bm{0}_{(\dim - \latentdim) \times \latentdim}.
\end{align*}
These constraints amount to say that the covariance matrix $\bS_k$ is block diagonal after being rotated by $\bD$. In other terms, writing $\bDelta_k = \bD^\top \bS_k \bD$, the \ref{eq:DLM} model assumes that:
\begin{align*}
\bDelta_k = \diag(\bSigma_{k}, \beta_{k} \bI_{\dim - \latentdim}) = 
\left(  
\begin{array}{c@{}c} 
\begin{array}{|ccc|}
\hline 
~~ & ~~ & ~~ \\  
& \bSigma_k &  \\  
& & \\ 
\hline 
\end{array} 
& \mathbf{0}\\ 
\mathbf{0} &  
\begin{array}{|ccc|}
\hline 
~~ & ~~ & ~~ \\  
& &  \\  
& \beta_k \bI_{\dim - \latentdim} & \\ 
& & \\
\hline 
\end{array} 
%		\begin{array}{|cccc|}
%			\hline \beta & & & 0\\ & \ddots & 	&\\  & & \ddots &\\ 0 & & & \beta_k \\ \hline 
%		\end{array} 
\end{array}
\right)  
\begin{array}{cc} 
\left.\begin{array}{c} \\\\\\ \end{array}\right\}  & d \leq K-1\vspace{1.5ex}\\ 
\left.\begin{array}{c} \\\\\\\\ \end{array}\right\}  & (p-d)
\end{array}.
\end{align*}

This hypothesis implies that the discriminative subspace contains the relevant clustering information, while the noise variance lies in the orthogonal directions. Denoting $\bmu = (\bmu_k)$ and $\bBeta = (\beta_k)$, the model parameters are then denoted as $\globalparam = (\bPi, \bmu, \bSigma, \bU)$, and the Fisher-EM proposed by \textcite{bouveyron2012simultaneous} decomposes in two steps. First, an EM-step is used to maximize the observed-data log-likelihood with respect to $(\bPi, \bmu, \bSigma, \bBeta)$. The resulting posterior probabilities $\tau_{ik} = \p(\rawclust_{ik} = 1 \mid \obs_i, \globalparam)$ are then used to replace $\rawclust_{ik}$ in order to compute the \textit{soft} within and between-class covariance matrices $\bS_W$ and $\bS_B$. The Fisher-step chooses the discriminative space $\bU$ as the one maximizing the Fisher criterion in \Cref{eq:FisherCriterion} with orthonormality constraints. These two steps are iterated over until convergence of the likelihood or a maximum number of iteration is reached. However, while efficient, this algorithm displays some instabilities due to poor conditioning of the scatter matrices, and can get stuck in poor local maxima in terms of clustering as demonstrated in \Cref{bfem:sec:NumericalExpe}.

\subsection{A Bayesian formulation and the family of submodels}
\label{bfem:subsec:BDLM}
\begin{figure}[ht!]
	\centering
	\includegraphics[width=0.6\linewidth]{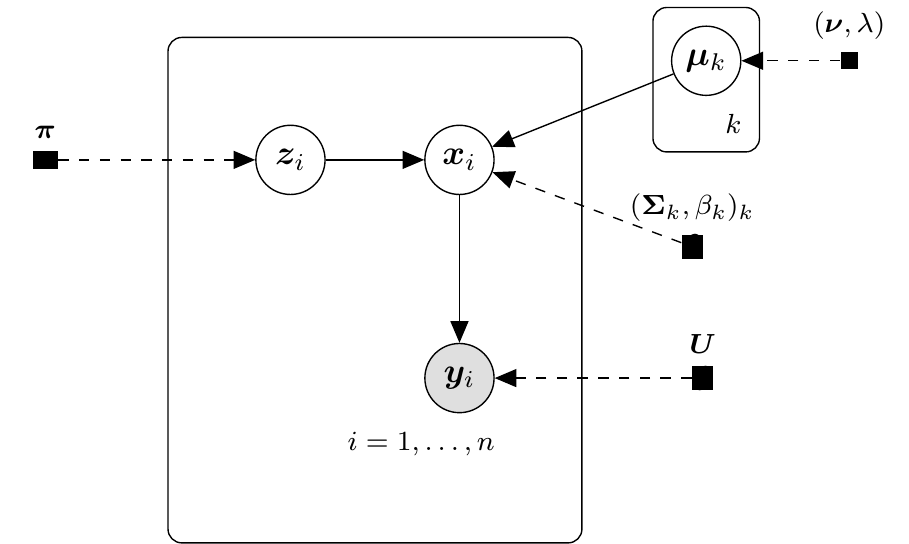}
	\caption{Graphical model representation of the Bayesian discriminative latent mixture model.}
	\label{bfem:fig:model-BFEM}
\end{figure}

We propose a Bayesian extension of the \ref{eq:DLM} model where a Gaussian prior distribution is put on $\bmu_k$ as in the standard Bayesian Gaussian mixture model:
\begin{equation}
\label{eq:BDLM}
\begin{aligned}
\bmu = (\bmu_k)_k, \quad  \bmu_{k} & \overset{i.i.d.}{\sim} \Gaussian_{ \latentdim}(\bnu, \lambda \bI_\latentdim), \\
\clust_i & \overset{\hphantom{i.i.d.}}{\sim}\Mult_{\K}(1, \bPi), \\
\scores_{i} \mid \{\rawclust_{ik} = 1\} &\overset{\hphantom{i.i.d.}}{\sim} \Gaussian_\latentdim(\bmu_{k}, \bSigma_{k}), \\
\obs_{i} & \overset{\hphantom{i.i.d.}}{=} \bU \scores_{i} + \bm{\epsilon}_i, & & \bm{\epsilon}_i \mid  \{\rawclust_{ik}=1\}  \sim \Gaussian_{\dim} (\bm{0}_{\dim}, \bPsi_k).
%\bm{\epsilon}_i \mid  \{\rawclust_{ik}=1\} & \sim \Gaussian_{\dim} (\bm{0}_{\dim}, \bPsi_k).
\end{aligned}
\tag{BDLM}
\end{equation}
%\begin{align}
%\p(\bmu \mid \bnu, \lambda) = \prod_{k=1}^{\K} \Gaussian_{ \latentdim}(\bmu_{k} \mid \bnu, \lambda \bI_\latentdim).
%\end{align}
Here, $\lambda$ is a hyper-parameter controlling the spreading of the $\bmu_{k}$'s in the latent space. The rest of the model and assumptions is unchanged and we refer to this Bayesian version as $\BDLM_{[\bSigma_{k}\beta_k]}$, which is represented as a graphical model in \Cref{bfem:fig:model-BFEM}. The set of parameters is then $\globalparam = (\bPi, \bSigma, \bU, \bBeta)$ of dimension
\begin{align*}
\dimparam = \K - 1 + \K \frac{\latentdim (\latentdim + 1)}{2} + \dim \latentdim - \frac{\latentdim (\latentdim+1)}{2} + \K,
\end{align*} 
and \Cref{bfem:sec:Inference} discusses inference and clustering.

Considering specific constraints on the matrix $\bDelta_k$, we can derive a family of submodels for the BDLM as in the original \ref{eq:DLM}. Akin to the spectral constraints of \textcite{banfield1993model}, we can assume a combination of hypotheses on the structure of the latent space covariance $\bSigma_k$ and on the noise covariance $\bPsi_k$, for a total of $12$ models. First homoscedasticity constraints of the type $\bSigma_k = \bSigma$ or $\bPsi_k = \bPsi$ may be considered, denoted as $\BDLM_{[\bSigma\beta_k]}$, $\BDLM_{[\bSigma_k\beta]}$, and $\BDLM_{[\bSigma\beta]}$ where the subscript $k$ denotes heteroscedasticity of the concerned parameter. Moreover, the covariance $\bSigma_k$ can be further assumed to be diagonal $\bSigma_{k} = \diag(\alpha_{k1}^2, \ldots, \alpha_{k\latentdim}^2)$ leaving $4$ possibilities depending on the homoscedasticity hypothesis denoted as $\BDLM_{[\alpha_{kh}\beta_k]}$, $\BDLM_{[\alpha_{h}\beta_k]}$, $\BDLM_{[\alpha_{kh}\beta]}$ and $\BDLM_{[\alpha_{h}\beta]}$. Finally, the latent covariance may be considered isotropic $\bSigma_k = \alpha^2_{k} \bI_{\latentdim}$, and the corresponding $4$ submodels are noted as $\BDLM_{[\alpha_{k}\beta_k]}$, $\BDLM_{[\alpha\beta_k]}$, $\BDLM_{[\alpha_{k}\beta]}$ and $\BDLM_{[\alpha\beta]}$. A comprehensive summary of these submodels, along with their number of free parameters is given in \Cref{bfem:tab:BDLMsubmodels}.
\begin{table}[!ht]
	\centering
	\renewcommand{\arraystretch}{1.5}\textsc{}
	\caption{The BDLM family of submodels with their associated number of free parameters, along with related model-based subspace clustering models. Here, $\omega = \K - 1 + \dim \latentdim - \frac{\latentdim (\latentdim+1)}{2}$ and the dimension of the latent space is fixed to $\latentdim = \K- 1$ in the last examples. A comparison with other Gaussian subspace clustering models in the same setting is available in \textcite[Table 3 to 6]{bouveyron2014model}}.
	\label{bfem:tab:BDLMsubmodels}
	\begin{tabular}{|m{0.25\linewidth}m{0.40\linewidth}m{0.15\linewidth}|}
		\hline
		Model & \centering{Number of free parameters $\dimparam$} & $\dim = 100$, $\K=4$  \\
		\hline
		Full-GMM & $\K-1 + \K \dim + \K \frac{\dim(\dim+1)}{2} $ & $20603$ \\
		Sphe-GMM & $\K-1 + 2\K \dim$ & $803$ \\
		\hline
		$\BDLM_{[\bSigma_k\beta_k]}$ & $\omega + \K \frac{\latentdim (\latentdim + 1)}{2} +  \K$ & $325$ \\
		$\BDLM_{[\bSigma_k \beta]}$ & $\omega + \K \frac{\latentdim (\latentdim + 1)}{2} +  1$ & $322$ \\
		$\BDLM_{[\bSigma \beta_k]}$ & $\omega + \frac{\latentdim (\latentdim + 1)}{2} +  \K$ & $307$ \\
		$\BDLM_{[\bSigma \beta]}$ & $\omega + \frac{\latentdim (\latentdim + 1)}{2} +  1$ & $304$ \\
		$\BDLM_{[\alpha_{kh}\beta_k]}$ & $\omega + \K \latentdim +  \K$ & $313$ \\
		$\BDLM_{[\alpha_{kh}\beta]}$ & $\omega + \K \latentdim +  1$ & $310$ \\
		$\BDLM_{[\alpha_{h}\beta_k]}$ & $\omega +  \latentdim +  \K$ & $304$ \\
		$\BDLM_{[\alpha_{h}\beta]}$ & $\omega +  \latentdim +  1$ & $301$ \\
		$\BDLM_{[\alpha_{k}\beta_k]}$ & $\omega +  \K +  \K$ & $305$ \\
		$\BDLM_{[\alpha_{k}\beta]}$ & $\omega +  \K +  1$ & $302$ \\
		$\BDLM_{[\alpha\beta_k]}$ & $\omega +  1 +  \K$ & $302$ \\
		$\BDLM_{[\alpha\beta]}$ & $\omega + 1 +  1$ & $299$ \\
		\hline 
		% 		 MCFA & $ $ & \\
		% 		 FMA & & \\
		% 		 PGMM (CUU) & & \\
		% 		 \hline
	\end{tabular}
\end{table}

\subsection{Link with Gaussian subspace clustering models}
\label{bfem:subsec:LinkWithMCFA}

As an extension of the DLM model, the BDLM model inherits its connections to the other model-based subspace clustering models. Indeed, they share the same \textit{global} linear Gaussian model formulations, with a common loading matrix $\bU$. The mixture of common factor analyzer \parencite[MCFA,][]{baek2009mixtures} model is the closest to the DLM model although the assumption on the nature of the subspace and the noise matrix $\bPsi_k$ are different. This model also shares deep ties with the heteroscedastic factor mixture analysis \parencite[HFMA,][]{montanari2010heteroscedastic} where the latent scores again follow a mixture of Gaussian distributions and a common loading matrix. However, the linear transformation $\bU$ is not constrained to be orthonormal here, and other identifiability constraints are put on $\bmu_{k}$ and $\bSigma_k$ such that the scores are standardized in the latent space. The PGMM family of \textcite{mcnicholas2008parsimonious} also contains 4 constrained models CUU, CCU, CUC and CCC, with common loadings across clusters $\bS_k = \bU \bU^\top + \bPsi_k$. Note that, in these models the matrix $\bU$ is not orthonormal and captures all the variance. As a consequence, the mixture is no longer in the latent space but in the observation one, which does not allow to put further constraints on $\bSigma_{k}$. Finally, \textcite{bouveyron2007high} proposed a family of 28 constrained Gaussian mixtures for high-dimensional data, with the decomposition $\bS_k = \bU_k \diag(\bLambda_{k}, \beta_{k} \bI_{\dim - \latentdim_k}) \bU_k^\top$ where $\bLambda_k = \diag(\lambda_{k1}, \ldots, \lambda_{k\latentdim_k})$. The submodels sharing the orientation matrices $\bU_k = \bU$ with same dimension $\latentdim_k = \latentdim$ are close to some BDLM submodels where $\bSigma_{k}$ is diagonal. However, once again, the subspace are not assumed to be discriminant and are estimate through maximum ikelihood estimation.

Related to our approach, a fully Bayesian extension of the MCFA model was proposed in \textcite{wei2013bayesian}, putting a Dirichlet prior on the mixture proportions $\bPi$, a standard Gaussian on each column of $\bU$ and a Gaussian-inverse-Wishart prior on $(\bmu,  \bSigma)$. However, the marginal distribution of $\scores_{i}$ in this model is now a mixture of Student t-distribution which differs from the mixture of Gaussian in our model. In addition, the factor loading matrix is no longer assumed to be column-orthonormal, and as in MCFA the subspace spanned by $\bU$ is not considered discriminant. The authors rely on a variational Bayes EM algorithm to approximate the posterior distribution of the parameters.
%\[
%(\bmu, \bSigma) \sim \prod_{k=1}^{\K} \Gaussian_{ \latentdim}(\bmu_{k} \mid \bnu_k, \lambda_k \bSigma_k) \InverseWishart_\latentdim(\bSigma_k \mid \bV_k, s_k),
%\]
%where $\InverseWishart$ denotes the inverse-Wishart distribution \parencite{robert2007bayesian}. Inference relies on a VEM algorithm. However, the distribution in the latent space is not the same in our model as one can see by looking at the conditional distribution of $\scores_{i} \mid \bSigma_{k}$, marginalizing on $\bmu_k$:
%\begin{align*}
%	 \scores_{i} \mid \bSigma_{k} \sim \sum_k \pi_k \Gaussian_{ \latentdim}(\bnu, \Sigma_{k} + \lambda \bI_{\latentdim}) \tag{BMCFA} \\
%	 (\textrm{BMCFA}) \;  \scores_{i} \mid \bSigma_{k} \sim \sum_k \pi_k \Gaussian_{ \latentdim}(\bnu, \bSigma_{k} + \lambda_k \bSigma_{k}) \tag{BMCFA}.
%\end{align*}
%Thus, equality of the two distributions is only achieved for $\BDLM_{[\alpha \beta]}$ with $\lambda_k = \lambda$ in the prior of BMCFA.

%The case of spectral constraints discussed in \Cref{chap2:subsec:ParsimoniousGMM} is also related. For the specific case of $\bS_k = \lambda_k \bD \bDelta_k \bD^\top$, maximum likelihood inference is hard and one needs to resort to numerical optimization \textcite{browne2014estimating} 

\section{Clustering with the Bayesian Fisher-EM algorithm}
\label{bfem:sec:Inference}
In the following, we propose a clustering algorithm based on the joint maximization of the Fisher criterion and the observed-data likelihood.  Contrary to the \ref{eq:DLM} model, the latter is intractable, and so is the posterior distribution of the latent variables $(\Clust, \bmu)$ given the data and the parameters. Thus, we propose to use variational inference, relying on a mean-field approximation of the posterior and the maximization of a lower bound of the log-likelihood \parencite{jaakkola2000bayesian}. First, we give the optimal form of the mean-field approximation in the variational E-step (VE-step). Then, we derive the expression of the variational lower bound as well as the M-step updates maximizing the latter with respect to the mixture parameters $(\bPi, \bSigma, \bBeta)$. Finally, following \textcite{bouveyron2012simultaneous}, we propose to choose $\bU$ as the best current discriminative subspace maximizing the Fisher criterion at each iteration. Thus, the proposed clustering algorithm for BDLM is named Bayesian Fisher-EM (BFEM) and alternates between $3$ steps:
\begin{itemize}
	\item The VE-step which finds an approximation of the posterior $\p(\Clust, \bmu \mid \Obs , \globalparam)$ in the mean-field family,
	\item The M-step where the latent space mixture parameters are estimated by maximizing the variational lower bound,
	\item The F-step where $\bU$ is chosen to maximize the current variational Fisher criterion.
\end{itemize}

%\subsection{Intractable posterior}

%\begin{equation}
%\p(\Obs,\bmu \mid \globalparam) = \p(\bmu \mid \bnu, \lambda) \times \prod_{i=1}^{\nb} \sum_{k=1}^K \pi_k \Gaussian_\dim(\obs_i , \bU\bmu_k, \bS_k).
%\end{equation}
%The first term of this product is the prior introduced in the new version, while the second one is the likelihood of the \ref{eq:DLM} model. 

\subsection{Variational approximation}
\label{bfem:subsec:VEM}

Similarly to the Bayesian formulation of standard GMM, the observed-data likelihood is no longer tractable. Indeed, the latter is written as:
\begin{align}
\label{eq:BFEMllhood}
\p(\Obs \mid \globalparam) &= \int_{\bmu} \p(\bmu) \prod_{i=1}^{\nb} \sum_{k=1}^K \pi_k \Gaussian_\dim(\obs_i \mid \bU\bmu_k, \bS_k) \dif \bmu .
\end{align}
Unfortunately, each $\bmu_k$ now appears in all $n$ factors of the integrand. Thus, the integral does not reduce to products and sums of $d$-dimensional trivial integrals over $\bmu_k$. Another way of seeing the difficulty is to swap the integrals over $\Clust$ and $\bmu$, leaving:
\begin{equation*}
\begin{aligned}
\p(\Obs \mid \globalparam) 
%	&= \sum_{\Clust} \int_{\bmu} \p(\Obs \mid \Clust, \bmu) \p(\Clust) 
=&  \sum_{\Clust} \p(\Clust) \prod_{k=1}^{\K} \int_{\bmu_k}  \p(\bmu_k)  \prod_{i : \rawclust_{ik} = 1} \Gaussian_\dim(\obs_i \mid \bU\bmu_k, \bS_k) \dif \bmu_k .
\end{aligned}
\end{equation*}
Now each integral may be computed thanks to Gaussian conjugacy. However, there are $\K^\nb$ possible configurations to sum over, which is not computationally feasible. 

Besides, the posterior $\p(\Clust, \bmu \mid \Obs)$ is not tractable either, and is only known up to its normalizing constant in \Cref{eq:BFEMllhood}, which prevents from calculating its moments. This fact is well known in Bayesian treatment of mixtures, and leads to either MCMC algorithms or approximate inference \parencite[section 5.2]{fruhwirth2019handbook}. 
%Here, we rely on the latter and derive the updates for the standard coordinate ascent variational inference algorithm \parencite[CAVI,][]{blei2017variational}. Introducing the mean-field approximation of the posterior as:
%\begin{align}
%\q(\bmu, \Clust) = \prod_{k=1}^{\K} \q(\bmu_k) \prod_{i=1}^{\nb} \q(\clust_{i}),
%\end{align}
%we aim at maximizing the following variational lower bound of the log-likelihood:
%\begin{align}
%\log \p(\Obs \mid \globalparam) \geq \J(\q, \globalparam) = \Expectation_{\q}\left[ \log \p(\Obs, \Clust, \bmu \mid \globalparam)\right] + \entropy(\q),
%\end{align}
%where $\entropy(\q)$ denotes the entropy of distribution $\q$. It is a well-known fact that maximizing $\J$ with respect to $\q$ amounts to minimize the Kullback-Leibler (KL) divergence between $\q$ and the posterior.   The CAVI algorithm consists in sequentially optimizing with respect to each individual distribution in $\q$, keeping the other fixed and cycling over until convergence. The following propositions give the optimal form of the CAVI updates for the BDLM model. Notice that, while no specific functional form were assumed on $\q$, the optimal distributions are $\q^\star(\clust_i)$ and $\q^\star(\bmu_k)$ happens to be multinomial and Gaussian as it is often the case with mean-field inference \parencite{blei2017variational}. In the VE-step, the updates of \Cref{eq:bfem:CAVIupdateZ,eq:bfem:CAVIupdateMu} are cycled over until a local maximum of $\J(\q)$ is reached.
Here, we rely on the latter and introducing a \textit{variational} distribution $\q(\bmu, \Clust)$, the classical identity \parencite[see \textit{e.g.}][Equation (14)]{blei2017variational} holds for any $\q$:
\begin{equation}
	\label{eq:VariationalIdentity}
	\log \p(\Obs \mid \globalparam) = \J(\q, \globalparam) + \KL{\q}{\p(\cdot \mid \Obs, \globalparam)} \geq  \J(\q, \globalparam),
\end{equation}
with
\begin{equation}
	\label{eq:LowerBound}
	 \J(\q, \globalparam) = \Expectation_{\q}\left[ \log \p(\Obs, \bmu , \Clust\mid \globalparam)\right] - \Expectation_\q\left[ \log \q(\bmu, \Clust)\right].
\end{equation}
Here, $\KLy$ denotes the Kullback-Leibler divergence between the variational distribution $\q$ and the posterior $\p(\cdot \mid \Obs, \globalparam)$:
\begin{equation*}
	\KL{\q}{\p(\cdot \mid \Obs, \globalparam)} = -  \int_{\bmu} \sum_{\Clust} \q(\bmu, \Clust) \log \frac{\p(\bmu, \Clust \mid \Obs, \globalparam)}{\q(\bmu, \Clust)} \dif \bmu  .
\end{equation*}
The latter being non-negative, \Cref{eq:LowerBound} is a lower bound of the observed-data log-likelihood in the left-hand side of \Cref{eq:VariationalIdentity}. Moreover, maximizing the bound with respect to $\q$ is equivalent to minimizing the $\KLy$ divergence of the latter to the posterior, the bound being tight when the two distributions are equal. However, the posterior being intractable, the $\KLy$ minimization problem is restricted to a certain family of distributions, hence the \textit{approximation} terminology. Here, we propose to rely on the classical mean-field variational family which posits a fully factorized distribution:
\begin{align}
\q(\bmu, \Clust) = \prod_{k=1}^{\K} \q(\bmu_k) \prod_{i=1}^{\nb} \q(\clust_{i}).
\end{align}
The main interest of the mean-field approximation is that it comes with the coordinate ascent variational inference \parencite[CAVI,][]{bishop2006pattern,blei2017variational} algorithm. The latter consists in sequentially optimizing the lower bound with respect to each individual distribution $\q(\clust_i)$ and $\q(\bmu_k)$ while keeping the other fixed, and to cycle over until convergence is reached like a fixed point algorithm. 

 The following propositions give the optimal form of the CAVI updates for the BDLM model. Notice that, while no specific functional forms were assumed on $\q$, the optimal distributions $\q^\star(\clust_i)$ and $\q^\star(\bmu_k)$ happen to be multinomial and Gaussian as it is often the case with mean-field variational inference \parencite{blei2017variational}. In the VE-step, the updates of \Cref{eq:bfem:CAVIupdateZ,eq:bfem:CAVIupdateMu} are cycled over until a local maximum of $\J(\q)$ is reached.
\begin{proposition}[Proof in \Cref{bfem:appendix:ProofCAVIupdateZ}] 
	\label{bfem:prop:CAVIupdateZ}
	The coordinate update for the variational distribution $\q(\clust_i)$ is
	\begin{align}
	\label{eq:bfem:CAVIupdateZ}
	\q^\star(\clust_i) &= \Mult_{\K}(\clust_i \mid 1, \bTau_i),
	\end{align}
	with $\forall i, k,$
	\begin{align*}
	\tau_{ik} & \propto \pi_k \exp \left\{ \Expectation_{\q^\star(\bmu_k)}\left[ \log \mathcal{N}_p(\obs_i \mid \bU \mu_k, \bS_k)\right]\right\}.
	\end{align*}
\end{proposition}
\begin{proposition}[Proof in \Cref{bfem:appendix:ProofCAVIupdateMu}] 
	\label{bfem:prop:CAVIupdateMu}
	The coordinate update for the variational distribution $\q(\bmu_k)$ is
	\begin{align}
	\label{eq:bfem:CAVIupdateMu}
	\q^\star(\bmu_k) &= \Gaussian_{ \latentdim}(\bmu_k \mid \bvarmean_k, \bvarcovar_k),
	\end{align}
	with $\forall k$,
	\begin{align*}
	\varn_k &= \sum_{i} \tau_{ik},\\
	\bvarcovar_k &= \left(\lambda^{-1} \bI_\latentdim + \varn_k \bSigma_{k}^{-1}\right)^{-1}, \\
	\bvarmean_k	&= \bnu + \bvarcovar_k \bSigma_k^{-1} \left(\bU^\top (\sum_i \tau_{ik} \obs_i ) - \varn_k \bnu \right).
	\end{align*}
\end{proposition}
Finally, note that the expression of $\tau_{ik}$ in \Cref{eq:bfem:CAVIupdateZ} involves an expectation in the observation space that may be reworked in order to avoid inverting the $\dim \times \dim$ matrix $\bS_k$. This is done by taking advantage of the specific block structure of the latter, hence only relying on the inverse of the $\latentdim \times \latentdim$ matrix $\bSigma_{k}$:
\begin{equation}
\begin{aligned}
%	\Expectation_{\q^\star(\bmu_k)}\left[ \log \mathcal{N}_p(\obs_i \mid \bU \mu_k, \bS_k) \right] = & - \frac{1}{2} \bigg\{\dim \log(2 \pi) + \log \vert \bS_k \vert + (\obs_i - \bU \bvarmean_{k})^\top \bU \bSigma_{k}^{-1} \bU^\top (\obs_i - \bU \bvarmean_{k}) \\
%	& \qquad \qquad + \frac{1}{\beta_k} (\Vert \obs_i \Vert_2 - \Vert \bU^\top \obs_i \Vert_2) + \Tr\left[\bvarcovar_k \bSigma_{k}^{-1}\right]\bigg\}.
\tau_{ik} & \propto \pi_k \exp \Bigg\{ \Expectation_{\bmu_k}\left[ \log \mathcal{N}_p(\obs_i \mid \bU \mu_k, \bS_k)\right] \Bigg\}, \\
& \propto  \pi_k \exp \Bigg\{ - \frac{1}{2} \bigg(\dim \log(2 \pi) + \log \vert \bS_k \vert + (\obs_i - \bU \bvarmean_{k})^\top \bU \bSigma_{k}^{-1} \bU^\top (\obs_i - \bU \bvarmean_{k}) \\
& \qquad \qquad \qquad \qquad \qquad \qquad+ \frac{1}{\beta_k} (\Vert \obs_i \Vert_2 - \Vert \bU^\top \obs_i \Vert_2) + \Tr\left[\bvarcovar_k \bSigma_{k}^{-1}\right] \bigg) \Bigg\},
\end{aligned}
\end{equation}
and
\begin{align*}
	\log \vert \bS_k \vert = \log \vert \bSigma_{k} \vert + (\dim -\latentdim) \log(\beta_k).
\end{align*}

\subsection{The M-step}
\label{bfem:subsec:Mstep}

In the M-step, the bound of \Cref{eq:LowerBound} is maximized with respect to the latent space mixture parameters $(\bPi, \bSigma, \bBeta)$. Note that $\bU$ is treated as a fixed, distinct parameter here, which will be dealt with in the next section. The following proposition gives the form of the lower bound as a function of $\globalparam$.
\begin{proposition}[Proof in \Cref{bfem:appendix:ProofELBO}]
	\label{bfem:prop:ELBO}
	In model $\BDLM_{[\bSigma_{k}\beta_{k}]}$, the variational lower bound as a function of $\globalparam$ may be written as:
	\begin{equation}
	\label{bfem:eq:ELBO}
	\begin{aligned}
	\J(\globalparam) = & \cst - \frac{1}{2} \sum_{k=1}^{\K} \varn_k \Bigg\{ -2 \log(\pi_k) + \log \vert \bSigma_{k} \vert + (\dim - \latentdim) \log(\beta_{k})   \\
	& \quad +  \Tr\left[ \bSigma_k^{-1} \bU^\top \bhatC_k \bU\right] + \frac{1}{\beta_k} \left( \Tr\left[\bhatC_k\right] -  \Tr\left[ \bU^\top \bhatC_k \bU \right] \right)  \Bigg\} ,
	\end{aligned}
	\end{equation}
	where 
	%	\[
	%	\bC_k = \frac{1}{\varn_k} \sum_{i=1}^{\nb} \tau_{ik} (\obs_{i} - \bU \bmu_k)(\obs_{i} - \bU \bmu_{k})^\top.
	%	\]
	\[
	\bhatC_k = \frac{1}{\varn_k} \sum_{i=1}^{\nb} \tau_{ik} (\obs_{i} - \bU\bvarmean_k)(\obs_{i} - \bU \bvarmean_{k})^\top + \bU \bvarcovar_k \bU^\top.
	\]
\end{proposition}

At iteration $(t)$, in the M-step, the mixture proportions are estimated classically as in other mixture models:
\begin{align}
\label{bfem:eq:updatePI}
\hat{\pi}_k^{(t)} = \frac{\varn_k^{(t)}}{\nb}.
\end{align}
The remaining parameters $(\bSigma_{k}, \beta_{k})$ depend on the chosen submodel and the following proposition details the estimates corresponding to each of the $12$ submodels.
\begin{proposition}[Proof in \Cref{bfem:appendix:ProofMstep}]
	\label{bfem:prop:M-step}
	The M-step estimates for $\bSigma_{k}$ and $\beta_{k}$ at iteration $(t)$ are:
	\begin{itemize}
		\item Model $\BDLM_{[\bSigma_{k}\beta_k]}$:
		\begin{align}
		\hat{\bSigma}_{k}^{(t)} = {\bU}^\top \bhatC_k^{(t)} \bU ,  & & \hat{\beta}_k^{(t)} = \frac{\Tr\left[\bhatC_k^{(t)}\right] -  \Tr\left[ {\bU}^\top \bhatC_k^{(t)} \bU \right]}{\dim - \latentdim}  .
		\end{align}
		
		\item Model $\BDLM_{[\bSigma_{k}\beta]}$:
		\begin{align}
		\hat{\bSigma}_{k}^{(t)} = {\bU}^\top \bhatC_k^{(t)} \bU ,  & & \hat{\beta}^{(t)} = \frac{\Tr\left[\bhatC^{(t)}\right] -  \Tr\left[ {\bU}^\top \bhatC^{(t)} \bU \right]}{\dim - \latentdim}  .
		\end{align}
		
		\item Model $\BDLM_{[\bSigma\beta_k]}$:
		\begin{align}
		\hat{\bSigma}^{(t)} = {\bU}^\top \bhatC^{(t)} \bU ,  & & \hat{\beta}_k^{(t)} = \frac{\Tr\left[\bhatC_k^{(t)}\right] -  \Tr\left[ {\bU}^\top \bhatC_k^{(t)} \bU \right]}{\dim - \latentdim}  .
		\end{align}
		
		\item Model $\BDLM_{[\bSigma\beta]}$:
		\begin{align}
		\hat{\bSigma}^{(t)} = {\bU}^\top \bhatC^{(t)} \bU ,  & & \hat{\beta}^{(t)} = \frac{\Tr\left[\bhatC^{(t)}\right] -  \Tr\left[ {\bU}^\top \bhatC^{(t)} \bU \right]}{\dim - \latentdim}  .
		\end{align}

		\item Model $\BDLM_{[\alpha_{kh}\beta_k]}$:
		\begin{align}
		\hat{\alpha}_{kh}^{(t)} = \bu_h^\top \bhatC_k^{(t)} \bu_h ,  & & \hat{\beta}_k^{(t)} = \frac{\Tr\left[\bhatC_k^{(t)}\right] -  \Tr\left[ {\bU}^\top \bhatC_k^{(t)} \bU \right]}{\dim - \latentdim}  .
		\end{align}
		
		\item Model $\BDLM_{[\alpha_{kh}\beta]}$:
		\begin{align}
		\hat{\alpha}_{kh}^{(t)} = \bu_h^\top \bhatC_k^{(t)} \bu_h ,  & & \hat{\beta}^{(t)} = \frac{\Tr\left[\bhatC^{(t)}\right] -  \Tr\left[ {\bU}^\top \bhatC^{(t)} \bU \right]}{\dim - \latentdim}  .
		\end{align}
		
		\item Model $\BDLM_{[\alpha_{h}\beta_k]}$:
		\begin{align}
		\hat{\alpha}_{h}^{(t)} = \bu_h^\top \bhatC^{(t)} \bu_h ,  & & \hat{\beta}_k^{(t)} = \frac{\Tr\left[\bhatC_k^{(t)}\right] -  \Tr\left[ {\bU}^\top \bhatC_k^{(t)} \bU \right]}{\dim - \latentdim}  .
		\end{align}
		
		\item Model $\BDLM_{[\alpha_{h}\beta]}$:
		\begin{align}
		\hat{\alpha}_{h}^{(t)} = \bu_h^\top \bhatC^{(t)} \bu_h ,  & & \hat{\beta}^{(t)} = \frac{\Tr\left[\bhatC^{(t)}\right] -  \Tr\left[ {\bU}^\top \bhatC^{(t)} \bU \right]}{\dim - \latentdim}  .
		\end{align}
		
		\item Model $\BDLM_{[\alpha_k \beta_k]}$:
		\begin{align}
		\hat{\alpha}_{k}^{(t)} = \frac{1}{\latentdim} \Tr \left[{\bU}^\top \bhatC_k^{(t)} \bU \right] ,  & & \hat{\beta}_k^{(t)} = \frac{\Tr\left[\bhatC_k^{(t)}\right] -  \Tr\left[ {\bU}^\top \bhatC_k^{(t)} \bU \right]}{\dim - \latentdim}  .
		\end{align}
		
		\item Model $\BDLM_{[\alpha_k \beta]}$:
		\begin{align}
		\hat{\alpha}_{k}^{(t)} = \frac{1}{\latentdim} \Tr \left[{\bU}^\top \bhatC_k^{(t)} \bU \right] ,  & & \hat{\beta}^{(t)} = \frac{\Tr\left[\bhatC^{(t)}\right] -  \Tr\left[ {\bU}^\top \bhatC^{(t)} \bU \right]}{\dim - \latentdim}  .
		\end{align}
		
		\item Model $\BDLM_{[\alpha \beta_k]}$:
		\begin{align}
		\hat{\alpha}^{(t)} = \frac{1}{\latentdim} \Tr \left[{\bU}^\top \bhatC^{(t)} \bU \right] ,  & & \hat{\beta}_k^{(t)} = \frac{\Tr\left[\bhatC_k^{(t)}\right] -  \Tr\left[ {\bU}^\top \bhatC_k^{(t)} \bU \right]}{\dim - \latentdim}  .
		\end{align}
		
		\item Model $\BDLM_{[\alpha \beta]}$:
		\begin{align}
		\hat{\alpha}^{(t)} = \frac{1}{\latentdim} \Tr \left[{\bU}^\top \bhatC^{(t)} \bU \right] ,  & & \hat{\beta}^{(t)} = \frac{\Tr\left[\bhatC^{(t)}\right] -  \Tr\left[ {\bU}^\top \bhatC^{(t)} \bU \right]}{\dim - \latentdim}  .
		\end{align}
		
	\end{itemize}
	
	Here, $\bu_h$ denotes the $h$-th column of $\bU$ which is computed in the F-step at iteration $(t)$ and:
	\begin{align}
	\bhatC_k^{(t)} = &\frac{1}{\varn_k^{(t)}} \sum_{i=1}^{\nb} \tau_{ik}^{(t)} (\obs_{i} - \bU\bvarmean_k^{(t)})(\obs_{i} - \bU \bvarmean_{k}^{(t)})^\top + \bU \bvarcovar_k^{(t)} {\bU}^\top, \\
	\bhatC^{(t)} = & \frac{1}{\nb}  \sum_{k=1}^{\K} \varn_k^{(t)}  \bhatC_k^{(t)}.
	\end{align}
	
\end{proposition}

\subsection{The Fisher step}
\label{bfem:subsec:Fisherstep}

As explained above, the subspace $\bU$ is supposed to be discriminative in the sense of the Fisher criterion. The partition $\Clust$ being unknown, the scatter matrices in \Cref{eq:FisherCriterion} cannot be formed. Following \textcite{bouveyron2012simultaneous} we propose to replace them by the soft within and between-class scatter matrices:
\begin{align*}
\tilde{\bmean}_k^{(t)}  &= \frac{1}{\varn_k^{(t)} } \sum_{i=1}^{\nb} \tau_{ik}^{(t)}  \obs_{i}, \\
\btildeS_W^{(t+1)} & = \frac{1}{\nb} \sum_{k=1}^{\K} \frac{1}{\varn_k^{(t)}} \sum_{i=1}^{\nb} \tau_{ik}^{(t)}  \left(\obs_{i} - \tilde{\bmean}_k^{(t)} \right) \left(\obs_{i} - \tilde{\bmean}_k^{(t)} \right)^\top , \\
\btildeS_B^{(t+1)} & = \frac{1}{\nb} \sum_{k=1}^{\K} \varn_k^{(t)}   \left(\tilde{\bmean}_k^{(t)}  - \bar{y}\right) \left(\tilde{\bmean}_k^{(t)}  - \bar{y}\right)^\top.
\end{align*}
Note that these matrices only involve the variational distribution of $\Clust$, although the latter also depends on $\q^{(t)}(\bmu)$ through the fixed point algorithm of the VE-step. Moreover, we recover the classical identity of linear discriminant analysis $\bS_T = \bS_W^{(t)}  + \bS_B^{(t)}$ at any iteration $(t)$, where $\bS_T = (1/\nb) \sum_{i=1}^{\nb} (\obs_{i} - \bar{y}) (\obs_{i} - \bar{y})^\top$ is the sample covariance matrix which does not depend on the clustering and is constant throughout the algorithm. 

Then, $\bU$ is supposed to maximize the following criterion:
\begin{align}
\label{bfem:eq:VariationalFisher}
\bU^{(t)} = \argmax_{\bU^\top \bU = \bI_{\latentdim}} \Fisher(\bU) = \Tr\left[(\bU^\top \bS_T \bU)^{-1} \bU^\top \btildeS_B^{(t)} \bU\right].
\end{align}
This criterion is slightly different from the one in \Cref{eq:FisherCriterion} since $\btildeS_W^{(t)}$ has been replaced by $\bS_T$. Working this this criterion is justified using the identity above since the problems of minimizing $\Tr[\bU^\top\bS_W\bU]$ or $\Tr[\bU^\top \bS_T\bU]$ are the same \parencite[chap. 10]{fukunaga1990introduction}. It is often used in practice \parencite{ye2005characterization}, and computationally efficient in this case since $\bS_T$ and its inverse need to be computed only once at the beginning of the algorithm. 

Without the orthonormality constraints, the problem in \Cref{bfem:eq:VariationalFisher} is directly solved by taking the leading $\latentdim$ eigenvectors of the generalized eigenvalue problem $\btildeS_B^{(t)} \bu_h = \gamma_h \bS_T\bu_h$  \parencite{ghojogh2019eigenvalue}. If $\bS_T$ is invertible, this can be done by computing the $d$ leading eigenvectors of $\bS_T^{-1} \btildeS_B^{(t)}$. However, since $\bS_T^{-1} \btildeS_B$ is not necessarily symmetric, the solution is not orthonormal with respect to the regular scalar product, but rather verifies $\bU^\top \bS_T \bU = \bI_{\latentdim}$. Unfortunately, there is no direct solution for the problem of \Cref{bfem:eq:VariationalFisher} with the constraint $\bU^\top \bU = \bI_{\latentdim}$. In the supervised context, algorithms have been derived to solve this problem which is called orthogonal LDA (OLDA). \textcite{foley1975optimal} proposed an iterative algorithm to successively find $\bu_1, \ldots, \bu_d$ in the $2$-class problem. It was later generalized for arbitrary values of $\K$ by \textcite{okada1985optimal}, and coined the orthonormal discriminant vectors (ODV) by \textcite{hamamoto1991note}. Note that simultaneous algorithms also exist to optimize with respect to $\bU$, based on successive eigen and QR-decompositions of carefully designed matrices \parencite[see.][]{ye2005characterization, lu2016new}. 

Relying on the ODV method, \textcite{bouveyron2012simultaneous} proposed an iterative algorithm starting from $\bu_1$, the leading eigenvector of $\bS_T^{-1}\btildeS_B^{(t)}$, and greedily maximizing the criterion by computing the $r$-th direction as the solution of the unconstrained problem in the orthogonal of the current subspace $\mathcal{B}_{r-1} = \vect(\bu_1, \ldots, \bu_{r-1})$.
%\begin{align*}
%	\frac{\bu_r^\top \btildeS_B^{(t)} \bu_r}{\bu_r^\top \bS_T \bu_r}, \quad \text{s.t.} \; \Vert \bu_r \Vert_2 = 1, \; \bu_r^\top \bu_h = 0, \forall h < r.
%\end{align*}
%\begin{align*}
%	\Tr\left[\bU^\top \bU = \bI_{\latentdim}} F_1(\bU) = \Tr\left[(\bU^\top \bS_T \bU)^{-1} \bU^\top \btildeS_B^{(t)}\right]
%\end{align*}
An orthogonal basis $V_r = (\bv_r, \ldots, \bv_\dim)$ of $\mathcal{B}_{r-1}^\bot$ can be found by the Gram-Schmidt procedure:
\begin{align*}
\bv_l = \alpha_l (\bI_{\dim} - \sum_{l'=1}^{l - 1} \bv_{l'}\bv_{l'}^\top) \psi_l, \quad \forall l = r, \ldots, \dim.
\end{align*}
where $\bv_l = \bu_l$ for $l=1,\ldots, r-1$, $\alpha_l$ is a normalization constant such that $\Vert \bv_l \Vert_2 = 1$ and $\psi_l$ are linearly independent vectors of $\bu_1, \ldots, \bu_{r-1}$. Then, the matrix $\bP_r = (\bv_r, \ldots, \bv_\dim)$ is used to project the scatter matrix in the orthogonal subspace $\mathcal{B}_{r-1}^\bot$:
\begin{align*}
\bS_{Tr} = \bP_r^\top \bS_T \bP_r, \\
\btildeS_{Br}^{(t)} = \bP_r^\top \btildeS_B^{(t)} \bP_r.
\end{align*}
Finally, the leading eigenvector $\ba_r$ of the generalized eigenvalue problem $\btildeS_{Br}^{(t)} \ba_r = \gamma_r \bS_{Tr} \ba_r$ is computed, and the $r$-th discriminant vector is chosen as 
\begin{align}
\bu_r = \frac{\bP_r \ba_r}{\Vert \ba_r \Vert_2}.
\end{align}
Thus, $\bu_r$ meets the constraints $\bu_r^\top \bu_h=0, \forall h < r$. This iterative procedure is repeated until $r=d$ discriminant vectors are found. 

\subsection{Estimation of the hyper-parameters $\bnu$ and $\lambda$}
The hyper-parameters $(\lambda, \bnu)$ may be set by the user and kept fixed during the whole procedure. For instance, when the data is centered, $\bar{y} = 0$, then $\bar{x} = \bU^\top \bar{y} = 0$ thus $\bnu$ could be set to $\bm{0}_\latentdim$. However, $\lambda$ controls the variance of $\bmu_{k}$ and setting it by hand can lead to poor performances. On the one hand, a too small value would not allow the space to be discriminant. On the other hand, when $\lambda \to +\infty$, the prior becomes non-informative. A quick asymptotic analysis of the variational distribution $\q(\bmu_{k})$ of \Cref{bfem:prop:CAVIupdateMu} confirms this as:
\begin{align*}
\bvarcovar_k \xrightarrow[\lambda \to +\infty]{} \frac{1}{\varn_k} \bSigma_{k},& &
\bvarmean_{k} \xrightarrow[\lambda \to +\infty]{} \bSigma_{k} \bSigma_{k}^{-1}  \frac{1}{\varn_k} \textstyle \sum_i \tau_{ik} \bU^\top \obs_{i} = \hat{\bmu}_k^{DLM}. 
\end{align*}
Thus, the variational posterior mean becomes the maximum-likelihood estimate of $\bmu_{k}$ in the frequentist formulation of BDLM. Under the hypothesis that $\varn_k \to +\infty$ as $n \to + \infty$, the variational approximation of the posterior becomes a Dirac mass at $\hat{\mu}_k^{DLM}$. This fact is somewhat similar to the well-known behavior of the posterior in Bayesian formulations of ridge regression when the prior becomes vague \parencite[p. 153]{bishop2006pattern}.

Here, we propose a parametric empirical Bayes approach \parencite{morris1983parametric}, using the variational bound as a surrogate for the type-II likelihood as it is commonly done in other well-known hierarchical Bayesian models \parencite{blei2003latent,airoldi2008mixed}.  The following proposition gives the form of the empirical Bayes estimates $(\hat{\bnu}, \hat{\lambda})$ maximizing $\J(\bnu, \lambda) \leq \log \p(\Obs \mid \globalparam, \bnu, \lambda)$.
\begin{proposition}[Proof in \Cref{bfem:appendix:ProofEmpiricalBayes}]
	\label{bfem:prop:EmpiricalBayes}
	The following updates maximize the variational lower bound with respect to $(\bnu ,\lambda)$:
	
	\begin{align}
	\hat{\bnu} &= \frac{\sum_{k=1}^{K} \bvarmean_k }{K}, \\
	\hat{\lambda} &= \frac{\sum_{k=1}^{\K}  \Vert \bvarmean_k - \hat{\bnu} \Vert_2^2 + \Tr\left[\bvarcovar_k \right]}{\latentdim \K}
	\end{align}
\end{proposition}

\subsection{Stopping criterion and properties}
Starting from a subspace $\bU^{(0)}$, the BFEM algorithm iterates over the VE-step, M-step and F-step updates, in this order. The algorithm is described in \Cref{bfem:alg:BFEM} and this section discusses initialization, convergence and useful properties of the algorithm. Let us begin by discussing the link to the original FEM algorithm.

\paragraph{Link to the original Fisher-EM} The proposed algorithm is largely inspired by the original FEM algorithm. However, note that the M-step updates of \textcite{bouveyron2012simultaneous} use: 
\begin{align}
\btildeC_k = \frac{1}{\varn_k} \sum_{i=1}^{\nb} \tau_{ik} (\obs_{i} - \tilde{\bmean}_k) (\obs_{i} - \tilde{\bmean}_k)^\top, \quad \text{with: } 	\tilde{\bmean}_k  &= \frac{1}{\varn_k } \sum_{i=1}^{\nb} \tau_{ik}  \obs_{i}.
\end{align}
The latter does not exactly correspond to the matrix $\bhatC_k$ of \Cref{bfem:prop:ELBO,bfem:prop:M-step}, since it uses $\tilde{\bmean}_k$ instead of $\hat{\bmean}_k = \bU \hat{\bmu}_k$. In particular, this has the consequence that the matrix $\btildeC_k$ does not directly depend on $\bU$, whereas it does in $\bhatC_k$. Therefore, our algorithm computes the true optimal updates in the M-step, while the FEM algorithm relies on the approximation $\bhatC_k \approx \btildeC_k$.

%\paragraph{Convergence} \textcite{bouveyron2012theoretical} showed that in the case $DLM_{[\alpha \beta]}$, the F-step in an M-step in $\bU$. However, preuve fausse je pense, nottament à cause de cette différence $\bhatC_k$ et $\btildeC_k$.

\paragraph{Convergence and stopping criterion} Since the F-step does not maximize the variational bound with respect to $\bU$, the latter is no longer monotonically increasing. This is also the case for the original FEM algorithm, hence we propose to rely on the same stopping criterion: Aitken's accelerated criterion \parencite[p. 145]{mclachlan2007algorithm}. The latter was introduced as an acceleration method for EM when the sequence of likelihood is linearly converging. Here, we replace the likelihood sequence with the variational bound $\{\J^{(t)} \}_t$. Then, define the Aitken accelerated estimate of $\J^\star$ is defined for $t \geq 2$ as:
\begin{align}
l_A^{(t+1)} = \J^{(t)} + \frac{1}{1- c^{(t+1)}} \left(\J^{(t+1)} - \J^{(t)} \right), \quad \text{with: } c^{(t+1)} = \frac{\J^{(t+1)} - \J^{(t)}}{\J^{(t)} - \J^{(t-1)}}.
\end{align}
The stopping criterion is defined as $\vert l_A^{(t+1)} - l_A^{(t)}\vert < \epsilon $, where $\epsilon$ is a user-defined tolerance parameter. Since there is no guarantee that the sequence $\{\J^{(t)}\}_t$ is increasing here, a maximum number of iterations is also provided by the user as an alternative stopping criterion, which is always done in standard implementation of VEM algorithms anyway.
%\Cref{bfem:sec:NumericalExpe} compares the performance of the two algorithms

Another possible stopping condition is the absolute change of the Fisher criterion of \Cref{bfem:eq:VariationalFisher} between two successive F-step: $ \vert \Fisher(\bU^{(t+1)}) - \Fisher(\bU^{(t)}) \vert \mathbin{/} \vert  \Fisher(\bU^{(t)}) \vert$. The latter was shown to have good performance for clustering applications \parencite{bouveyron2012theoretical}. 

\paragraph{Initialization} The BFEM algorithm needs an initial variational distribution $\q^{(0)}$ defined by its starting variational parameters $(\bTau, \bvarmean, \bvarcovar)$, an initial set of parameters $(\bSigma, \bBeta, \bPi)$ and an initial subspace $\bU^{(0)}$. We recommend initializing by setting $\bTau^{(0)} = \Clust^{(0)}$, a partition obtained by any suitable clustering algorithm, \textit{e.g.} random or $k$-means partitions. Then, the matrix $\btildeS_B^{(0)}$ can be formed to solve the problem in \Cref{bfem:eq:VariationalFisher}, giving an initial $\bU^{(0)}$. Next, the initial parameters $(\bSigma, \bBeta, \bPi)$ are obtained by using the frequentist M-step of \textcite{bouveyron2012simultaneous} with $\bU^{(0)}$. The remaining variational parameters $(\bvarmean_{k}, \bvarcovar_k)$ can then be set using \Cref{bfem:prop:CAVIupdateMu} with $\bTau^{(0)}$, $\bU^{(0)}$ and $\globalparam^{(0)}$. As for the hyper-parameters, we initialize $\bnu$ as $\bnu^{(0)} = (1/\nb) \sum_{i=1}^{\nb} {\bU^{(0)}}^\top \obs_i$, and set $\lambda^{(0)} = 10^{3}$ as a vague prior for the first iteration, which is refined by empirical Bayes estimation throughout the algorithm.

Naturally, as in every algorithm with non-convex objective, the procedure can fall into poor local maxima of the bound. Thus, we recommend several restarts with different initializations. In the experiments of \Cref{bfem:sec:NumericalExpe}, we try several $k$-means initializations and take the one achieving the greatest variational lower bound.

\paragraph{Computational complexity} The ODV method necessitates $\latentdim (\dim -1)$ Gram-Schmidt operations overall, and $\latentdim$ generalized eigenvalue problems to solve, although only the first leading eigenvector needs to be found which can be done efficiently \parencite{ge2016efficient}. Since $\latentdim \leq \K -1$ is supposed to be small compared to $\dim$, this is not too computationally expensive. Moreover, this has to be compared to the computational cost of maximizing the lower bound of \Cref{bfem:prop:ELBO} with respect to $\bU$ as in a traditional M-step. Indeed, there is no closed-form solution for this problem and relying on gradient descent can rapidly become cumbersome since it necessitates relying on the steepest descent in the Stiefel manifold $St(\dim, \latentdim) = \{\bU \in \R^{\dim \times \latentdim}, \; \bU^\top \bU = \bI_{\latentdim}\}$.

\begin{figure}[ht!]
	\centering
	\includegraphics[width=1\linewidth]{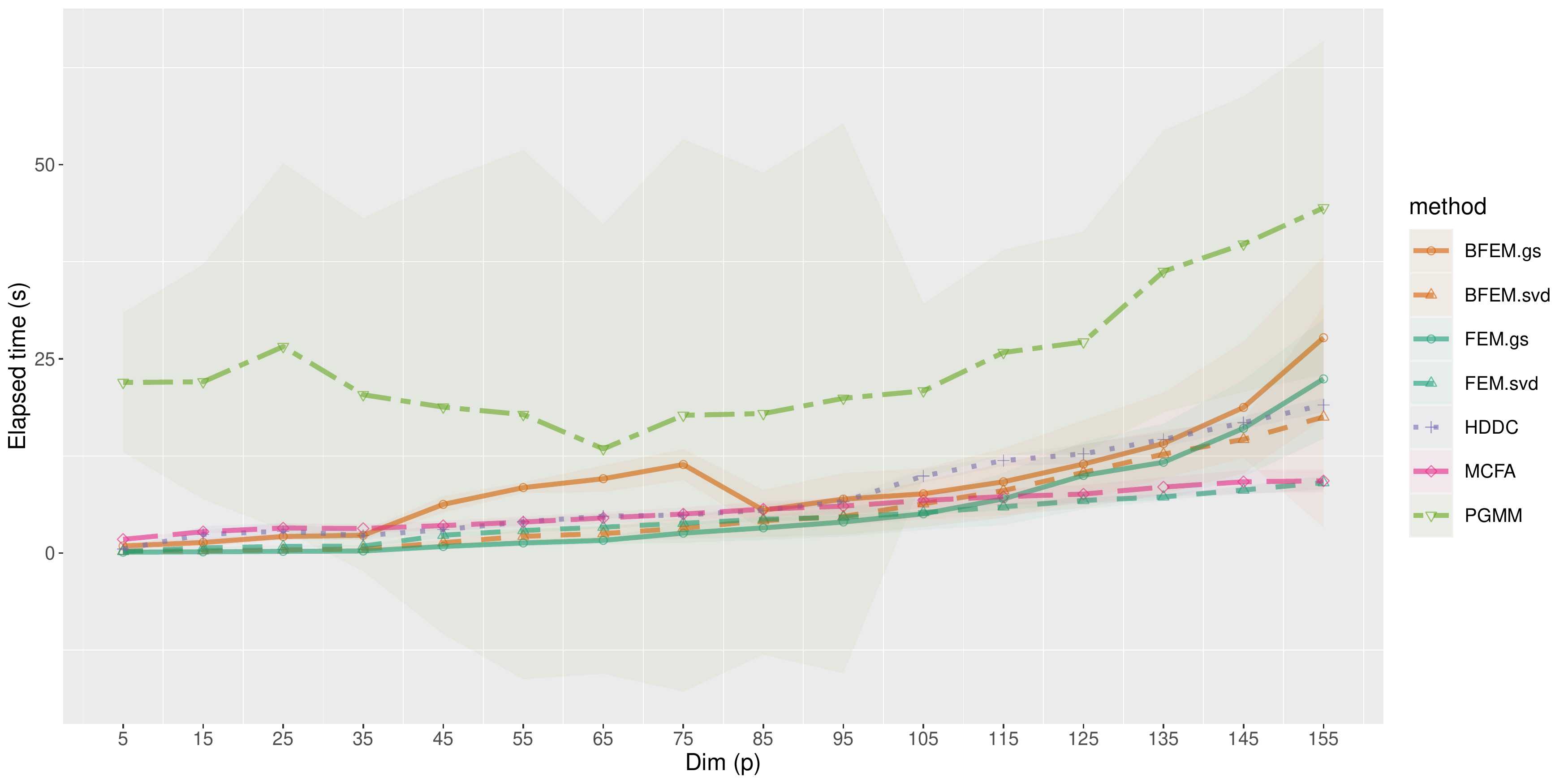}	
	\caption{Mean elapsed time in seconds for one run of each method, computed on $100$ datasets for each level $\dim$.}
	\label{bfem:fig:TimeComplexity}
\end{figure}

Still, \Cref{bfem:subsec:AlternativeFisherCriterion} introduces an alternative Fisher criterion which only necessitates performing one singular value decomposition at each step $(t)$. The empirical computing time of the BFEM algorithm and competing methods are shown in \Vref{bfem:fig:TimeComplexity} of the experimental settings of \Cref{bfem:sec:NumericalExpe}.

\SetKw{KwBreak}{Break}
\SetKw{KwReturn}{Return}

\begin{algorithm}[ht!]
	\KwData{$\Obs$}
	\KwResult{A clustering $\Clust$ and a discriminative subspace $\bU$}
	\KwIn{$\K$, $\bZ^{(0)}$, $\epsilon_{VE}$, $\epsilon_M$, $T_{M}$, $T_{VE}$, $\Umethod$, $\lambda^{(0)}$}
	\BlankLine
	\tcp{Initialization}
	Set $\bTau \gets \bZ^{(0)}$ \\
	Compute $\bS_{T}$, $\btildeS_B^{(0)}$ and subspace $\bU$ with $\Umethod$ \\
	Compute initial parameters $(\bPi, \bSigma, \bBeta)$ with frequentist M-step \\
	Compute variational parameters $(\bvarmean, \bvarcovar)$ with \Cref{bfem:prop:CAVIupdateMu} \\
	Set $\bnu \gets \frac{1}{\nb} \sum_{i=1}^{\nb} \bU^\top \obs_i$ and $\lambda \gets \lambda^{(0)}$\\
	Set $\llhood[0] \gets \J( \globalparam, \q, \bnu, \lambda)$ \\
	\BlankLine
	\tcp{Optimization}
	\For{$t \gets 1$ \KwTo $T_{M}$}{
		\BlankLine
		\tcp{F-step}
		Compute $\btildeS_B^{(t)}$ \\
		Update $\bU$ with $\Umethod$ \\
		\BlankLine
		\tcp{VE-step (fixed point algorithm)} 
		\For{$v \gets 1$ \KwTo $T_{VE}$}{
			Set $\temp \gets \J(\q)$  \\
			Update $\bTau$ using \Cref{bfem:prop:CAVIupdateZ} \\
			Update $(\bvarmean, \bvarcovar)$ using \Cref{bfem:prop:CAVIupdateMu} \\
			\BlankLine
			\lIf{$ \vert (\J(\q) -  \temp)  \mathbin{/} \J(\q) \vert  < \epsilon_{VE}$}{\KwBreak}
		}
		\BlankLine
		\tcp{M-step}
		Update $(\bPi, \bSigma, \bBeta)$ with \Cref{bfem:eq:updatePI} and \Cref{bfem:prop:M-step} \\
		\BlankLine
		\tcp{Empirical Bayes}
		Update $(\bnu, \lambda)$ with \Cref{bfem:prop:EmpiricalBayes}\\
		\BlankLine
		\tcp{Compute the variational lower bound} 
		$\llhood[t] \gets \J( \globalparam, \q, \bnu, \lambda)$ \\
		\If{$t \geq 2$}{
			$c \gets \frac{\llhood[t] - \llhood[t-1]}{\llhood[t-1] - \llhood[t-2]}$ \\
			$\aitken[t] \gets \llhood[t-1] + \frac{1}{1 - c} \left( \llhood[t] - \llhood[t-1] \right)$
			\BlankLine
			\lIf{$\vert \aitken[t] - \aitken[t-1] \vert < \epsilon_{M}$}{\KwBreak}
		}
	}
	\KwReturn{$\q$, $\globalparam = (\bU, \bPi, \bSigma, \bBeta)$}
	\caption{Pseudo code of the BFEM algorithm}
	\label{bfem:alg:BFEM}
\end{algorithm}

\subsection{Model selection}
\label{bfem:subsec:ModelSelection}

\paragraph{Choosing the latent space dimension} As in the supervised case, the rank of $\btildeS_B^{(t)}$ is at most $\K-1$, hence $d \leq \K-1$. We recommend setting it to $d= \K - 1$ for inference, as it is preferable to have redundant information than to lose discriminant directions. This presents the advantage to report the problem of selecting $\latentdim$ to the one of selecting $\K$.

Moreover, following \textcite{okada1985optimal}, the discriminant vectors found by the ODV procedure may be ordered according to the value of their $1$-dimensional fisher criterion
\[
\Fisher(\bu_r) = \frac{\bu_r^\top \btildeS_B^{(t)} \bu_r }{ \bu_r^\top \bS_T \bu_r} = \gamma_r,
\]
where $\gamma_r$ is the largest eigenvalue of the $r$-th problem solved, leaving:
\begin{align}
\Fisher(\bu_1) \geq \ldots \geq \Fisher(\bu_{\latentdim}).
\end{align}
Thus, for visualization purpose, we can choose $d=2$ or $d=3$, and project the data onto the corresponding subspace. Another solution, as is commonly done in PCA, is to show several combinations of discriminant axes, for instance with a matrix of $2$-dimensional scatter plots. 
%Naturally, the visualization quality is dependent on the clustering quality, thus a poor visualization can guide the user to restart analysis with a different initialization. 

\paragraph{Choosing the number of clusters} In a clustering perspective, we propose to rely on the integrated classification likelihood \parencite[ICL,][]{biernacki2000assessing} to choose $\K$ and the submodel. Recall that the criterion of \textcite{biernacki2000assessing} is defined as:
\begin{align}
\label{bfem:eq:ICL}
\ICLbic(\Model, \K) =  \log \p(\Obs, \hat{\Clust} \mid \hat{\globalparam}, \Model, \K) - \frac{\gamma_{\Model, \K}}{2} \log(\nb),
\end{align}
where we take $\hat{\globalparam}$ to be the parameter estimates at the end of BFEM. Although the marginal likelihood is intractable as explained in \Cref{bfem:subsec:VEM}, the first term above is the classification likelihood which is tractable in the $\BDLM$ models. Actually, it can be computed with the variational lower bound $\J$, replacing $\tau_{ik}$ by $\hat{\rawclust}_{ik}$ in the formulas of \Cref{bfem:prop:CAVIupdateZ,bfem:prop:CAVIupdateMu}. A detailed proof of this fact is given in \Cref{bfem:appendix:ProofModelSelection}.
%the variational lower bound $\J(\q_{\ICL} , \hat{\globalparam})$ where 
%\[
%\q_{\ICL}(\Clust, \bmu) = \delta_{\hat{\Clust}} (\Clust) \q(\bmu)
%\]
%Since the first term is intractable, we rely on the variational bound after convergence, as in \Cref{chap:MMPCA}. 
%\begin{proposition}
%	An ICL criterion for the $\BDLM$ model is given as:
%	\begin{align}
%			\ICLbic(\Model, \K) = \J_{\Model, \K}(\q^\star , \globalparam^\star) - \frac{\gamma_{\Model, \K}}{2} \log(\nb). 
%	\end{align}
%	Where $\gamma_{\Model, \K}$ is given in \Cref{bfem:tab:BDLMsubmodels} for the $12$ submodels.
%\end{proposition}

\subsection{An alternative Fisher criterion}
\label{bfem:subsec:AlternativeFisherCriterion}

The ODV method proposed to maximize the criterion in \Cref{bfem:eq:VariationalFisher} can be viewed as a greedy method, sequentially solving $1$-dimensional problems $\bu_r = \argmax_{\bu} \Fisher(\bu)$ under the growing set of constraints $\{\bu \top \bu_h =0, \forall h < r\}$. Such a method is not guaranteed to globally converge as emphasized in \textcite{hamamoto1991note}. Moreover, relying on the Gram-Schmidt procedure can sometimes lead to numerical instabilities in the BFEM algorithm. Thus, \textcite{bouveyron2011estimation} proposed an alternative Fisher criterion, searching for the orthogonal projection matrix $\bU \in \R^{\dim \times \latentdim}$ minimizing the following reconstruction error:
\begin{align}
\label{bfem:eq:modifiedFisher}
\bU^{(t)}	= \argmin_{\bU^\top \bU = \bI_{\latentdim}} \Vert \bS_T^{-1} \btildeS^{(t)}_B - \bU \bU^\top \bS_T^{-1} \btildeS^{(t)}_B \Vert_F^2 
%	= \argmax_{\bU^\top \bU = \bI_{\latentdim}} \Tr\left[\bU^\top \left(\bS_T^{-1} \btildeS^{(t)}_B\right)^2 \bU  \right]
\end{align}
This optimization problem has a somewhat PCA like flavor except that the matrix we wish to reconstruct is not the original data $\Obs$ but the measure of class separability: $\bS_T^{-1} \btildeS^{(t)}_B$. The classical results still holds, and the optimal $\bU^{(t)}$ is given as the leading $\latentdim$ left singular-vectors of $\bS_T^{-1} \btildeS^{(t)}_B = \bU \bLambda \bV$. Note that, since the product of these two symmetric matrices is not symmetric, the singular value decomposition is different from its spectral decomposition.

This modified F-step can be used to replace the ODV procedure at step $(t)$. Thus, we only have to perform a partial singular-value decomposition (SVD) on $\bS_T^{-1} \btildeS^{(t)}_B$ at each step. Since $\bS_T^{-1}$ is computed only once, this is particularly efficient.

\section{Numerical experiments}
\label{bfem:sec:NumericalExpe}
This section compares the performance of different subspace clustering on simulated and classical real data benchmarks. We considered $6$ different algorithms:
\begin{enumerate}
	\item The proposed BFEM algorithm, with the ODV procedure described in \Cref{bfem:subsec:Fisherstep} for the F-step. The results are also displayed for the alternative Fisher criterion of \Cref{bfem:subsec:AlternativeFisherCriterion}, which we refer to as the SVD procedure,
	%	\item BFEM-svd (ours)
	\item The Fisher-EM algorithm of \textcite{bouveyron2012simultaneous}, implemented in the \pkg{FisherEM} R Package \parencite{FisherEM}. Again, we also show results for the F-step using the ODV procedure as well as the SVD,
	\item The EM algorithm for the PGMM of \textcite{mcnicholas2008parsimonious} with model CCU, corresponding to the low-rank constraint $\bS_k = \bU \bU + \bPsi$ of the covariance matrix, without orthonormality constraints on $\bU$. An implementation is available in the eponymous R package \pkg{pgmm} \parencite{pgmm},
	\item The EM algorithm for MCFA of \textcite{baek2009mixtures} and implemented in the \pkg{EMMIXmfa} R package \parencite{EMMIXmfa},
	\item The EM algorithm for the HDDC model of \textcite{bouveyron2007high} with model $[\cdot QD]$, so that the learned subspaces are common, as in the other methods. We used the implementation available in the \pkg{HDclassif} R package \parencite{HDclassif},
	%	\item HFMA \pkg{FactMixtAnalysis} \parencite{FactMixtAnalysis}
	\item A $k$-means baseline.
	%	\item k-means + PCA ($90\%$)
\end{enumerate}
A total of $8$ distinct algorithms are tested, since both models BFEM and FEM may have two distinct F-step procedures. In \Cref{bfem:subsec:BehaviorWithDim,bfem:subsec:SNR}, we use colors to distinguish between models, and line marker types to differentiate between the ODV and SVD method. For the sake of readability, we do not show the results of the HFMA model of \textcite{montanari2010heteroscedastic} in the following figures since it did not perform well on our experimental settings. This might be due to the different constraints on the subspace means and covariances, making it more distant to the BDLM model than other subspace clustering methods.

Throughout the rest of this section, unless stated otherwise, each method has the same $10$ restarts consisting of $10$ different $k$-means results. The one achieving the greatest likelihood is kept, and the clustering is done with a MAP estimate over the posterior of $\Clust$. The maximum of iterations is set to $100$ everywhere and the same absolute tolerance of $10^{-6}$ is used. The fixed point algorithm in the VE-step has a tolerance of $10^{-6}$ but a maximum number of iterations set to $3$.

Concerning the choice of $\latentdim$, it is set to $\K-1$ for both FEM and BFEM and to the true value $\latentdim^\star$ whenever it is known in the simulations. When $\latentdim$ is unknown, the HDDC model has an internal heuristic to choose the best intrinsic dimension $\latentdim_k$ of each cluster, and we use the Bayesian Information Criterion for the MCFA and PGMM as suggested in their original papers. The clustering results are reported using the Adjusted Rand Index \parencite[ARI,][]{hubert1985comparing}, a label independent measure of statistical similarity between two partitions. An ARI of $0$ means that the two partitions are statistically independent, while identical partitions (up to label switching) give an ARI of $1$. Hence, the higher the ARI, the better.
% of the increased computational time compared to other methods. 

\subsection{An introductory example}
\label{bfem:subsec:IntroductoryExample}
In order to illustrate the interest of discriminative subspaces, we begin with the numerical setting of \textcite{chang1983using} discussed in the introduction and \Cref{fig:chang1983}. There are $\nb = 300$ observations and $\K = 2$ clusters in the data, defined as follows:
\begin{align*}
\obs_i = -0.5 \bm{r} + \bm{r} \mathds{1}_{\{\rawclust_{ik} = 1\}} + \Gaussian_{\dim}(\bm{0}, \bS), 
\end{align*}
with $\forall j = 1, \ldots, 15$:
\begin{align*}
&  r_j= 0.95 - 0.05j, \\
&  \bS_{jj} = 1  \text{ and } \forall j' \neq j, \bS_{jj'} = -0.13f_j f_{j'}  \text{ with } f_{j} = \left\{ \begin{array}{ll}
-0.9 & j \leq 8 \\
0.5 &  j > 8 
\end{array}\right.  .
\end{align*}
Thus, it is a $2$-component Gaussian mixture in dimension $\dim = 15$, with $\bmean_1 = - 0.5 \bm{r}$ and $\bmean_2 = 0.5 \bm{r}$ and homoscedastic covariance $\bS_1 = \bS_2 = \bS$. We emphasize that this simulation is not favoring any of the tested methods, except maybe for the standard GMM since the simulation is according to this model.

\begin{figure}[!ht]
	\centering
	\includegraphics[width=\linewidth]{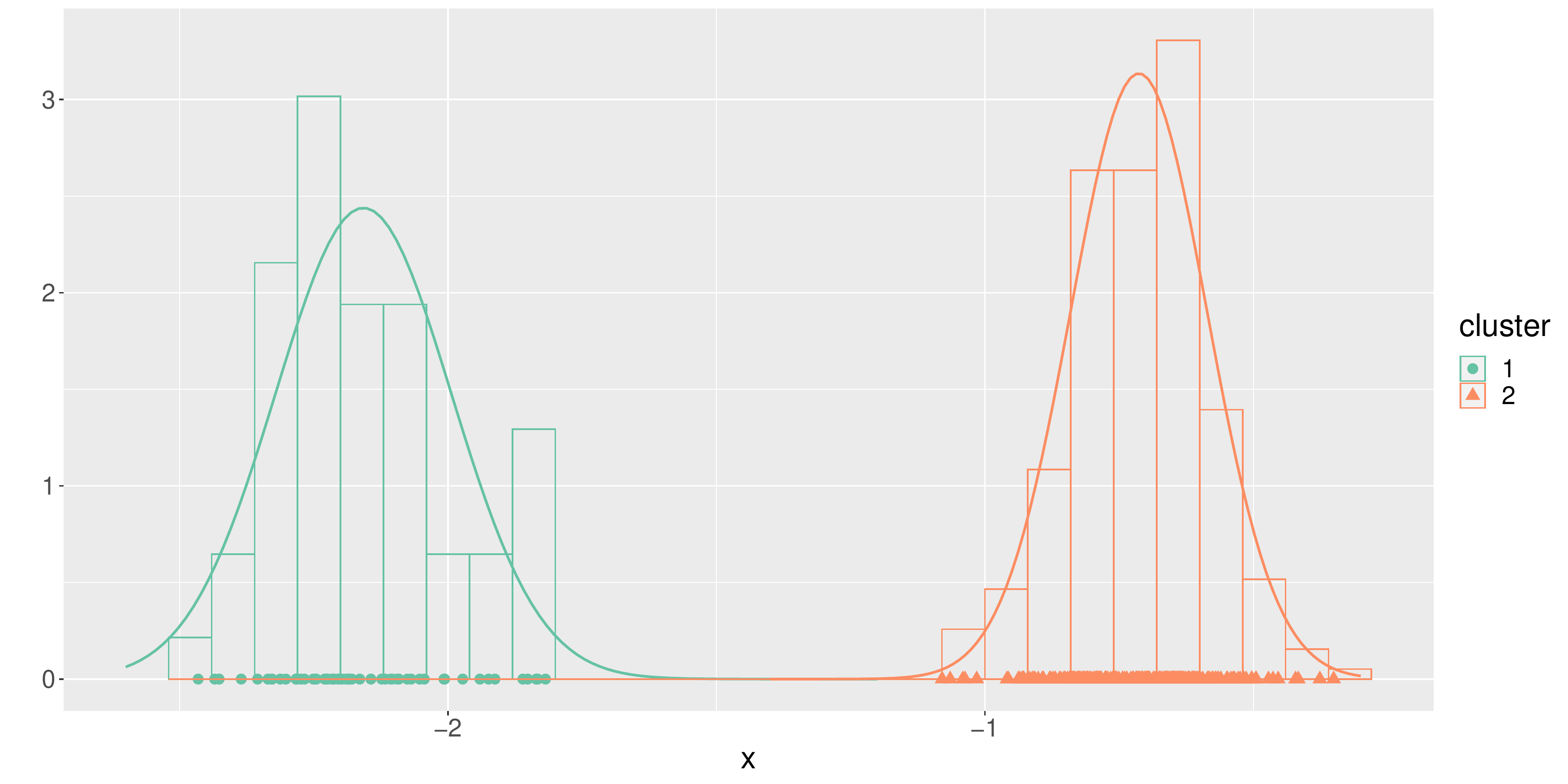}
	\caption{Projection of Chang's data in the $1$-dimensional subspace found by BFEM, colors indicate the estimate cluster memberships. Solid lines represent the learned within-cluster Gaussian distributions, while the histograms represent the empirical ones.}
	\label{bfem:fig:BFEMchang1983}
\end{figure}

We ran each method with the true number of clusters $\K=2$, and used model selection for the choice of $\latentdim$ for the concerned methods. The average results over $100$ simulated datasets are represented for each method in \Cref{bfem:tab:Chang1983}. We do not distinguish the ODV and SVD methods here, since they lead to the same results on this simple example. One can see that the proposed discriminative subspace approach yields a better clustering in this setting, with a slight advantage over the frequentist version. In particular, extensions of pPCA like MCFA or PGMM do not allow to recover the correct partitions. This highlights the interest of discriminative subspaces even in different scenarios. The HDDC algorithm exhibits the same performance as BFEM. However, we point out that it selects intrinsic dimensions $\latentdim_k = 14$ to do so, which are the maximum values in this model. In contrast, BFEM works with $\latentdim=1$ enabling to visualize the latent space in \Cref{bfem:fig:BFEMchang1983}, with a clear separation of the two classes. Finally, since the dimension $\dim$ is still reasonable compared to $\nb$, a standard GMM with spectral constraints may be fitted and performs well. Here, the BIC criterion selects the EEE model which means ellipsoidal, equal volume, shape and orientation $\bS_k = \lambda \bD \bDelta \bD$, and corresponds to the true model.
\begin{table}[ht!]
	\centering
	\caption{Mean ARI and standard errors for BFEM and competing methods over $100$ simulations of \textcite{chang1983using}'s setting.}
	\label{bfem:tab:Chang1983}
	\begin{tabular}{c|c|c|c|c|c|c}
		Kmeans & BFEM & FEM & HDDC & MCFA  & PGMM & Mclust \\ 
		\hline
		$0.24 \pm 0.1$ & $\bm{1 \pm 0}$ & $0.98 \pm 0.11$ & $\bm{1 \pm 0}$ & $0.62 \pm 0.07$  & $0.42 \pm 0.22$ & $0.97 \pm 0.12$ 
	\end{tabular}
	\vspace*{5pt}
\end{table}

\begin{figure}[!ht]
	\centering
	\includegraphics[width=\linewidth]{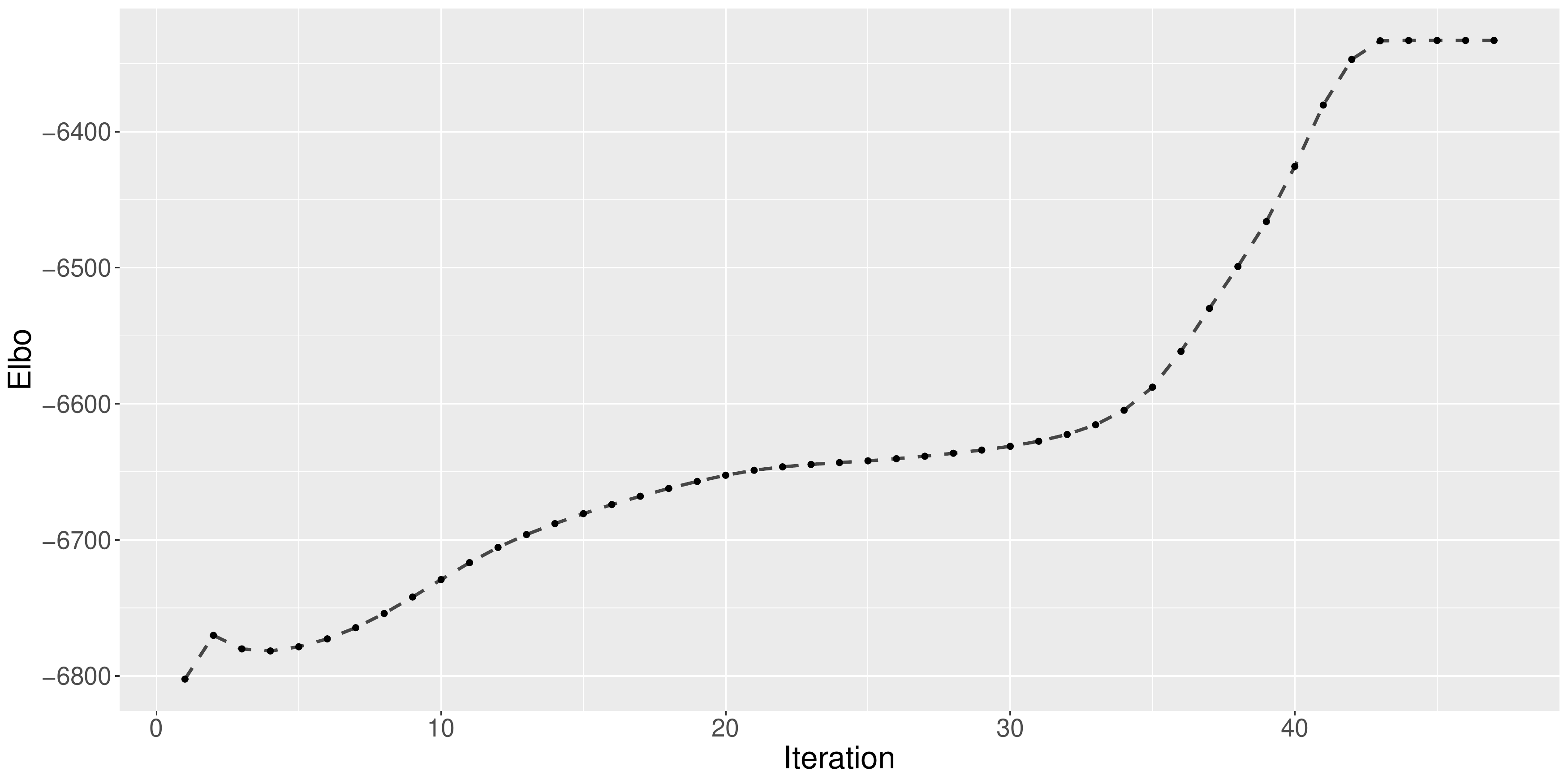}
	\caption{Evolution of the evidence lower bound during a run of the BFEM algorithm with ODV procedure on Chang's dataset.}
	\label{bfem:fig:boud_evolution_chang1983}
\end{figure}

\Cref{bfem:fig:BFEMchang1983} shows the $1$-dimensional discriminative subspace found by BFEM on one simulation, with colors indicating cluster membership. Moreover, the solid lines represent the Gaussian density in each cluster $\p(\scores \mid \bvarmean_{k}, \hat{\bSigma}_k)$, along with the empirical within-cluster distribution as a histogram. Unsurprisingly, the subspace induces well-separated clusters and the empirical within-cluster distribution has a Gaussian shape fitting the theoretical one. Finally, the evolution of the evidence lower bound during the BFEM algorithm is displayed in \Cref{bfem:fig:boud_evolution_chang1983}. As expected, it is not monotonically increasing, especially in the first step, although the evolution is quite smooth. In addition, convergence happens before the limit of $100$ iterations is reached.

\subsection{Sensitivity to the dimension}
\label{bfem:subsec:BehaviorWithDim}
\begin{figure}[ht!]
	\centering
	\includegraphics[width=0.8\linewidth]{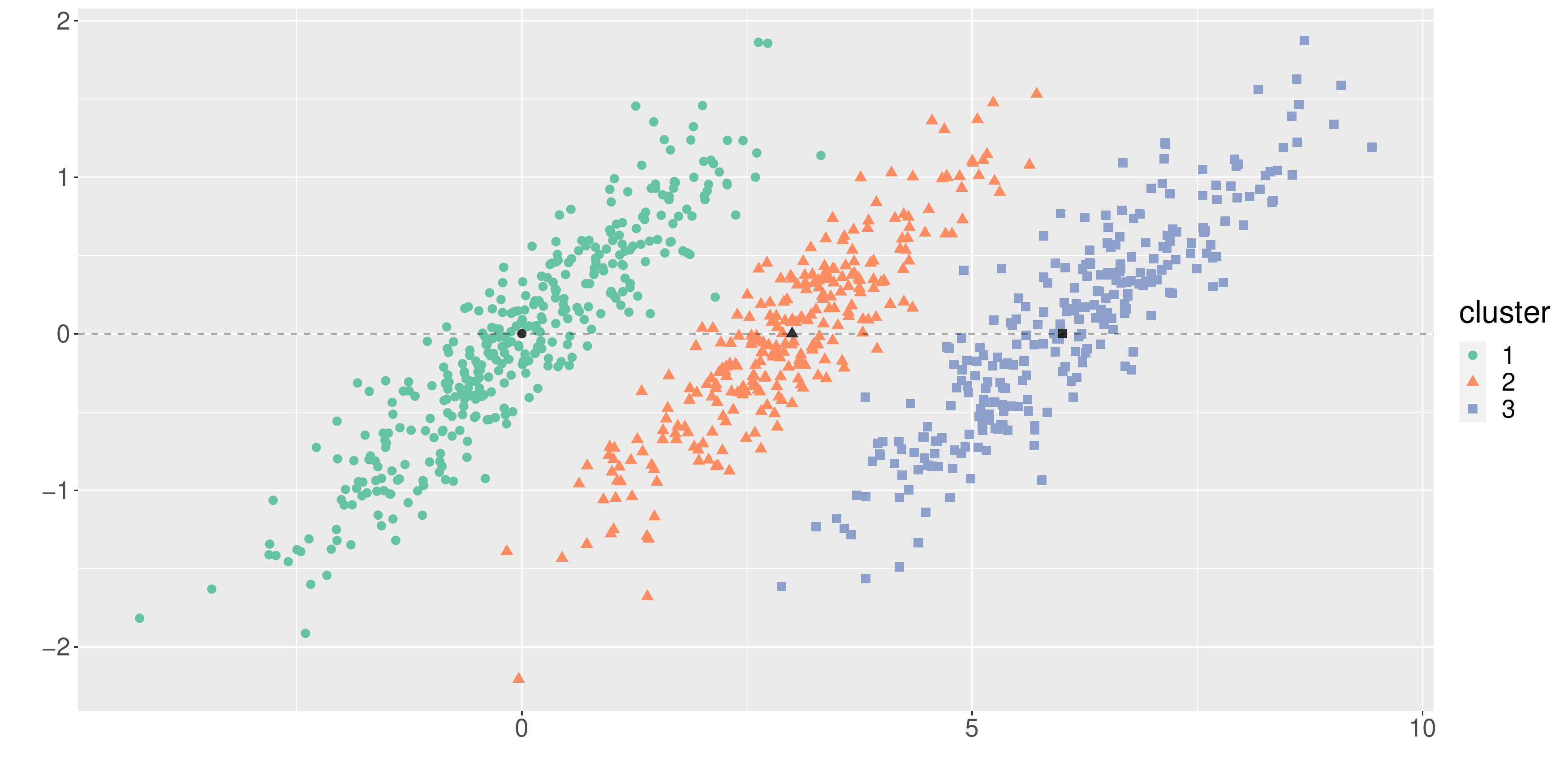}	
	\caption{Simulation of the $3$-components GMM of \Cref{bfem:subsec:BehaviorWithDim}.}
	\label{bfem:fig:HighDimSetting}
\end{figure}

We now propose to investigate the behavior of subspace clustering methods on increasingly high-dimensional problems. In this setting, we simulate $\Scores$ from $K=3$ Gaussian components in dimension $\latentdim=2$. The respective means and covariance matrices are 
\begin{align*}
\bmu_k = 3 \; (0, k)^\top  & &  \bSigma_{k} = \begin{pmatrix}
1.5 & 0.75 \\
0.75 & 0.45
\end{pmatrix} & & \bPi = (0.4, 0.3, 0.3)^\top.
\end{align*}
\Cref{bfem:fig:HighDimSetting} illustrates a particular simulation of $n=900$ data points. As can be seen, it corresponds to the particular case where the clusters are parallel Gaussian ellipses, differentiated with a mean-shift along the x-axis.
%(Un peu hors sujet) A similar settings was considered in \textcite{witten2010framework}.

\begin{figure}[ht!]
	\centering
	\includegraphics[width=0.8\linewidth]{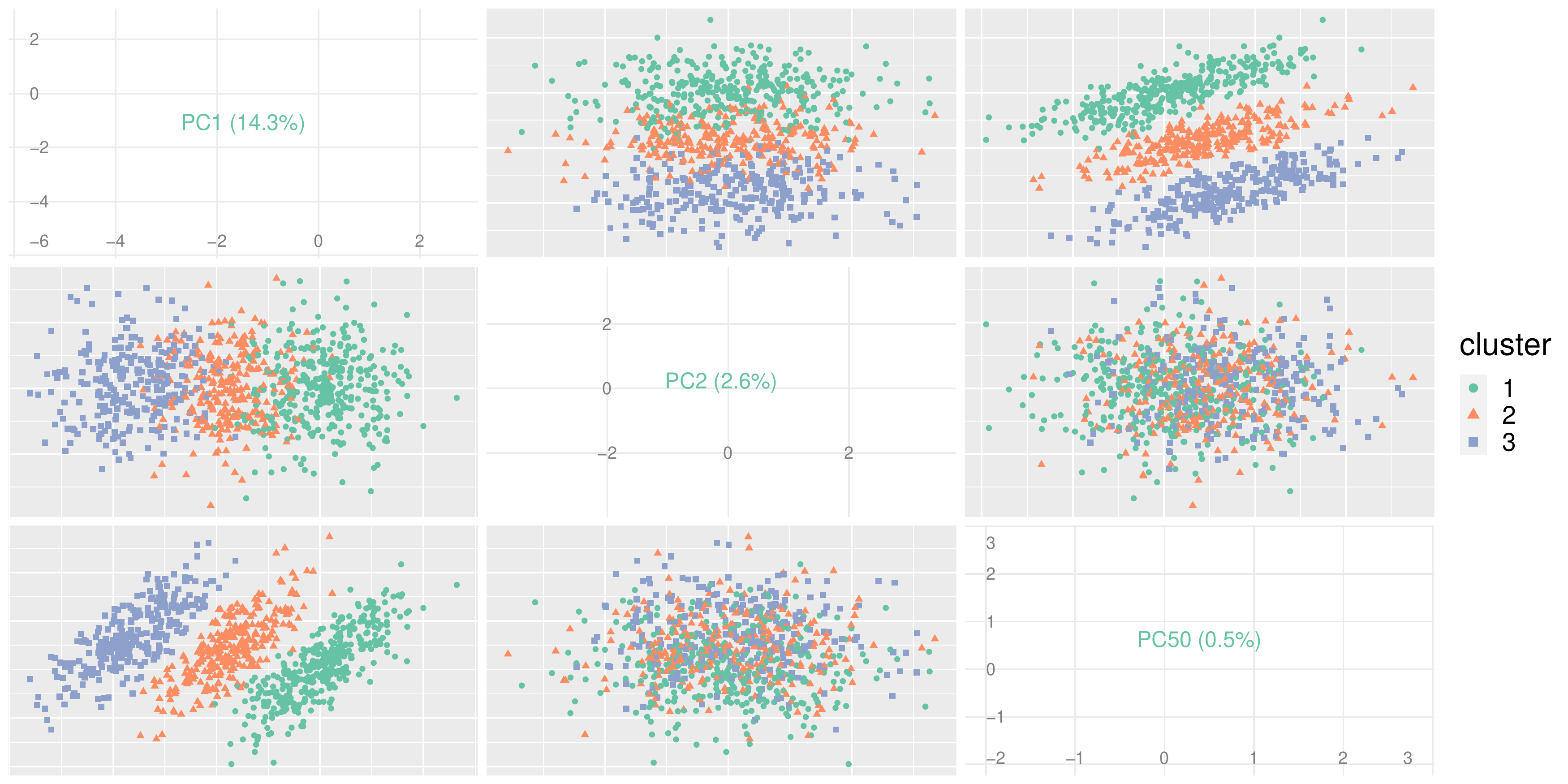}
	\caption{Subspaces found by PCA on a simulation with $\dim = 50$, color and shape indicate the true cluster membership. Once again, the most discriminative subspace is given by the first and last components, the other corresponding to noisy directions.}
	\label{bfem:fig:PCAHighDimSetting}
\end{figure}

Next, we propose to simulate according to the DLM model. First, $\bD = [\bU, \bV]$ is simulated from the Q-matrix of the QR decomposition of the matrix $\bA \in \R^{\dim \times \dim}$, the latter being itself simulated according to $\bA \sim \Gaussian_{\dim^2}(\bm{0}, 100 \bI)$. Then, for each observation $i$, a $(\dim - \latentdim)$ dimensional standard Gaussian noise $\bm{\epsilon}_i$ is simulated. Finally, the data points are created as the linear transformation $\obs_i = \bD (\scores_i^\top, \bm{\epsilon}_i^\top)^\top$. The first principal components are expected to behave poorly in terms of class separation in this scenario, which is illustrated in \Cref{bfem:fig:PCAHighDimSetting} for $\dim=50$. Indeed, the directions of greatest variations include noisy directions that contribute more to the variance than the second signal dimension.
%This setting is placed on a limiting case of Gaussian mixtures with means aligned along the x-axis and density ellipses perpendicular . 
%We propose to simulate from a specific Gaussian mixture model with aligned covariance

\begin{figure}[ht!]
	\centering
	\includegraphics[width=1\linewidth]{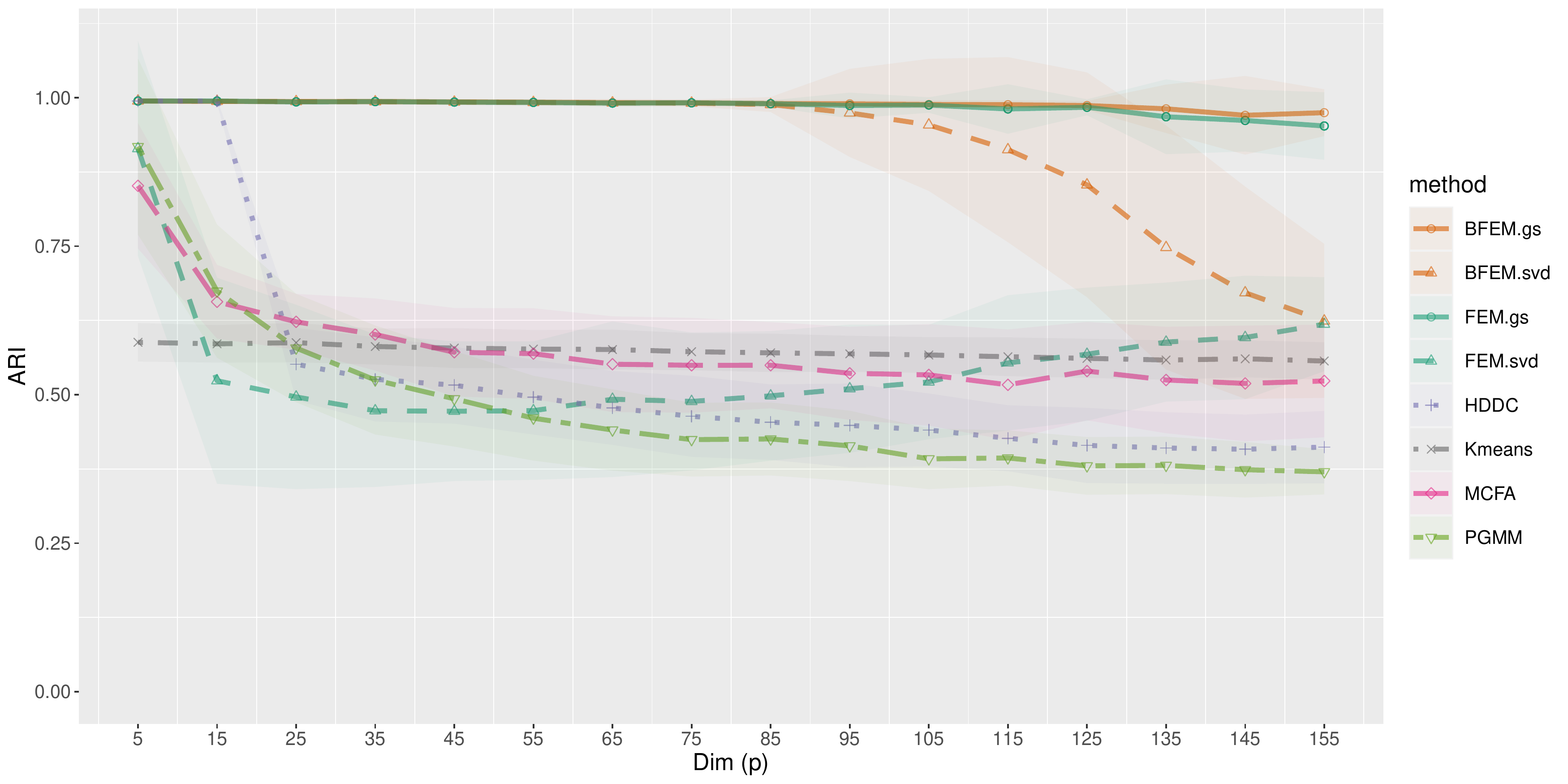}	
	\caption{Evolution of the mean ARI over $100$ different runs for each method, with an increasing dimensionality $\dim$ and $\nb = 900$.}
	\label{bfem:fig:ARI_HighDimSetting}
\end{figure}

We investigate the behavior of each method as the dimension $\dim$ increases from $5$ to $155$. Mean ARI and standard variations were computed on a $10$-spaced linear grid, for $100$ simulated datasets at each level $\dim$. As stated above, we use colors to differentiate between the BFEM and FEM, and line marker types to distinguish between OVD (solid) and SVD (dashed) for both algorithms. Concerning MCFA and PGMM, the subspace dimension was set to the true value $\latentdim = 2$. The results are displayed in \Cref{bfem:fig:ARI_HighDimSetting} and shows several things. First, the BFEM and FEM with the ODV method are very robust in this scenario, with a perfect recovery at each level $\dim$. Other subspace clustering methods quickly decrease beyond $\dim=15$ with performances comparable, or below $k$-means. Thus, it underlines a limit of likelihood-based approach, as noisy dimensions are being fitted in the subspace when $\dim$ increases. The discriminative approach, injecting clustering information in the search for the optimal subspace, is robust in this context as the optimal subspace is not necessarily aligned with greatest variance directions. Another interesting fact is the sensitivity to noise of the Fisher-EM with the SVD procedure, which displays a rather unstable behavior. We note that BFEM with the SVD procedure, while suboptimal, is still displaying a strong stability in high-dimensional settings, with an ARI decreasing only after $\dim = 85$

%\begin{figure}[ht!]
%	\centering
%	\includegraphics[width=1\linewidth]{figures/BehaviorWithDim_100datasets_plot_chapterThese.pdf}
%	\caption{Evolution of the mean ARI over $100$ different runs for each method, with an increasing dimensionality $\dim$ and $\nb = 900$.}
%	\label{bfem:fig:ARI_HighDimSetting}
%\end{figure}

%\Cref{bfem:fig:ARI_HighDimSetting_unbalanced} shows the results for the same simulations except now unbalanced class are more pronounced: $\bPi = (0.6, 0.2, 0.2)^\top$ and heteroscedasticity in introduced with $\bSigma_{k} = \sigma_k \begin{pmatrix}
%1 & 0.5\\
%0.5 & 0.3
%\end{pmatrix}$ and $\bm{\sigma} = (1, 0.2, 1.5)$ controls the volume of the ellipses.
%\begin{figure}[ht!]
%	\centering
%%	\includegraphics[width=1\linewidth]{figures/BehaviorWithDim_finAout_unbalanced.pdf}
%%	\includegraphics[width=1\linewidth]{figures/BehaviorWithDim_finAout_unbalanced_raw.pdf}
%	\includegraphics[width=1\linewidth]{figures/BehaviorWithDimm_cst3_100datasets_unbalanced_heterosced.pdf}
%	\caption{Unbalanced and heteroscedastic setting. Evolution of the mean ARI over $100$ different runs for each method, with an increasing dimensionality $\dim$ and $\nb = 900$.}
%	\label{bfem:fig:ARI_HighDimSetting_unbalanced}
%\end{figure}

\subsection{Signal-to-noise ratio}
\label{bfem:subsec:SNR}
We place ourselves in the same setting as \cref{bfem:subsec:BehaviorWithDim}, only this time the dimension is fixed to a high-dimensional scenario $p = 150$. We propose to investigate the impact of the noise, with $\bm{\epsilon}_i$ now drawn from a centered Gaussian distribution with covariance $\beta \bI_{\dim - \latentdim}$. The latter may be interpreted as controlling the signal-to-noise ratio (SNR) which can be defined as the ratio between signal variance in the subspace of dimension $\latentdim = 2$ and the noise variance\footnote{One could use $(p-d) \beta$ as the actual variance of the signal, taking into account the fact that there are $(p-d)$ noisy directions. However, since $\dim - \latentdim$ is fixed here, it only acts as a scaling factor for the SNR.} $\beta$. Since all clusters have the same subspace covariance, we define the signal variance as $\Tr[\bSigma]$, the inertia of a cluster cloud point in the latent subspace. The SNR is best expressed in decibels (dB), which corresponds to ten times the decimal logarithm of the variance ratio.
\begin{align*}
SNR = 10 \times \log_{10} \left( \frac{\Tr\left[ \bSigma \right]}{\beta} \right) .
\end{align*} 
A value of $0$ means that both variances are equal, and an increase (resp. decrease) of $3$ dB means that the variance of the noise was divided (resp. multiplied) by $2$. For example, a SNR of $3$ dB means that the variance of the signal is $2$ times that of the noise ($\beta\approx 1$), and a SNR of $-6$ means that $\beta$ is $4$ times greater than the signal variance ($\beta \approx 8$).

\begin{figure}[ht!]
	\centering
	\includegraphics[width=1\linewidth]{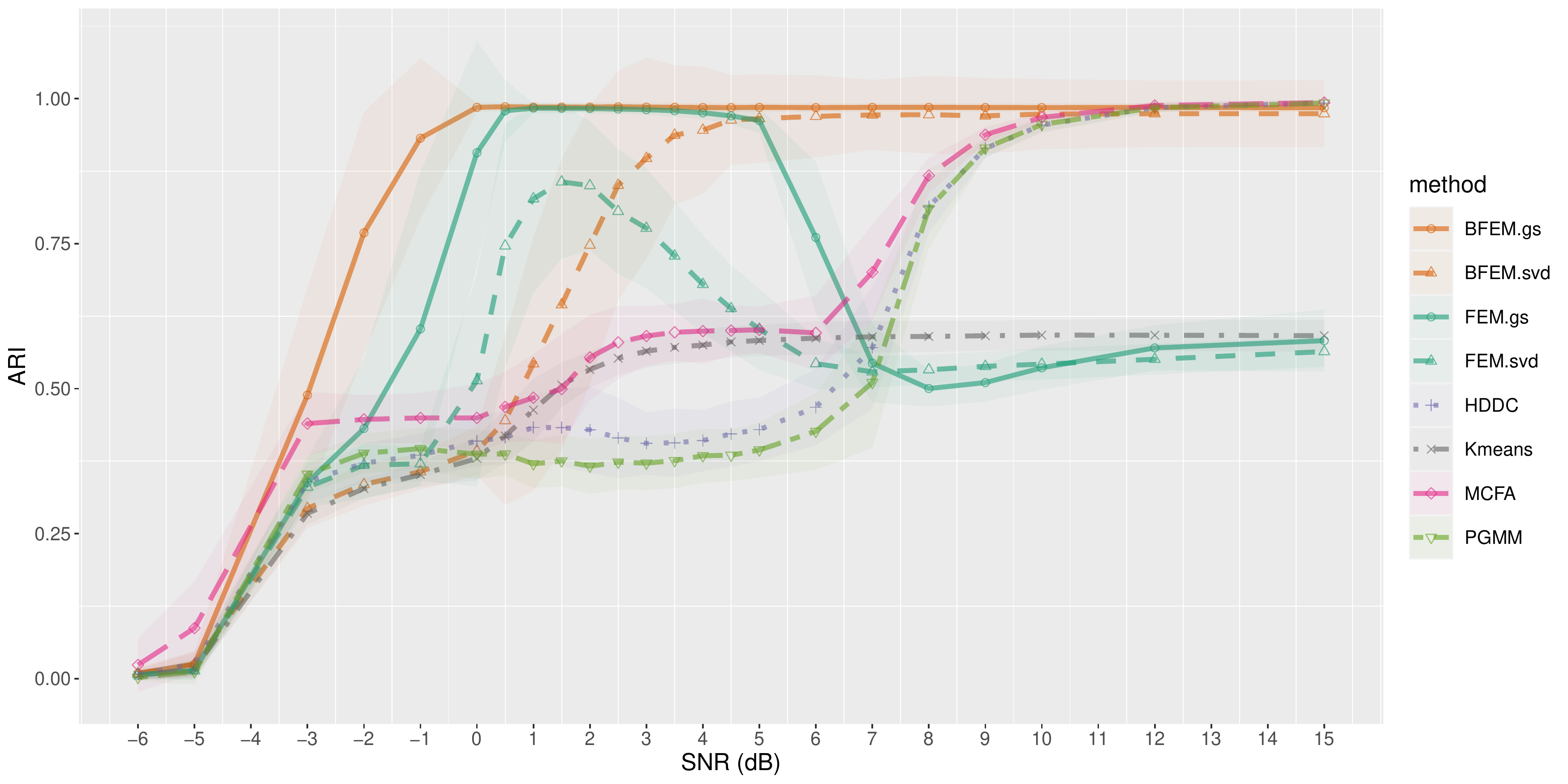}
	\caption{Evolution of the mean ARI over $100$ different runs for an increasing signal-to-noise ratio, $\dim = 150$ and $\nb = 900$.}
	\label{bfem:fig:ARI_SNR}
\end{figure}

\Cref{bfem:fig:ARI_SNR} shows the mean ARI and standard deviations for an increasing SNR from $-6$ to $15$ with a $0.5$-spaced linear grid. Again $100$ datasets are simulated for each level. Several comments are in order. First, for high values of the SNR, which we refer to as the noiseless regime, the clustering problem of \Cref{bfem:subsec:BehaviorWithDim} becomes trivial, except for $k$-means which is disadvantaged by the non-spherical shapes of clusters and Fisher-EM which seems to display a surprising instability in the noiseless limit. From the preceding section, we know that the Fisher-EM algorithm with the SVD procedure is not robust to high-dimension. However, this shows that the ODV procedure also suffers from instability in the frequentist setting. This may be due to poor conditioning of the soft between-class scatter matrix arising in this case. Apart from this somewhat surprising fact, the behaviors of other subspace clustering methods such as HDDC, MCFA and PGMM are expected. However, their performances quickly decrease, even for reasonable values of the SNR where the noise variance is orders of magnitude below the signal. A contrario, the BFEM displays a strong stability, and the SVD method seems to be applicable as long as the noise variance remains reasonably below the signal. In addition, BFEM with the ODV procedure is the most stable of all, with perfect recoveries even when the SNR is $0$ dB, \textit{i.e.} the equal variance case. Eventually, no clustering structure can be recovered below $0$ dB, as the signal is completely overwhelmed by noisy directions, and the ARI of each method quickly decreases to $0$.
%Traditional subspace clustering models such as HDDC, MCFA, or PGMM are able to handle the clustering problem of 

\subsection{Model selection}
\label{bfem:subsec:ExpeModelSelection}

Here, we investigate the ability of the ICL criterion of \Cref{bfem:subsec:ModelSelection} to choose both the number of clusters and the model. We use the setting of \Cref{bfem:subsec:BehaviorWithDim}, which corresponds to $\dim = 150$, $\K=3$ and a model $\BDLM_{[\bSigma \beta]}$. Two different levels of SNR are tried: $3$dB ($\beta = 1$), which corresponds to the setting of \Cref{bfem:fig:ARI_HighDimSetting}, and $-2$ dB ($\beta \approx 3$), which is a more complicated case for BFEM as shown in \Cref{bfem:fig:ARI_SNR}. The results are shown in \Cref{bfem:tab:modelSel_p50}, we see that the performance of model selection are perfect for the first setting, and still very satisfying in the more difficult scenario with $90\%$ of correct selection of the pair $(\K, \Model)$, and $98\%$ of correct selection of $\K$.

\begin{table}[ht!]
	\renewcommand{\arraystretch}{1.5}
	\centering
	\caption{Percentage of correct model selection for BFEM with $\dim = 150$ and varying SNR. The true model is $\BDLM_{[\Sigma \beta]}$ with $\K=3$.}
	\label{bfem:tab:modelSel_p50}
	\newcolumntype{s}{>{\columncolor{lightgray}} c}
	\begin{minipage}{0.9\linewidth}
		\centering
		$\bm{SNR = 3}$ \\
		\begin{tabular}{|r|cccscc|}
			\hline
			$\K$ \textbackslash $\Model$ & $[\Sigma_k\beta]$ & $[\alpha_{kh} \beta]$ & $[\alpha_{k} \beta]$  & $\bm{[\Sigma \beta]}$  & $[\alpha \beta]$  & $[\alpha_{h} \beta]$  \\ 
			\hline
			2 & 0 & 0 & 0 & 0 & 0 & 0 \\ 
			\rowcolor{lightgray}
			\textbf{3} & 0 & 0 & 0 & \textbf{100\%} & 0 & 0 \\ 
			4 & 0 & 0 & 0 & 0 & 0 & 0 \\ 
			5 & 0 & 0 & 0 & 0 & 0 & 0 \\ 
			6 & 0 & 0 & 0 & 0 & 0 & 0 \\ 
			7 & 0 & 0 & 0 & 0 & 0 & 0 \\ 
			\hline
		\end{tabular}
	\end{minipage} \\ \vspace*{5mm}
	\begin{minipage}{0.9\linewidth}
		\centering
		$\bm{SNR = -2}$ \\
		\begin{tabular}{|r|cccscc|}
			\hline
			$\K$ \textbackslash $\Model$ & $[\Sigma_k\beta]$ & $[\alpha_{kh} \beta]$ & $[\alpha_{k} \beta]$  & $\bm{[\Sigma \beta]}$  & $[\alpha \beta]$  & $[\alpha_{h} \beta]$  \\ 
			\hline
			2 & 0 & 0 & 2\% & 0 & 0 & 0 \\ 
			\rowcolor{lightgray}
			\textbf{3} & 8\% & 0 & 0 & \textbf{90\%} & 0 & 0 \\ 
			4 & 0 & 0 & 0 & 0 & 0 & 0 \\ 
			5 & 0 & 0 & 0 & 0 & 0 & 0 \\ 
			6 & 0 & 0 & 0 & 0 & 0 & 0 \\ 
			7 & 0 & 0 & 0 & 0 & 0 & 0 \\ 
			\hline
		\end{tabular}
	\end{minipage} 

\end{table}

\subsection{Real data benchmarks}
\begin{figure}[ht!]
	\centering
	\subfloat[Iris]{
		\includegraphics[width=0.49\linewidth]{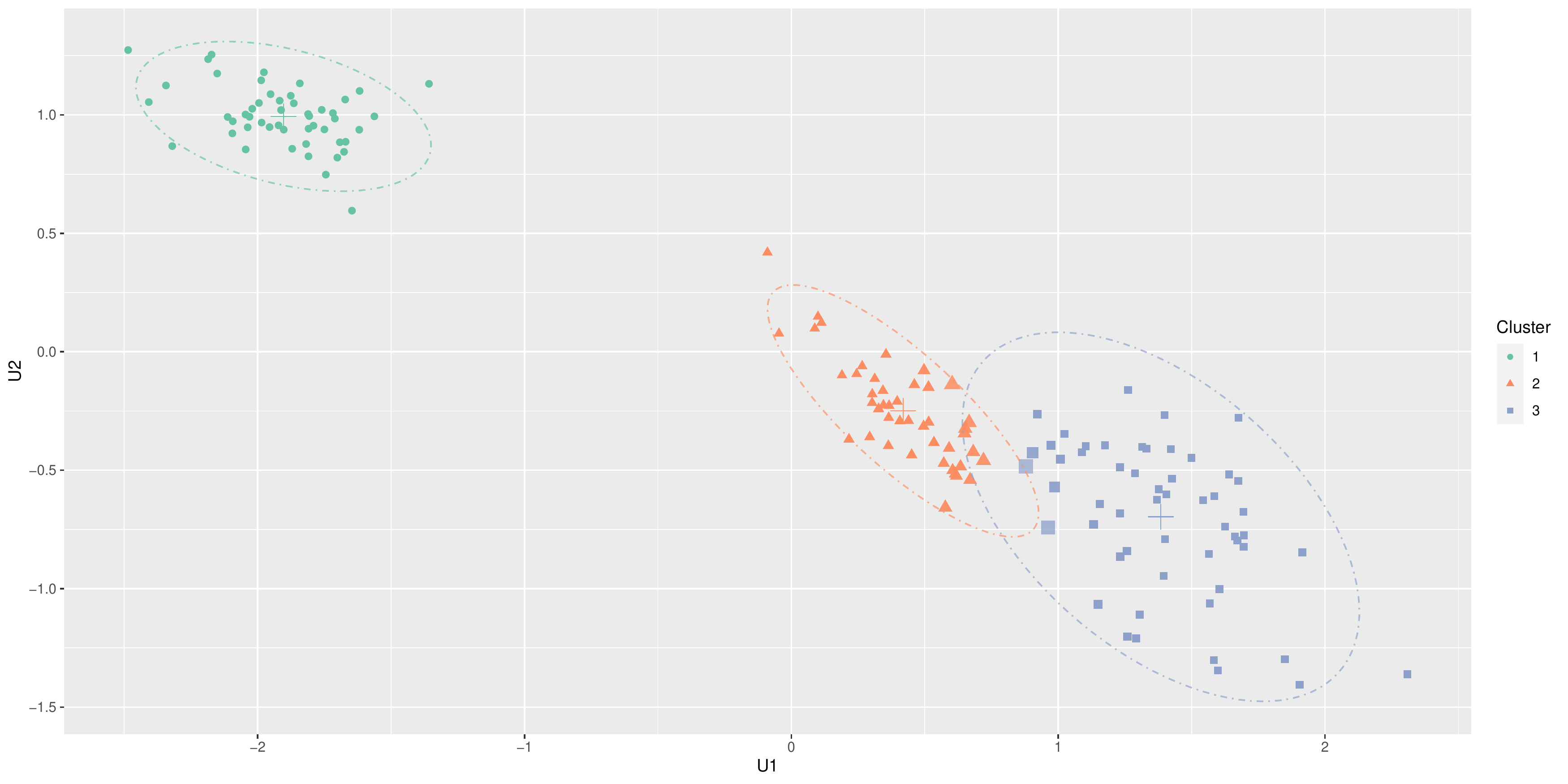}
		\label{sub:Iris}
	}
	\subfloat[Wine27]{
		\includegraphics[width=0.49\linewidth]{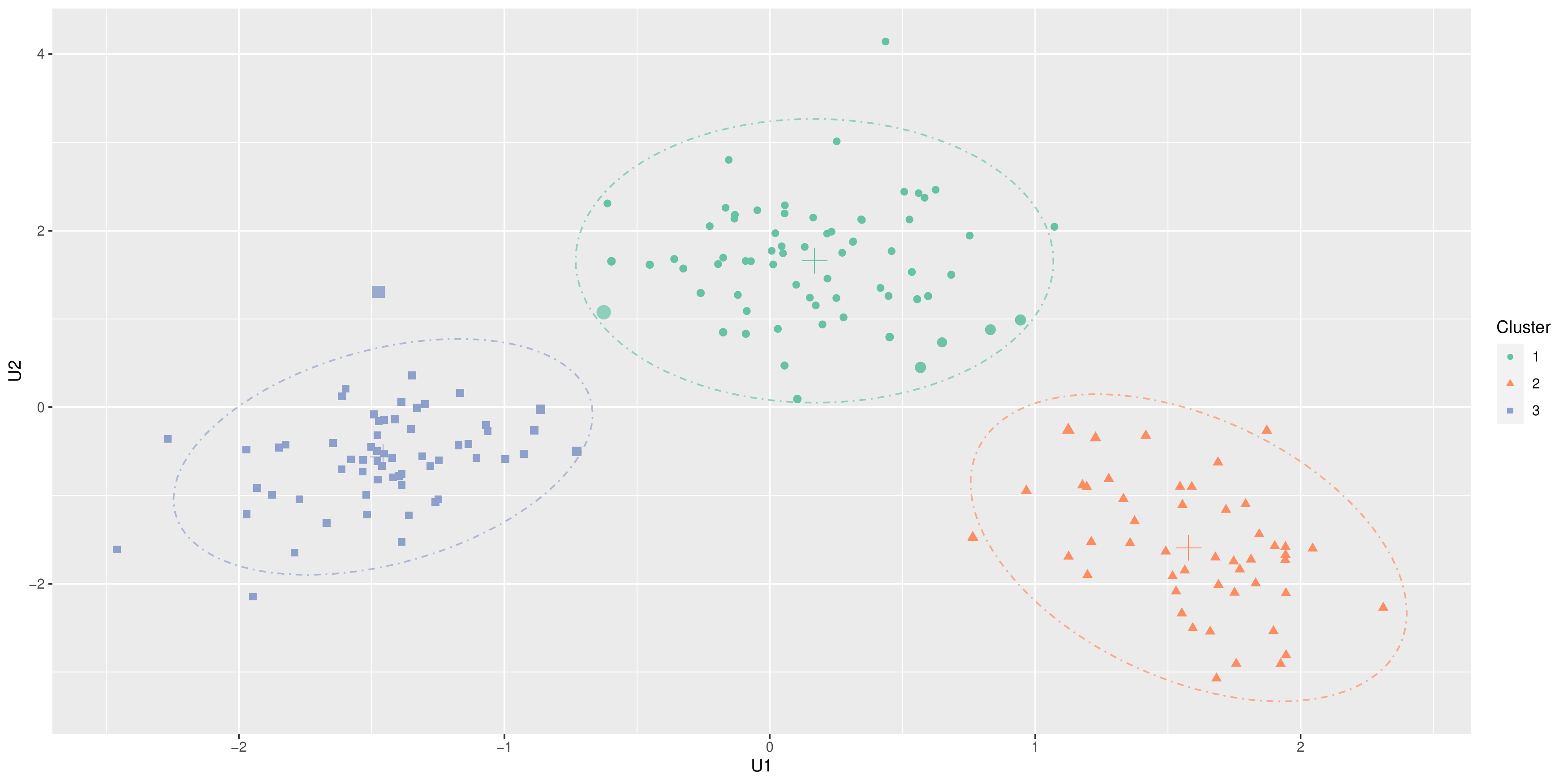}
		\label{sub:Wine27}
	} \\
	\subfloat[USPS358]{
		\includegraphics[width=0.49\linewidth]{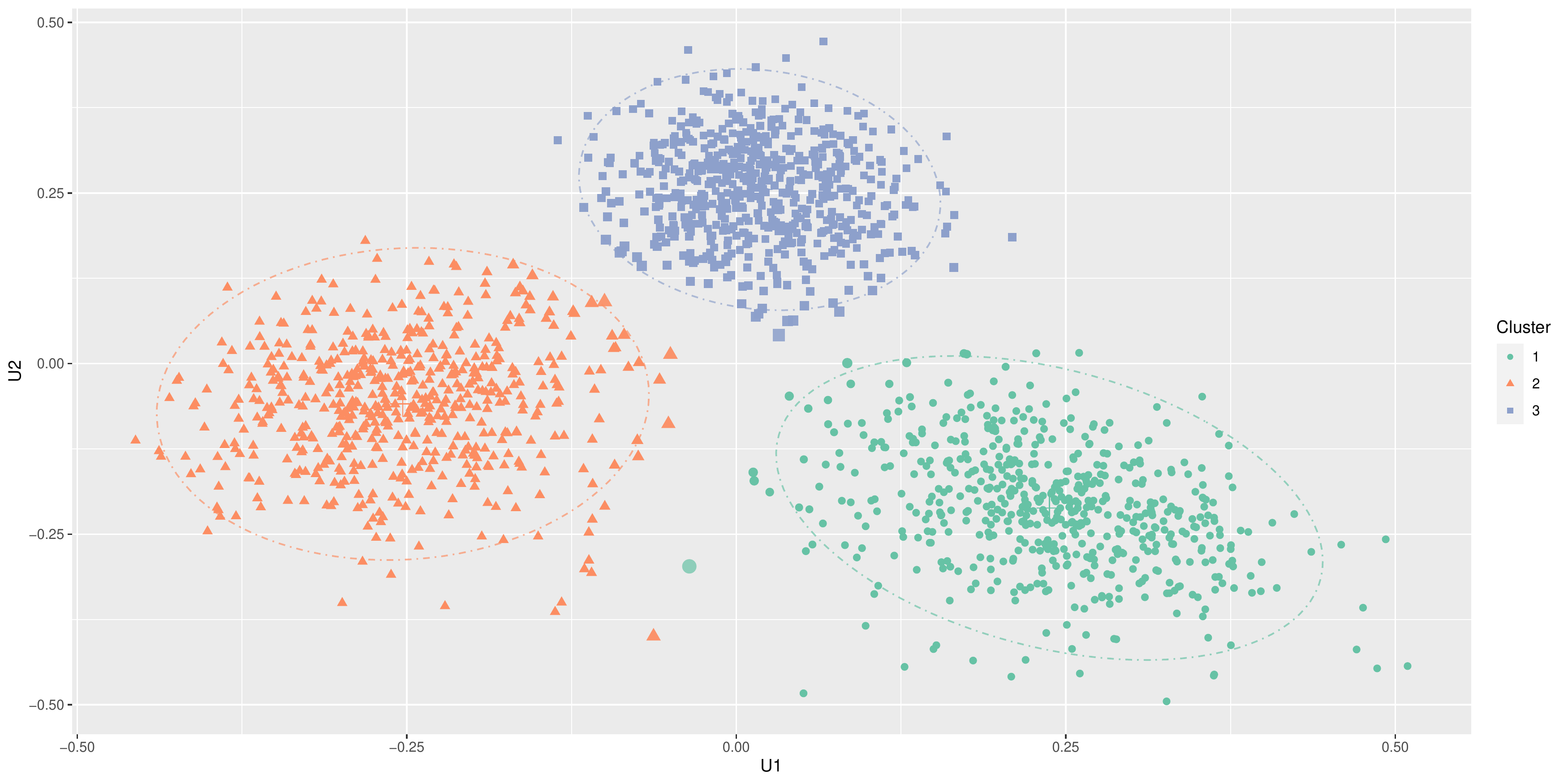}
		\label{sub:USPS358}
	}
	\caption{Iris, Wine27 and USPS358 datasets projected onto the two-dimensional subspace learned by BFEM.}
	\label{bfem:fig:RealDataSubspace}
\end{figure}

Here, we consider classical real-data benchmarks studied in the Gaussian subspace clustering literature:
\begin{itemize}
	\item Fisher's \textit{iris} is a traditional real dataset used to assess clustering algorithms, although it cannot be deemed as a high-dimensional problem. It consists in $150$ observations of $3$ iris species, $50$ each, described by $4$ variables. 
	\item The \textit{Italian Wine} dataset contains the description of 178 wines 27 variables related to \textit{e.g.} color or alcohol \parencite{von1986multivariate}. There are $\K=3$ types of wines, and we wish to know to which extent the $\dim 27$ variables can help relate to the type of wine. This dataset is also a famous introductory dataset for subspace clustering method \parencite{mcnicholas2008parsimonious, bouveyron2019model}.
	\item The \textit{Satellite} dataset consists in a satellite image decomposed into small sub-areas $\obs_i$ of $3 \times 3$ pixels. For each of these pixels, we have access to their value in $4$ spectral bands, leaving a total of $\dim = 4 \times 9 = 36$ values describing an area. The goal is to recover the classification of the central pixel in $\obs_i$ which can be one of the $\K=7$ different type of soils. 
	\item The \textit{USPS358} dataset is a more realistic example of high-dimensional data clustering. It is a subset of the US postal dataset from UCI, which originally contained $16 \times 16$ images of scanned digits from 0 to 9, with only digits 3, 5 and 8 known to be the most difficult to discriminate. There are $\nb= 1,756$ images, described by pixels values indicating gray level in dimension $\dim = 16 \times 16 = 256$ . In this scenario, we want to recover the 3 different digit classes.
\end{itemize}
These datasets are available in the \pkg{MBCbook} R package, except for the \textit{Satellite} data which is from the UC Irvine repository\footnote{Available at this address \url{https://archive.ics.uci.edu/ml/datasets/Statlog+(Landsat+Satellite)}}. Among them, only the \textit{Satellite} and \textit{USPS358} are high-dimensional problem in the sense that standard Gaussian mixtures fail to recover any relevant information. We ran each method with the true number of clusters of the corresponding dataset. The latent dimension of PGMM and MCFA was chosen by BIC with $\latentdim \in \{1,2,3\}$ for \textit{Iris} and $\latentdim \in \{1, \ldots, 20\}$ for \textit{Wine27}, \textit{USPS358} and \textit{Satellite}. All submodels were allowed for HDDC, including the one with different subspaces, and we used the ICL as the model selection criterion. \Cref{bfem:tab:RealDataBenchmarks} shows the results and we can see that, while PGMM performs better on \textit{Iris} and \textit{Wine27}, BFEM displays a real interest on the high-dimensional settings of \textit{Staellite} and \textit{USPS358} in dimension $\dim = 256$, achieving the top performance. Moreover, for the \textit{Wine27} dataset, MCFA and PGMM respectively chose $\latentdim = 6$ and $\latentdim = 4$ while HDDC selected a $[a_{j}bQd]$ model with $\latentdim_k = \latentdim=4$ in each class. A contrario, the FEM and BFEM works with $\latentdim=2$ and the discriminative subspace can therefore be plotted entirely in a two-dimensional graphic, as shown in \Cref{bfem:fig:RealDataSubspace}. Once again, the data are well separated Gaussian ellipses according indicating the clustering quality. For the \textit{Satellite} data, the problem is harder since there are $7$ classes, the PGMM model and MCFA models select $\latentdim = 20$ which is the maximum value, while HDDC selects the $[a_{kj}b_kQ_kd]$ model with $d=7$. 
\begin{table}[!ht]
	\renewcommand{\arraystretch}{1.4}
	\centering
	\caption{Comparison of the BFEM algorithm with other methods on common real datasets benchmarks.}
	\label{bfem:tab:RealDataBenchmarks}
%	\begin{tabular}{|l||cc|cc|}
%		\hline
%		Method &  \multicolumn{2}{c|}{Non-HD} &  \multicolumn{2}{c|}{HD} \tabularnewline
%		& Iris & Wine27 & Satellite  & USPS358 \\ \hline
%		BFEM   & 0.90                           & 0.95                          & \textbf{0.56} & \textbf{0.76}  \\
%		FEM    & 0.88                           & 0.93                            & 0.53                           & 0.66                          \\
%		HDDC   & 0.90                           & 0.93                          & 0.45                            & 0.35                           \\
%		MCFA   & 0.92                           & 0.96                          & 0.43                            & 0.28                           \\
%		PGMM   & \textbf{0.94} & \textbf{0.98} & 0.51                           & 0.38                           \\
%		\hline
%		Mclust & 0.90                           & 0.93                           & 0.36                           & 0                              \\ 
%		Kmeans & 0.73                           & 0.90                   & 0.53                                   & 0.64                           \\
%		\hline
%	\end{tabular}
	\begin{tabular}{|ll||cc|ccccc|}
		\hline
		 & & \multicolumn{2}{c|}{Non-HD models} & \multicolumn{5}{c|}{HD models} \\
%		 \hline
		Dataset & $\dim$& $k$-means & Mclust & HDDC & MCFA & PGMM & FEM & BFEM \\
		\hline
		\textit{Iris} & 4 &0.73 & 0.90 &	0.90 & 0.92 & \textbf{0.94} &   0.88 & 0.90  \\
		\hline
		\textit{Wine 27} & 27 & 0.90 & 0.93  & 0.95 & 0.96 & \textbf{0.98}  & 0.93 & 0.93 \\
		\hline
		\textit{Satellite} & \textbf{36} & 0.53 & 0.36  & 0.45 & 0.43 & 0.56 & 0.53 & \textbf{0.64} \\  
		\hline
		\textit{USPS358} & \textbf{256} & 0.64 & 0 & 0.35 & 0.28 & 0.38 & 0.66 & \textbf{0.76} \\
		\hline 
	\end{tabular}
\end{table}

%
%
%
%\section{Real data application}
%\label{bfem:sec:RealData}
%Pas le temps pour Curie.

\section{An application to patch-based Gaussian single image denoising}
\label{bfem:sec:application}
In this section, we present an authentic unsupervised application of high-dimensional model-based clustering, contrasting with common applications with supervised datasets.
\subsection{Single image denoising with the BFEM algorithm}
 Single image denoising consists in the restoration of an image degraded by an additive noise modeling the fluctuations induced by the physical process underlying recording devices such as camera \parencite{buades2005review}. While classical methods include local filters, Fourier or Wavelet transforms in the frequency domain, or sparse representations, patch-based methods have been successfully applied to the denoising problem \parencite{buades2011non}. Decomposing an image into a set of small sub-images or patches, of size $f \times f$, similar patches are supposed to be \textit{i.i.d} from the same distribution and can be averaged to reduce their noise in a non-local fashion. The restored image pixels can then be reconstructed by averaging over all the denoised patches they belong to. Denoting by $\bpatch_i \in \R^{\dim}$ a flattened original patch in dimension $\dim = f^2$, and $\obs_i$ its noisy observation, then patch-based image denoising assumes the following model:
\begin{equation}
\label{eq:NoiseModel}
\obs_i = \bpatch_i + \bm{\epsilon}_i, \qquad \bm{\epsilon}_i \sim \Gaussian_{\dim}(\bm{0}_{\dim}, \sigma^2 \bm{I}_{\dim}).
\end{equation}
Here, $\sigma^2$ denotes the variance of the additive Gaussian white noise, which is common over patches. In this context, Gaussian mixture models have been used extensively \parencite{yu2011solving, zoran2011learning} as they provide a sound probabilistic framework modeling each patch as:
\begin{equation}
	\p(\bpatch_i) = \sum_{k=1}^\K \pi_k \Gaussian_{\dim}( \bpatch_i \mid \bmean_{k}, \bm{\Phi}_k).
\end{equation} 
Thus, marginalizing on $\bpatch_i$ leaves the distribution:
\begin{equation*}
	\p(\obs_i) = \sum_{k=1}^\K \pi_k \Gaussian_{\dim}( \obs_i \mid \bmean_{k}, \bS_k ), \quad \bS_k = \bm{\Phi}_k + \sigma^2 \bI_{\dim},
\end{equation*}
and the conditional expectation of $\bpatch_i$ given $\obs_i$ may be used as the denoised patch estimate minimizing the expected mean squared error:
\begin{equation}
	\label{bfem:eq:PatchReconstruction}
		\hat{\bpatch}_i = \Expectation\left[ \bpatch_i \mid \obs_i \right] = \sum_{k=1}^{\K} \tau_{ik} \Expectation \left[\bpatch_i \mid \obs_i, \rawclust_{ik} = 1 \right].
\end{equation}
Here, $\tau_{ik} = \p(\rawclust_{ik} = 1 \mid \obs_i)$ denotes the posterior probabilities of the cluster membership $\clust_i$,
% or the variational approximation of \Cref{bfem:prop:CAVIupdateZ} in a Bayesian formulation.
and, ($\bpatch_i, \obs_i)$ being jointly Gaussian given $\clust_i$, the conditional expectation of a partitioned Gaussian vector leaves an exact formula for the right-hand side:
\begin{equation}
\label{bfem:eq:ConditionalPatch}
\Expectation \left[\bpatch_i \mid \obs_i, \rawclust_{ik} = 1 \right] = \bmean_{k} +  \left(\bS_k - \sigma^2 \bI_{\dim}\right) \bS_k^{-1} (\obs_i - \bmean_k).
\end{equation}
Thus, the final patch estimate $\hat{\bpatch}_i$ is a patch-dependent combination of $\K$ linear \textit{filters} (in the image processing terminology) of the noisy patch, and the denoising task amounts to compute the model parameters. However, the dimension being fixed by the patch size $\dim = f^2$, relatively high-dimensional problems arise even for common patch sizes found in the literature, \textit{e.g.} $f=8$ and $\dim = 64$. Thus, Gaussian subspace mixture models have been proposed to circumvent this issue. In their S-PLE algorithm, \textcite{wang2013sure} used the mixture of principal component analyzers \parencite[MPPCA]{tipping1999mixtures} which amounts to the low-rank constraint $\bS_k = \bU_k \bU_k^\top + \sigma^2 \bm{I}_\dim$. The authors also relaxed the original MPPCA model assumptions to allow each subspace $\bU_k$ to have its possibly different dimension $\latentdim_k$, and fixed the latter to be either $1$, $\dim/2$ or $\dim -1$. More recently, the high dimensional mixture model for image denoising \parencite[HDMI,][]{houdard2018high} used a slightly modified version of the HDDC model \parencite{bouveyron2007high}, considering the specific decomposition $\bS_k = \bU_k \bLambda_k \bU_k^\top + \sigma^2 \bI_{\dim}$, with $\bLambda_{k} = \diag(\lambda_{k1}, \ldots, \lambda_{k\latentdim_k})$ and $\bU_k \in \R^{\dim \times \latentdim_k}$ column-orthonormal. This approach yields state-of-the-art performances in denoising applications, and includes a procedure for selecting each dimension $\latentdim_k$ and the optimal level of noise through model selection. Here, we show that the BDLM model is also applicable in this context and that it yields satisfactory results compared to standard approaches. 

\paragraph{Denoising with the BFEM algorithm} Keeping the notations of \Cref{bfem:sec:BDLM}, we introduce a low-dimensional patch representation $\scores_{i} \in \R^d$ and the observed patch is supposed to be observed from model~\eqref{eq:NoiseModel} with the modification:
\begin{equation}
	\obs_{i} = \bU \scores_{i} + \bm{\epsilon}_i, \quad \scores_{i} \sim \Gaussian_{ \latentdim}(\bmu_k, \bLambda_{k}), \quad \bU^\top \bU = \bI_{\latentdim}.
\end{equation}
Thus, a noisy patch $\obs_i$ is supposed to be observed from the $\BDLM_{[\bSigma_{k}, \sigma^2]}$ model, with the decomposition:
\begin{equation}
	 \bS_k = \bD \bDelta_k \bD^\top, \textrm{ with: }\quad  \bD = [\bU, \bV], \quad \bDelta_k = \diag(\bSigma_k, \sigma^2 \bI_{\dim - \latentdim}).
\end{equation}
Note that this is a slightly modified version of the BDLM  model since the matrix $\bSigma_{k} = \bLambda_k + \sigma^2 \bI_{\latentdim}$ now depends on the noise variance. This modification was also proposed for the HDDC model in the HDMI algorithm, assuming that $\bLambda_{k}$ is diagonal but of variable dimension, and a specific EM was designed to keep the noise variance known and fixed to $\beta_k=\sigma^2$ throughout the estimation. Here, we propose to take an even simpler approach and to use the BFEM algorithm for the estimation of $\bU$ and $\bSigma_{k}$. Having derived an estimate of these quantities, the true noise level $\sigma^2$ can be used for supervised denoising in \Cref{bfem:eq:ConditionalPatch}. Moreover, this formula may be further simplified in order to inverse the block matrix $\bDelta_k$ instead of $\bS_k$ as the following proposition describes.
\begin{proposition}[Proof in \Cref{bfem:appendix:ProofPatchExpectation}]
	\label{bfem:prop:PatchExpectation}
	Under the BDLM model, the conditional patch expectation in cluster $k$ may be written:
	\label{bfem:prop:Denoising}
	\begin{equation}
	\Expectation \left[\bpatch_i \mid \obs_i, \rawclust_{ik} = 1 \right] = \bU \bvarmean_{k} +  \tilde{\bD} \left( \bI_{\dim} - \sigma^2 \bDelta_k^{-1} \right) \tilde{\bD}^\top (\obs_i - \bU \bvarmean_k).
	\end{equation}
	where $\tilde{\bD} = [\bU, \bm{0}_{\dim \times (\dim - \latentdim)}]$, $\bDelta_k = \diag(\hat{\bSigma}_k, \sigma^2 \bI_{\dim - \latentdim})$ and $\sigma^2$ is provided by the user.
\end{proposition}
Finally, we emphasize that even for basic image resolution, the number of observations $\nb$ can be quite large. For example a $512 \times 512$ image divided in patch size $f=8$, leaves us with $\nb = 255025$ patches in dimension $\dim = 64$. While this is good for estimation purposes, computations may be sped up by estimating the model parameters on a portion of the $\nb$ initial patches only. Indeed, the variational posterior cluster membership distribution, $\tau_{new,k}$, of a new data point $\obs_{new}$, may be directly computed from \Cref{bfem:prop:CAVIupdateZ} and used in the patch denoising formula in \Cref{bfem:eq:PatchReconstruction}.

\subsection{Denoising results on natural images} 

We now proceed to illustrate the performances of BFEM for image denoising with the proposed methodology. The standard quantitative way to measure the quality of a restored image $\hat{I}$ is the peak signal-to-noise ratio (PSNR), defined as:
\begin{align*}
	PSNR(I, \hat{I}) = 10 \times \log_{10} \left( \frac{255}{ \frac{1}{\vert \bm{\Omega} \vert} \sum_{x \in \bm{\Omega}} (I(x) - \hat{I}(x) )^2 }\right), \quad \bm{\Omega} = \{1,\ldots, n_r\} \times \{1, \ldots , n_c\},
\end{align*}
where $n_r$ and $n_c$ are the number of row and column pixels respectively. The denominator in the formula is the mean squared error and the higher the PSNR, the better the reconstruction. 

\begin{figure}[!ht]
	\centering
	\subfloat[Alley]{
		\includegraphics[width=0.33\linewidth]{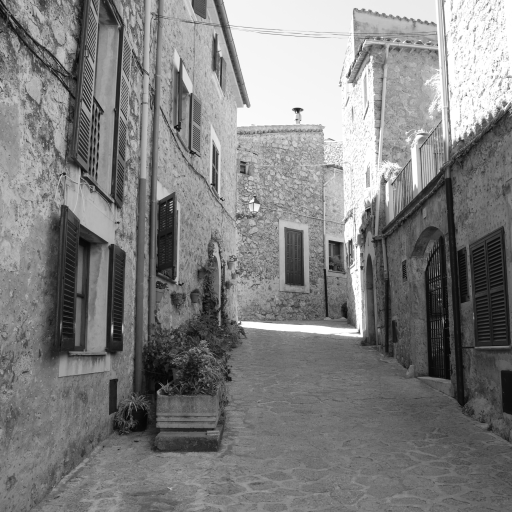}
	}
	\subfloat[Barbara]{
	\includegraphics[width=0.33\linewidth]{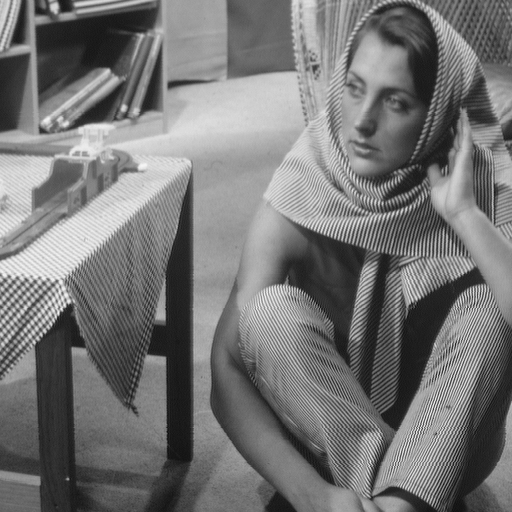}
	}
	\subfloat[Simpson]{
	\includegraphics[width=0.33\linewidth]{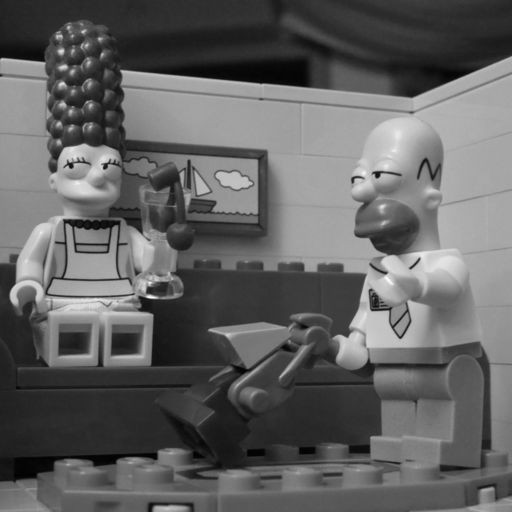}
	}
	\caption{The three grayscale benchmark images used to illustrate the denoising algorithms.}
	\label{fig:BenchmarkImages}
\end{figure}

As a point of comparison with other Gaussian subspace clustering methods in single image denoising, we report the results of \textcite[Table 4]{houdard2018high} on $3$ natural grayscale images of resolution $512 \times 512$: \textit{Alley}, \textit{Barbara} and \textit{Simpson} which are displayed in \Cref{fig:BenchmarkImages}. The first two are benchmark images in the denoising literature as they capture natural scenes with highly structured regions (the brick in \textit{Alley}, the tablecloth and scarf motives in \textit{Barbara}). The last one was introduced in the HDMI paper. The methods considered are Non-Local Bayes \parencite[NL Bayes]{lebrun2013nonlocal} which uses an unrestricted Gaussian model, the S-PLE \parencite{wang2013sure} and the HDMI \parencite{houdard2018high} described above. We emphasize that these methods are core-designed for single image denoising, with pre-processing and post-processing steps for NLBayes and S-PLE, which we do not include in our BFEM denoising methodology. As a matter of comparison, we also include a denoising with a standard HDDC clustering, which differs from HDMI since the noise $\beta_k$ is not fixed in the EM.
\Cref{tab:benchmark} gives the PSNR obtained for all the considered methods and two levels of standard error $\sigma \in \{ 20, 30\}$. In order to be consistent with the results taken from the HDMI paper, we use the same patch size of $f=10$, leaving a problem in dimension $\dim = 100$. We tried similar values as the one used in the denoising literature, and we present the results for $\K=40$ and $\K=90$. The models are estimated on the whole images, and the BFEM algorithm uses a latent dimension fixed to $\latentdim = 39$. The BFEM results are of the same order of magnitude that state-of-the-art approaches, albeit not being core designed for denoising. Moreover, the denoising results also display qualitative properties. 
 
\begin{table}[!ht]
	\centering
	\caption{PSNR results comparison for HDDC and BFEM denoising with other methods for Gaussian single image denoising methods NLBayes \parencite{lebrun2013nonlocal} with and without flat patch tricks, S-PLE \parencite{wang2013sure}.}
	\label{tab:benchmark} 
	\renewcommand{\arraystretch}{1.1}\textsc{}
%	\hbox to \textwidth{\hss
%		\begin{tabular}{lc|ccccc||cccc|}
%			\multicolumn{2}{l|}{} & \multicolumn{5}{c||}{Table 4 of \textcite{houdard2018high}} & \multicolumn{4}{c|}{New results} \tabularnewline
%			\hline 
%			Image & $\sigma$ & \multicolumn{2}{c}{NL-Bayes} & S-PLE &  \multicolumn{2}{c||}{HDMI$_{sup}$} & \multicolumn{2}{c}{HDDC} & \multicolumn{2}{c|}{BFEM} \tabularnewline
%			&  & \emph{original} & \emph{no flat} &  & {\footnotesize{}$K=40$} & {\footnotesize{}$K=90$} & {\footnotesize{}$K=40$} & {\footnotesize{}$K=90$} & {\footnotesize{}$K=40$} & {\footnotesize{}$K=90$} \tabularnewline
%			\hline 
%			\multirow{2}{*}{\emph{Barbara}} 
%			& 20 & 31.52 & 31.29  &  30.37 & 31.32 & 31.61 & 30.71 & 30.45 & 30.77 & 31.07 \tabularnewline
%			& 30 & 29.72 & 29.44  & 28.22 &  29.31 & 29.49 & 28.62 & 28.66 & 28.73 & 28.40 \tabularnewline
%			\hline 
%			\multirow{2}{*}{\emph{Simpson}} 
%			& 20  & 34.74 & 33.72  & 34.08 & 35.05  & 34.91 & 33.71 & 33.86 & 34.13 & 34.26 \tabularnewline
%			& 30  & 32.53 & 31.54  & 31.53 &  32.33 & 32.50 & 31.25 & 31.34 & 31.35 & 31.46  \tabularnewline
%			\hline 
%			\multirow{2}{*}{\emph{Alley}} 
%			& 20 & 29.10 & 29.07  & 28.67  &  29.03 & 29.07 & 28.07 & 28.11 & 28.26 & 28.37 \tabularnewline
%			& 30  & 27.43 & 27.37  & 26.92 &  27.31 & 27.39 & 25.61 & 26.09 &26.20 & 26.17 \tabularnewline
%			\hline 
%		\end{tabular}
%		\hss}
	\hbox to \textwidth{\hss
	\begin{tabular}{lc|ccccc||cccc|}
		\multicolumn{2}{l|}{} & \multicolumn{5}{c||}{Table 4 of \textcite{houdard2018high}} & \multicolumn{4}{c|}{New results} \tabularnewline
		\hline 
		Image & $\sigma$ & \multicolumn{2}{c}{NL-Bayes} & S-PLE &  \multicolumn{2}{c||}{HDMI$_{sup}$} & \multicolumn{2}{c}{HDDC} & \multicolumn{2}{c|}{BFEM} \tabularnewline
		&  & \emph{original} & \emph{no flat} &  & {\footnotesize{}$K=40$} & {\footnotesize{}$K=90$} & {\footnotesize{}$K=40$} & {\footnotesize{}$K=90$} & {\footnotesize{}$K=40$} & {\footnotesize{}$K=90$} \tabularnewline
		\hline 
		\multirow{2}{*}{\emph{Barbara}} 
		& 20 & 31.52 & 31.29  &  30.37 & 31.32 & \textbf{31.61} & 30.93 & \textbf{31.35} & 29.88 & 30.82 \tabularnewline
		& 30 & \textbf{29.72} & 29.44  & 28.22 &  29.31 & 29.49 & 28.40 & 28.91 & 28.46 & \textbf{28.95} \tabularnewline
		\hline 
		\multirow{2}{*}{\emph{Simpson}} 
		& 20  & 34.74 & 33.72  & 34.08 & 35.05  & \textbf{34.91} & 34.45 & 34.17 & 34.32 & \textbf{34.54} \tabularnewline
		& 30  & \textbf{32.53} & 31.54  & 31.53 &  32.33 & 32.50 & 31.77 & 32.02 & 31.95 & \textbf{32.06}   \tabularnewline
		\hline 
		\multirow{2}{*}{\emph{Alley}} 
		& 20 & \textbf{29.10} & 29.07  & 28.67  &  29.03 & 29.07 & 28.55 & 27.95& 28.51 &  \textbf{28.75} \tabularnewline
		& 30  & \textbf{27.43} & 27.37  & 26.92 &  27.31 & 27.39 & 26.71 & 27.06 & 27.05 & \textbf{27.16} \tabularnewline
		\hline 
	\end{tabular}
	\hss}
\end{table}

\begin{figure}[p]
	\centering
	\subfloat[Original]{\includegraphics[width=0.33\linewidth]{alley_512.png}}
	\subfloat[Noisy, $\sigma=30$]{\includegraphics[width=0.33\linewidth]{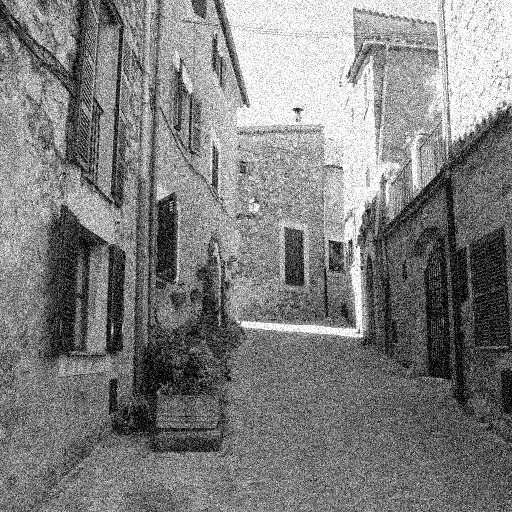}}
	\subfloat[NLBayes]{	\includegraphics[width=0.33\linewidth]{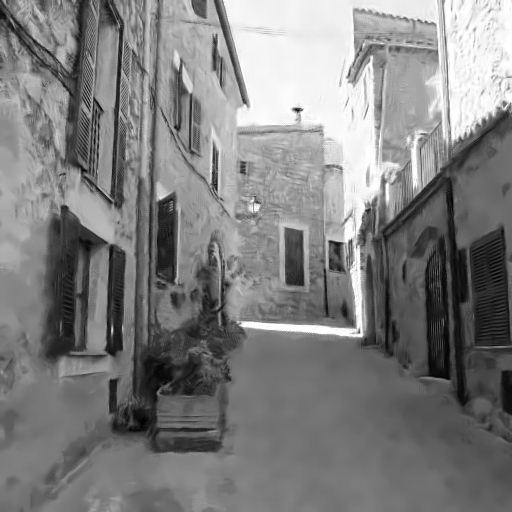}
	} \\
	\subfloat[S-PLE]{		\includegraphics[width=0.33\linewidth]{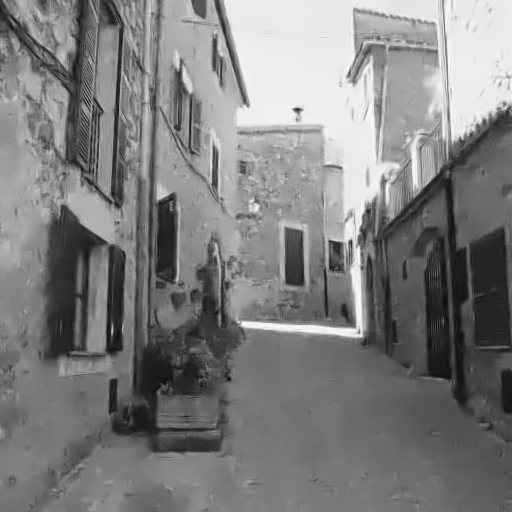}} % \\
	\subfloat[HDMI]{			\includegraphics[width=0.33\linewidth]{alley_30K_90_HDMI_sigma_denoised.png}}
%	\subfloat[HDDC]{				\includegraphics[width=0.33\linewidth]{alley-stdv30-denoisedHDDC-K90.png}}
	\subfloat[BFEM]{							\includegraphics[width=0.33\linewidth]{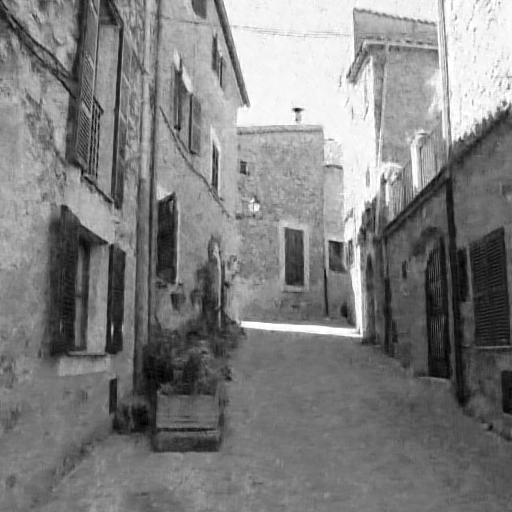}}		\hfill

	\caption{Qualitative results of each denoising methods at $\sigma=30$ on the \textit{Alley} gray-scale image. Images should be seen at full resolution on the electronic version of the paper.}
	\label{fig:AlleyResults}
\end{figure}

While the PSNR is a good indicator of the reconstruction error, the denoising results may vary widely in term of restoration quality. \Cref{fig:AlleyResults} shows the reconstructed images by each of the considered methods for the \textit{Alley} image. We can see that the highly textured motives like the pavement of the alley are blurred, and high-frequency location such as the electric cables in the white sky are lost by S-PLE and NLBayes. A contrario, BFEM reconstructs these motives really well, with visual results comparable to HDMI denoising which achieves state-of-the-art performances in this context. This denoising is obtained without knowing the noise level and we emphasize that a refinement of the algorithm could be easily designed for denoising, keeping the noise variance fixed to $\beta_{k} = \sigma^2$ during estimation. The methodology also easily extends to color images, stacking the patch pixels in the red, green and blue channel and leaving a problem in dimension $\dim = 3 f^2$. 

\section{Conclusion and perspectives}

In this paper, we introduced a new clustering algorithm for high-dimensional continuous data. Relying on a Bayesian formulation of the discriminative latent mixture model, it relies on variational inference for the estimation of the model parameters. Building on an unsupervised version of Fisher discriminant analysis, the discriminative subspace is estimated iteratively throughout the algorithm with soft scatter matrices updated with the posterior membership probabilities at each step. An empirical Bayes strategy was proposed for setting the hyper-parameter $\lambda$, controlling between-class separation in the latent space, and an ICL criterion was derived for model selection. The simulations assesed the stability of the method with respect to both high-dimension and noise, and showed its significant improvement over the frequentist version as well as other Gaussian subspace clustering methods. Finally, we trust that the image denoising application provides an interesting setting that could inspire the model-based clustering community in the design of specific algorithms.
% favorably with respect to the frequentist Fisher-EM and other state-of-the-art Gaussian subspace clustering algorithms.

In further works, we plan to consider two types of extensions for the BFEM algorithm. First, variable selection could be added in the model through sparsity constraints on the matrix $\bU$. A sparse version of the BFEM could be designed, in the line of \textcite{bouveyron2014discriminative}, based on the regression formulation of the Fisher criterion with a Lasso-like penalty \parencite{qiao2009sparse}. Second, the trace of ratio formulation in \Cref{eq:FisherCriterion} is itself a simplification of the ratio of trace problem:
\[
\max_{\bU^\top \bU = \bI_\latentdim} \frac{\Tr[\bU^\top \bS_B \bU]}{\Tr[\bU^\top \bS_W \bU]},
\] 
which does not accept a closed-form solution and necessitate iterative algorithms to solve. Recent works highlight the better discriminative power of this new formulation in the supervised context, and new iterative algorithms have been designed  \parencite{guo2003generalized, wang2007trace, kokiopoulou2011trace}. Based on these empirical results, it would be interesting to modify the F-step in order to maximize such a criterion. Given the diversity of possible algorithms and their need for calibration, we leave a detailed study to future work.

\section*{Aknowledgments}

 The authors wish to thank Antoine Houdard for the images and helpful discussions on image denoising. 

 This work has benefited from the support of the French government, through the 3IA Côte d’Azur Investment in the Future project managed by the National Research Agency (ANR) with the reference number ANR-19-P3IA-0002. This work was also supported by a DIM MathInnov grant from Région Ile-de-France. The authors are finally thankful for the support from fédération F2PM, CNRS FR 2036, Paris.

% Authors must disclose all relationships or interests that 
% could have direct or potential influence or impart bias on 
% the work: 
%
\section*{Conflict of interest}

The authors declare that they have no conflict of interest.

\clearpage

\appendix

\section{Proofs}

\subsection{Optimization of \texorpdfstring{$\q(\Clust)$}{q(Z)}}
\label{bfem:appendix:ProofCAVIupdateZ}

\begin{proof}[Proof of \Vref{bfem:prop:CAVIupdateZ}]
	A classical result of the CAVI algorithm, see \textit{e.g.} \textcite[Equation (18)]{blei2017variational}, states that the lower bound $\J$ is maximized with respect to $\q(\clust_i)$ by:
	\begin{align}
	\q^\star(\clust_{i}) \propto \exp\left\{ \mathbb{E}_{\clust_{-i}, \bmu} \left[\log \p(\Obs, \bmu, \clust_{i}, \clust_{-i} \mid \globalparam) \right]\right\}.	\end{align}
	Here, the expectation is taken with respect to all the variables in $\bmu$ and $\Clust$ except for $\clust_i$. Taking the log of this expression and leaving out everything that does not depend on $\clust_{i}$ leads to the following functional form:
	\begin{align}
	\log \q^\star(\clust_{i}) &= \mathbb{E}_{\bmu_k }\left[\sum_k \rawclust_{ik} \left[\log(\pi_k) + \log \mathcal{N}_p(\obs_i \mid \bU \bmu_k, \bS_k) \right]\right]+ \textrm{C}, \\
	&=\sum_k \rawclust_{ik} \left[\log(\pi_k) + \mathbb{E}_{\bmu_k }\left[\log \mathcal{N}_p(\obs_i \mid \bU \bmu_k, \bS_k) \right]\right]+ \textrm{C}.
	\end{align}
	Here we recognize the functional form of a multinomial distribution:
	\begin{equation}
	\q^\star(\clust_{i}) = \Mult_{\K}(\clust_{i} \mid 1, \bTau_i),
	\end{equation}
	with:
	\begin{align}
	\tau_{ik} & \propto \pi_k \exp \{ \mathbb{E}_{\bmu_k}\left[ \log \mathcal{N}_p(\obs_i \mid \bU \mu_k, \bS_k)\right]\}.
	\end{align}
\end{proof}

\subsection{Optimization of \texorpdfstring{$\q(\bmu)$}{q(mu)}}
\label{bfem:appendix:ProofCAVIupdateMu}
\begin{proof}[Proof of \Vref{bfem:prop:CAVIupdateMu}]
	Let $k \in \{1, \ldots, K\}$, again the optimal $\q(\bmu_k)$ verifies:
	\begin{align}
	\q^\star(\bmu_k) \propto \exp\left\{ \mathbb{E}_{\Clust, \bmu_{-k}} \left[\log \p(\Obs, \bmu_k, \bmu_{-k}, \Clust) \mid \globalparam \right]\right\},
	\end{align}
	where the expectation is taken with respect to all variables in $\Clust$ and $\bmu$ expect for $\bmu_k$
	Taking the logarithm of this expression and leaving out everything that does not depend on $\bmu_k$ leads to the following functional form:
	\begin{align}
	\log \q^\star(\bmu_k) & = \log \p(\bmu_k) + \sum_{i=1}^{n} \mathbb{E}_{\clust_{i}, \bmu_{-k}} \left[\log \p(\obs_i \mid \clust_{i}, \bmu_{-k}) \right] + \textrm{C}_1, \\
	&= \log \mathcal{N}_d(\bmu_k \mid \bnu, \lambda \bI_{\latentdim}) + \sum_{i=1}^{n} \tau_{ik} \log \mathcal{N}_p(\obs_i \mid \bU \bmu_k, \bS_k) + \textrm{C}_2 \\
	&=  - \frac{1}{2} \left[ \lambda^{-1} (\bmu_k - \bnu)^\top (\bmu_k - \bnu) + \sum_{i=1}^n \tau_{ik} (\obs_i - \bU \bmu_k)^\top \bS_k^{-1} (\obs_i - \bU \bmu_k) \right] + \textrm{C}_3, \\
	&= - \frac{1}{2} \left[ \lambda^{-1} \bmu_k^\top \bmu_k + \sum_{i=1}^n \tau_{ik} \bmu_k^\top \bU^\top \bS_k^{-1} \bU \bmu_k \right. \nonumber  \\
	& \left. \qquad \qquad \qquad \qquad - 2 \left(\sum_{i=1}^n \tau_{ik}\obs_i^\top \bS_k^{-1} \bU \bmu_k + \lambda^{-1} \bnu^\top \bmu_k \right)  \right] + \textrm{C}_4, \\
	&= - \frac{1}{2} \left[\bmu_k^\top ( \lambda^{-1} \bI_{\latentdim} + \varn_k \bU^\top \bS_k^{-1} \bU) \bmu_k - 2 \left(\sum_{i=1}^n \tau_{ik}\obs_i^\top \bS_k^{-1} \bU + \lambda^{-1} \bnu^\top\right) \bmu_k \right] + \textrm{C}_4.
	\end{align}
	Putting:
	\begin{align}
	\bvarcovar_k &= \left(\lambda^{-1} \bI_{\latentdim} + \varn_k \bU^\top \bS_k^{-1} \bU\right) ^{-1}, \\
	\bvarmean_k &=   \bvarcovar_k \left( \sum_{i=1}^{n} \tau_{ik} \bU^\top \bS_k^{-1} \obs_i + \lambda^{-1} \bnu \right)
	\end{align}
	we can then write:
	\begin{align}
	\log \q^\star(\bmu_k) =  - \frac{1}{2} \left[  (\bmu_k - \bvarmean_k)^\top \bvarcovar_k^{-1}(\bmu_k - \bvarmean_k)  \right] + \cst.
	\end{align}
	We recognize the logarithm of a Gaussian density with mean $\bvarmean_k$ and covariance $\bvarcovar_k$.	Moreover, the expressions of $\bvarmean_{k}$ and $\bvarcovar_k$ can be simplified, since $\bS_k = \bD \bDelta_k \bD^\top $ with $\bD = [\bU, \bV]$ and $\bV$ is the orthogonal complement of $\bU$. Thus, $\bS_k^{-1} = \bD \bDelta_k^{-1} \bD^\top$ and:
	\begin{align}
	\bU^\top \bS_k^{-1}	& = \bU^\top \bD \bDelta_k^{-1} \bD^\top = \bSigma_k^{-1} \bU^\top,\\
	\bU^\top \bS_k^{-1} \bU & =  \bSigma_k^{-1}.
	\end{align}
	Thus, a use of Woodbury's identity gives
	\begin{align*}
	\bvarcovar_k \lambda^{-1} \bnu =&  \left(\bI_{\latentdim} + \varn_k \bI_{\latentdim} \lambda\bI_{\latentdim}\bSigma_k^{-1}\right)^{-1} \bnu,  \\
	=& \left[ \bI_{\latentdim} - \bI_{\latentdim} \varn_k \left(\lambda^{-1} \bI_{\latentdim} + \varn_k \bSigma_{k}^{-1}\right)^{-1} \bSigma_{k}^{-1} \bI_{\latentdim} \right] \bnu,  \\
	=& \bnu - \bvarcovar_k \bSigma_{k}^{-1} \varn_k \bnu,
	\end{align*}
	and we finally get:
	\begin{align}
	\bvarmean_k &=   \bvarcovar_k \left( \sum_{i=1}^{n} \tau_{ik} \bU^\top \bS_k^{-1} \obs_i + \lambda^{-1} \bnu \right), \nonumber \\
	&= \bvarcovar_k \bSigma_{k}^{-1} \left( \sum_{i=1}^{n} \tau_{ik} \bU^\top \obs_i \right) + \bvarcovar_k \lambda^{-1} \bnu ,  \nonumber \\
	&= \bnu + \bvarcovar_k \bSigma_{k}^{-1} \left( \sum_{i=1}^{n} \tau_{ik} \bU^\top \obs_i -  \varn_k \bnu \right).
	\end{align}
\end{proof}

\subsection{Variational lower bound}
\label{bfem:appendix:ProofELBO}
We recall
\[
\J(\q, \globalparam) =    \Expectation_{\q}\left[ \log \p(\Obs, \bmu , \Clust\mid \globalparam)\right] - \Expectation_\q\left[ \log \q(\bmu, \Clust)\right]. \\
\]
Now that $\q^\star$ have been derived in \Cref{bfem:prop:CAVIupdateZ,bfem:prop:CAVIupdateMu}, we can compute the variational lower bound explicitly:
\begin{equation}
\label{bfem:eq:FullELBO}
\begin{aligned}
\J(\q, \globalparam)
%= &\mathbb{E}_{\Clust, \bmu} \left[ \log \p(\Obs \mid \Clust, \bmu , \globalparam)\right] +  \mathbb{E}_{\q(\Clust) }\left[\log\p(\Clust) \right] + \mathbb{E}_{\q(\bmu) }\left[\log\p(\bmu) \right]  + \entropy(\q), \\
= & -\frac{1}{2} \sum_{i=1}^{n} \sum_{k=1}^{K} \tau_{ik}\left\{ p \log(2\pi) + \log \vert \bS_k \vert + \mathbb{E}_{\bmu_k}\left[ (\obs_i - \bU \bmu_k)^\top \bS_k^{-1} (\obs_i - \bU \bmu_k)  \right] \right\} \\
&+ \sum_{k=1}^{K} \varn_k \log(\pi_k) \\
& -\frac{1}{2} \sum_{k=1}^{K}  d \log(2\pi) + d \log(\lambda) + \frac{1}{\lambda} \mathbb{E}_{\bmu_k}\left[ (\bmu_k - \bnu)^\top (\bmu_k - \bnu) \right] \\
& - \sum_i \sum_k \tau_{ik} \log(\tau_{ik}) \\
& + \frac{Kd}{2} (\log(2\pi) + 1) + \frac{1}{2} \sum_k \log \vert \bvarcovar_k \vert,
\end{aligned}
\end{equation} 
with:
\begin{equation}
\label{eq:UsefulQuantities}
\begin{aligned}
\Expectation_{\q^\star}[\bmu_{k}] &= \bvarmean_{k}, \\
\Expectation_{\q^\star}[\bmu_{k}\bmu_{k}^\top] &= \bvarmean_{k} \bvarmean_{k}^\top + \bvarcovar_k, \\
\log \vert \bS_k \vert & = \log \vert \bSigma_{k} \vert + (\dim - \latentdim) \log (\beta_{k}).
\end{aligned}
\end{equation}

\begin{proof}[Proof of \Vref{bfem:prop:ELBO}] 
	Recall the form of the bound given in \Cref{bfem:prop:ELBO}:
	\begin{equation}
	\begin{aligned}
	\J(\globalparam) = & \cst - \frac{1}{2} \sum_{k=1}^{\K} \varn_k \Bigg\{ -2 \log(\pi_k) + \log \vert \bSigma_{k} \vert + (\dim - \latentdim) \log(\beta_{k})   \\
	& \quad +  \Tr\left[ \bSigma_k^{-1} \bU^\top \bhatC_k \bU\right] + \frac{1}{\beta_k} \left( \Tr\left[\bhatC_k\right] -  \Tr\left[ \bU^\top \bhatC_k \bU \right] \right)  \Bigg\} .
	\end{aligned}
	\end{equation}
	Only the first two lines of \Cref{bfem:eq:FullELBO} depend on $\globalparam$. Focusing on the first line:
	\begin{align*}
	& \Expectation_\q \left[\log \p (\Obs \mid \Clust, \bmu , \globalparam)\right] \\
	= &  \sum_{i=1}^{n} \sum_{k=1}^{K} \tau_{ik}\left\{ p \log(2\pi) + \log \vert \bS_k \vert + \mathbb{E}_{\bmu_k}\left[ (\obs_i - \bU \bmu_k)^\top \bS_k^{-1} (\obs_i - \bU \bmu_k)  \right] \right\} \\
	= & \gamma + \sum_{k=1}^{K} \varn_k \left\{ \log \vert \bS_k \vert + \frac{1}{\varn_k} \mathbb{E}_{\bmu_k}\left[ \sum_i \tau_{ik} (\obs_i - \bU \bmu_k)^\top \bS_k^{-1} (\obs_i - \bU \bmu_k)  \right] \right\},
	\end{align*}
	where $\gamma = \nb \dim \log(2\pi)$. The terms inside the expectation may be rearranged using the usual \textit{trace trick} to make the empirical cluster covariance matrices appear:
	\begin{align}
	\label{eq:TrSkCk}
	& \frac{1}{\varn_k} \sum_i \tau_{ik} (\obs_i - \bU \bmu_k)^\top \bS_k^{-1} (\obs_i - \bU \bmu_k), \nonumber\\
	= &  \frac{1}{\varn_k} \sum_i \tau_{ik} \Tr \left[ (\obs_i - \bU \bmu_k)^\top \bS_k^{-1} (\obs_i - \bU \bmu_k)\right],\nonumber \\
	=& \Tr \left[\bS_k^{-1}  \frac{1}{\varn_k} \sum_i \tau_{ik} (\obs_i - \bU \bmu_k) (\obs_i - \bU \bmu_k)^\top\right],\nonumber \\
	=& \Tr \left[\bS_k^{-1}  \bC_k\right],
	\end{align}
	where
	\begin{equation*}
	\label{eq:Ck}
	\bC_k = \frac{1}{\varn_k} \sum_i \tau_{ik} (\obs_i - \bU \bmu_k) (\obs_i - \bU \bmu_k)^\top,
	\end{equation*}
	is the empirical covariance matrices of cluster $k$. This trace may be further decomposed using the following lemma which relies on the particular form of $\bS_k^{-1}$ in the $\BDLM$ model.
	\begin{lemma}
		\label{lemma:TrSkA}
		For any square matrix $\bA \in \mathcal{M}_{p\times p}(\mathbb{R})$, the following identity holds:
		\begin{align}
		\label{eq:TrSkA}
		\Tr\left[\bS_k^{-1} \bA \right] = \Tr\left[ \bSigma_k^{-1} \bU^\top \bA \bU \right] + \frac{1}{\beta} \left(\Tr\left[\bA\right] - \Tr\left[\bU^\top \bA \bU\right]\right)
		\end{align} 
	\end{lemma}
	\begin{proof}
		We can split $\bS_k^{-1}$ in two parts depending on the discriminative subspace and its orthogonal complement: take $\bD = \tilde{\bD} + \bar{\bD}$, with $\tilde{\bD} = [\bU, \bm{0}_{\dim \times (\dim - \latentdim)}]$ and $\bar{\bD} = [\bm{0}_{\dim \times \latentdim}, \bV]$. Then, $\bS_k^{-1} = \tilde{\bD} \bDelta_k^{-1} \tilde{\bD}^\top + \bar{\bD} \bDelta_k^{-1} \bar{\bD}^\top$, and:
		\begin{align*}
		\Tr\left[\bS_k^{-1} \bA \right] &= \Tr\left[ \bDelta_k^{-1} \tilde{\bD}^\top \bA \tilde{\bD} \right] +  \Tr\left[ \bDelta_k^{-1} \bar{\bD}^\top \bA \bar{\bD} \right] , \\
		&= \Tr\left[ \bSigma_k^{-1} \bU^\top \bA \bU \right] +  \frac{1}{\beta} \Tr\left[ \bV^\top \bA \bV \right] 
		\end{align*}
		Moreover, $\bD \bD^\top = \bD^\top \bD = I_p$ and $\bD = \tilde{\bD} + \bar{\bD}$, hence:
		\begin{align*}
		\Tr\left[\bA\right] & = \Tr\left[\bD^\top \bA \bD \right] = \Tr\left[\bU^\top \bA \bU\right] + \Tr\left[\bV^\top \bA \bV \right]
		\end{align*}
		This concludes \Cref{lemma:TrSkA}'s proof.
	\end{proof}
	\noindent Applying \Cref{lemma:TrSkA} to \Cref{eq:TrSkCk} with $\bA = \mathbb{E}_{\bmu_k}\left(\bC_k\right)$ leaves:
	\begin{equation}
	\label{eq:DLMpart2}
	\begin{aligned}
	\Expectation_\q \left[\log \p (\Obs \mid \Clust, \bmu , \globalparam)\right] =& \gamma + \sum_{k=1}^{K} \varn_k \Bigg\{ \log \vert \bSigma_k \vert + (p-d) \log(\beta_ k)  + \Tr\left[ \bSigma_k^{-1} \bU^\top \mathbb{E}(\bC_k) \bU\right] \\
	& + \frac{1}{\beta} \left(\Tr\left[\mathbb{E}(\bC_k)\right] -  \Tr\left[ \bU^\top \mathbb{E}(\bC_k) \bU \right] \right) \Bigg\}.
	\end{aligned}
	\end{equation}
	The matrix $\mathbb{E}_{\bmu_k}\left(\bC_k\right)$ is denoted as $\bhatC_k$ and, using \Cref{eq:UsefulQuantities}, one gets:
	\begin{align}
	\bhatC_k & =\mathbb{E}_{\bmu_k}\left(\bC_k\right), \nonumber \\
	&= \frac{1}{\varn_k} \sum_i \tau_{ik} \Expectation \left[(\obs_i - \bU \bmu_k) (\obs_i - \bU \bmu_k)^\top\right], \nonumber \\
	&= \frac{1}{\varn_k} \sum_i \tau_{ik}\left( \obs_i\obs_i^\top - \bU  \bvarmean_{k} \obs_{i}^\top - \obs_{i} (\bU \bvarmean_{k})^\top + \bU \Expectation\left[\bmu_k \bmu_k^\top\right] \bU^\top \right) , \\ \nonumber
	&= \frac{1}{\varn_k} \sum_{i=1}^{\nb} \tau_{ik} (\obs_{i} - \bU\bvarmean_k)(\obs_{i} - \bU \bvarmean_{k})^\top + \bU \bvarcovar_k \bU^\top.
	\end{align}
	Finally, the second line of \Cref{bfem:eq:FullELBO} is simply 
	\begin{align}
	\Expectation_\q[\log\p(\Clust \mid \bPi)] = -\frac{1}{2} \sum_{k=1}^{\K} -2 \varn_k \log(\pi_k).
	\end{align}
	This concludes the proof.
\end{proof}

\subsection{M-step}
\label{bfem:appendix:ProofMstep}

\begin{proof}[Proof of \Vref{bfem:prop:M-step}]
	Although there are $12$ different submodels, a lot of the proofs are the same. 
	\begin{proof}[Optimization of $\bBeta$]
		Let us start with the two possible cases for $\bBeta$, which are common regardless of the constraint on the latent covariance matrices:
		\begin{itemize}
			\item Model $\BDLM_{[(\cdot) \beta_k]}$: In this case, the variational bound as a function of $\beta_{k}$ is:
			\begin{align*}
			\J(\beta_{k}) = - \frac{1}{2} \varn_k \left[(\dim - \latentdim) \log(\beta_{k}) + \frac{1}{\beta_k} \left( \Tr\left[\bhatC_k\right] -  \Tr\left[ \bU^\top \bhatC_k \bU \right] \right) \right].
			\end{align*}
			Thus, its only stationary point is:
			\begin{align*}
			\nabla_{\beta_{k}} \J(\hat{\beta_{k}}) = 0 \iff \hat{\beta_{k}} = \frac{\Tr\left[\bhatC_k\right] -  \Tr\left[ \bU^\top \bhatC_k \bU \right]}{p-d}.
			\end{align*}
			
			\item Model $\BDLM_{[(\cdot) \beta]}$: In this case, the variational bound as a function of $\beta$ is:
			\begin{align*}
			\J(\beta) &= - \frac{1}{2} \sum_{k=1}^{\K}\varn_k \left\{(\dim - \latentdim) \log(\beta) + \frac{1}{\beta} \left( \Tr\left[\bhatC_k\right] -  \Tr\left[ \bU^\top \bhatC_k \bU \right] \right) \right\}, \\
			&= - \frac{1}{2} \nb \left\{ (\dim - \latentdim) \log(\beta) + \frac{1}{\beta} \left( \Tr\left[ \frac{1}{\nb} \sum_{k=1}^{\K}\varn_k \bhatC_k\right] -  \Tr\left[ \bU^\top ( \frac{1}{\nb} \sum_{k=1}^{\K}\varn_k\bhatC_k) \bU \right] \right) \right\}.
			\end{align*}
			And, again, its only stationary point is:
			\begin{align*}
			\nabla_{\beta} \J(\hat{\beta}) = 0 \iff \hat{\beta} = \frac{\Tr\left[\bhatC\right] -  \Tr\left[ \bU^\top \bhatC \bU \right]}{p-d},
			\end{align*}
			with $\bhatC = \frac{1}{\nb} \sum_{k=1}^{\K}\varn_k \bhatC_k$.
		\end{itemize}
	\end{proof}
	
	\begin{proof}[Optimization of $\bSigma$]
		There are now $6$ cases to treat, which are the full, diagonal and isotropic covariance matrices where each case can be with or without homoscedasticity. We will need the two following formulas concerning matrix derivation. For any invertible matrix, $\bA \in \R{^{\dim \times \dim}}$ we have:
		\begin{equation}
		\label{bfem:eq:derivativelogdet}
		\nabla_{\bA} \log \vert \bA \vert = \bA^{-1},
		\end{equation}
		\begin{equation}
		\label{bfem:eq:derivativetraceinverse}
		\nabla_{\bA} \Tr \left[ \bA^{-1} \bB  \right] = - (\bA^{-1} \bB \bA^{-1})^\top.
		\end{equation}
		
		\begin{itemize}
			\item Model $\BDLM_{[\bSigma_k (\cdot)]}$: We rewrite the bound of \Cref{bfem:eq:ELBO} as a function of $\bSigma_{k}$:
			\begin{equation}
			\begin{aligned}
			\J(\bSigma_{1},\ldots, \bSigma_{\K}) = & - \frac{1}{2} \sum_{k=1}^{\K} \varn_k \Bigg\{ \log \vert \bSigma_{k} \vert + \Tr\left[ \bSigma_k^{-1} \bU^\top \bhatC_k \bU\right] \Bigg\} + \cst.
			\end{aligned}
			\end{equation}
			Thus, using \Cref{bfem:eq:derivativelogdet,bfem:eq:derivativetraceinverse} with $\bA = \bSigma_{k}$ we get:
			\begin{equation*}
			\nabla_{\bSigma_{k}} \J(\bSigma_{k}) = -\frac{\varn_k}{2} \left(\bSigma_{k}^{-1} -  \bSigma_k^{-1}  \bU^\top \bhatC_k \bU\bSigma_k^{-1}\right).  
			\end{equation*}
			Then, a first order condition gives
			\begin{align}
			\nabla_{\bSigma_{k}} \J(\bSigma_{k}) = 0 \iff \bSigma_k^{-1} = \bSigma_k^{-1}  \bU^\top \bhatC_k \bU\bSigma_k^{-1} ,
			\end{align}
			and we obtain the M-step estimate $\hat{\bSigma}_{k}$ by multiplying left and right by $\bSigma_{k}$:
			\begin{align}
			\hat{\bSigma}_{k} =  \bU^\top \bhatC_k \bU.
			\end{align} 
			\item Model $\BDLM_{[\bSigma (\cdot)]}$: In this case, the variational bound can be rewritten as:
			\begin{equation}
			\begin{aligned}
			\J(\bSigma) = & - \frac{\nb}{2} \Bigg\{ \log \vert \bSigma \vert +  \Tr\left[ \bSigma^{-1} \bU^\top  (\frac{1}{\nb} \sum_{k=1}^{\K}\varn_k\bhatC_k)\bU\right] \Bigg\} + \cst.
			\end{aligned}
			\end{equation}
			And finding the root of the gradient leads to:
			\begin{equation}
			\hat{\bSigma} =  \bU^\top \bhatC \bU.
			\end{equation}
			\item Model $\BDLM_{[\alpha_{kh} (\cdot)]}$: In this model, the bound writes as a function of $\balpha = (\balpha_1, \ldots, \balpha_\K)$:
			\begin{align}
			\J(\balpha) = & - \frac{1}{2} \sum_{k=1}^{\K} \varn_k \Bigg\{ \sum_{h=1}^{\latentdim} \log ( \alpha_{kh}) +  \frac{\bu_{h}^\top \bhatC_k \bu_{h}}{\alpha_{kh}} \Bigg\} + \cst .
			\end{align}
			Thus, the partial derivative with respect to $\alpha_{kh}$ is given as:
			\begin{equation*}
			\nabla_{\alpha_{kh}} \J(\alpha_{kh}) = - \frac{\varn_k}{2} \left( \frac{1}{\alpha_{kh}} - \frac{\bu_{h}^\top \bhatC_k \bu_{h}}{\alpha_{kh}^2} \right),
			\end{equation*}
			and finding its root gives:
			\begin{equation}
			\hat{\alpha}_{kh} = \bu_h^\top \bhatC_k \bu_h .
			\end{equation}
			\item Model $\BDLM_{[\alpha_{h} (\cdot)]}$: Put $\balpha = (\alpha_1, \ldots, \alpha_\latentdim)$ and we have,
			\begin{equation}
			\J(\balpha) = - \frac{\nb}{2}  \Bigg\{ \sum_{h=1}^{\latentdim} \log ( \alpha_{h}) +  \frac{\bu_{h}^\top (\frac{1}{\nb} \sum_{k=1}^{\K}\varn_k\bhatC_k) \bu_{h}}{\alpha_{h}} \Bigg\} + \cst .
			\end{equation}
			Its gradient with respect to $\alpha_h$ is computed in the same manner as above, and a first-order condition gives:
			\begin{equation}
			\hat{\alpha}_{h} = \bu_h^\top \bhatC \bu_h 
			\end{equation}
			\item Model $\BDLM_{[\alpha_{k} (\cdot)]}$: Introducing $\balpha = (\alpha_1, \ldots, \alpha_\K)$, the bound is written as:
			\begin{equation}
			\J(\balpha) =  - \frac{1}{2} \sum_{k=1}^{\K} \varn_k \Bigg\{ \latentdim \log (\alpha_{k}) +  \frac{1}{\alpha_{k}} \Tr\left[\bU^\top \bhatC_k \bU \right] \Bigg\} + \cst .
			\end{equation}
			Again, its gradient with respect to $\alpha_k$ is easily computed as:
			\begin{equation*}
			\nabla_{\alpha_k} \J(\alpha_k) = - \frac{\varn_k}{2} \left( \frac{\latentdim}{\alpha_k} - \frac{\Tr\left[\bU^\top \bhatC_k \bU \right] }{\alpha_k^2} \right).
			\end{equation*}
			Finding it $0$ point leaves the following M-step update for $\hat{\alpha}_k$
			\begin{equation}
			\hat{\alpha}_{k} = \frac{1}{\latentdim} \Tr \left[{\bU}^\top \bhatC_k \bU \right]. 
			\end{equation}
			\item Model $\BDLM_{[\alpha (\cdot)]}$: For this final model, $\alpha$ is a positive scalar and the bound is:
			\begin{equation}
			\J(\alpha) = - \frac{\nb}{2}  \Bigg\{ \latentdim \log (\alpha) +  \frac{1}{\alpha} \Tr\left[\bU^\top (\frac{1}{\nb} \sum_{k=1}^{\K}\varn_k\bhatC_k) \bU \right] \Bigg\} + \cst .
			\end{equation}
			\begin{equation}
			\hat{\alpha} = \frac{1}{\latentdim} \Tr \left[{\bU}^\top \bhatC \bU \right].
			\end{equation}
		\end{itemize}

	\end{proof}
	
\end{proof}

\subsection{Hyper-parameter estimation}
\label{bfem:appendix:ProofEmpiricalBayes}
To prove \Cref{bfem:prop:EmpiricalBayes}, we maximize the variational bound with respect to $(\bnu, \lambda)$. From \Cref{bfem:eq:FullELBO} we get:
\begin{align*}
\J(\bnu, \lambda) &=-\frac{1}{2} \sum_{k=1}^{K}  d \log(2\pi) + d \log(\lambda) + \frac{1}{\lambda} \mathbb{E}_{\bmu_k}\left[ \Vert \bmu_k - \bnu \Vert_ 2^2 \right] , \\
&= -\frac{1}{2} \sum_{k=1}^{K}  d \log(2\pi) + d \log(\lambda) + \frac{1}{\lambda} \left[ \Vert \bvarmean_k - \bnu \Vert_2^2 + \Tr\left[\bvarcovar_k \right] \right] .
\end{align*}

\begin{proof}[Optimization with respect to  $\bnu$]
	\begin{align}
	\nabla_{\bnu} \J(\bnu) = -\frac{1}{\lambda} \sum_{k=1}^{K} ( \bvarmean_k - \bnu)
	\end{align}
	Hence,
	\begin{align*}
	& \nabla_{\bnu} \J(\hat{\bnu}) = 0 , \\
	\iff & \hat{\bnu} = \frac{\sum_{k=1}^{K} \bvarmean_k }{K}
	\end{align*}
	Since $\J$ is a concave function of $\bnu$ with a negative definite Hessian $\frac{-1}{2\lambda} \bI_d$, this concludes the proof
\end{proof}

\begin{proof}[Optimization with respect to $\lambda$]
	\[
	\nabla_{\lambda} \J(\lambda) = - \frac{1}{2} \sum_{k=1}^{\K} \frac{\latentdim}{\lambda} - \frac{1}{\lambda^2} \left[ \Vert \bvarmean_k - \bnu \Vert_2^2 + \Tr\left[\bvarcovar_k \right] \right] .
	\]
	Thus, the first-order condition gives:
	\begin{align*}
	&	\nabla_{\lambda} \J(\hat{\lambda}) = 0,\\
	& \hat{\lambda} = \frac{\sum_{k=1}^{\K}  \Vert \bvarmean_k - \bnu \Vert_2^2 + \Tr\left[\bvarcovar_k \right]}{\latentdim \K}.
	\end{align*}
	The second-order derivative gives us a condition for $\hat{\lambda}$ to be a maximum. Indeed, the following must hold for $\lambda = \hat{\lambda}$:
	\begin{align*}
	\nabla^2_{\lambda} \J(\lambda) = \frac{ \left(\sum_{k=1}^{\K}  \Vert \bvarmean_k - \bnu \Vert_2^2 + \Tr\left[\bvarcovar_k \right]\right) \lambda - \latentdim}{2 \lambda^3} < 0.
	\end{align*}
	Hence, for positive $\lambda$, the latter is negative if and only if $\lambda < \frac{\sum_{k=1}^{\K}  \Vert \bvarmean_k - \bnu \Vert_2^2 + \Tr\left[\bvarcovar_k \right]}{\latentdim } = \K \hat{\lambda}$. Obviously, $\hat{\lambda}$ verifies this condition for $\K \geq 2$ and is thus a maximum of $\J$.
	
\end{proof}

\subsection{Model selection}
\label{bfem:appendix:ProofModelSelection}
In this section, we briefly recall the key elements for the derivation of the $\ICLbic$ of \textcite{biernacki2000assessing}, given in \Vref{bfem:eq:ICL}. Then, we detail the computation of the latter via the variational lower bound. For the sake of notations, we drop the dependencies in $\Model$ and $\K$ here, since the discussion is independent of these quantities.

\paragraph{Derivation of the ICL} Taking a Bayesian perspective on the model parameters $\globalparam = (\bPi, \bSigma, \bBeta, \bU)$, we posit a prior distribution $\p (\globalparam)$ and the ICL computes the following quantity:
\begin{align*}
	\log \p ( \Obs, \Clust ) = \log \int_{\globalparam}  \log \p ( \Obs, \Clust, \globalparam ) \dif \globalparam .
\end{align*}
Considering a factorized prior $\p (\globalparam) = \p (\bPi) \p (\bSigma, \bBeta, \bU)$, we can rewrite the ICL as the sum of two terms:
\begin{align*}
	\log \p ( \Obs, \Clust ) = \log \p (\Obs \mid \Clust) + \log \p (\Clust).
\end{align*}
\textcite{biernacki2000assessing} proposed to approximate the first term with a Laplace approximation:
\begin{align*}
	\log \p (\Obs \mid \Clust) \approx \max_{\bSigma, \bBeta, \bU} \log \p (\Obs \mid \Clust, \bSigma, \bBeta, \bU) - \frac{\gamma(\bSigma, \bBeta, \bU)}{2} \log(\nb),
\end{align*}
where  $\gamma$ is the number of free parameters in $(\bSigma, \bBeta, \bU)$. Then, putting a symmetric Dirichlet prior on $\bPi \sim \Dir_\K (\balpha = (\alpha, \ldots, \alpha))$, the second term can be computed exactly, involving Gamma functions. Fixing $\alpha = 1/2$, and assuming that each $n_k = \sum_i \rawclust_{ik}$ is large with $\nb$, a Stirling approximation on the Gamma functions is used to get:
\begin{align*}
	\log \p (\Clust) \approx \max_{\bPi} \log \p (\Clust \mid \bPi) - \frac{\K -1}{2} \log (\nb).
\end{align*}
Finally, summing these two approximations, we get a BIC-like approximation of the integrated classification likelihood:
\begin{align*}
	\ICLbic(\Model , \K) = \max_{\globalparam} \log \p (\Obs, \Clust \mid \globalparam) - \frac{\gamma_{\Model, \K}}{2} \log(\nb).
\end{align*}

\paragraph{Computing the classification likelihood with the variational bound} We now turn into the problem of computing $\log \p(\Obs, \hat{\Clust} \mid \hat{\globalparam})$ with the variational bound $\J$. We consider the parameters $\hat{\globalparam}$ to be estimated via the BFEM algorithm and fixed. Consider $\Clust = \hat{\Clust}$ fixed and known. The result stems from the fact that the conditional posterior $\p(\bmu \mid \hat{\Clust}, \Obs)$ is tractable in the $\BDLM$ model and it is equal to $\q(\bmu_k)$ if $\tau_{ik} =\hat{\rawclust}_{ik}$. Thus, the variational bound in is tight and equals the classification likelihood.

Formally, we want to show that the variational bound of \Cref{bfem:eq:FullELBO} is equal to the classification likelihood $\log \p(\Obs, \hat{\Clust} \mid \hat{\globalparam})$ when $\tau_{ik} = \hat{\rawclust}_{ik}$ in \Cref{bfem:prop:CAVIupdateZ,bfem:prop:CAVIupdateMu}. When $\bTau \gets \hat{\Clust}$, we have:
\begin{align}
\label{bfem:eq:VariationalDistributionICL}
\q(\bmu, \Clust \mid \hat{\Clust}) = \q\left(\bmu \mid \bvarmean(\hat{\Clust}), \bvarcovar(\hat{\Clust})\right) \times \delta_{\hat{\Clust}}(\Clust),
\end{align}
with $\delta_x$ the Dirac mass at $x$, and $(\bvarmean, \bvarcovar)$ computed with $\hat{\Clust}$ instead of $\bTau$:
\begin{align*}
\hat{\nb}_k &= \sum_i \hat{\rawclust}_{ik}, \\
\bvarcovar_k(\hat{\Clust}) &= \left( \lambda^{-1} \bI_{\latentdim} + \hat{n}_k \bSigma_{k}^{-1}\right)^{-1}, \\
\bvarmean_k(\hat{\Clust}) &= \bnu + \bvarcovar_k \bSigma_{k}^{-1} \left(\bU^\top \sum_i \hat{\rawclust}_{ik} \obs_i - \hat{n}_k \bnu\right).
\end{align*}

It happens that the conditional posterior distribution of $\bmu$: $\p(\bmu \mid \hat{\Clust}, \Obs)$ is tractable as a product of $\K$ distributions since:
\begin{align*}
\p(\bmu \mid \hat{\Clust}, \Obs, \hat{\globalparam}) & \propto \p(\bmu) \p( \Obs \mid  \hat{\Clust}, \bmu, \hat{\globalparam}),\\
& \propto \prod_{k=1}^{\K} \Gaussian_{ \latentdim}(\bmu_{k} \mid \bnu, \lambda) \times \prod_{i=1}^{\nb} \prod_{k=1}^{\K} \Gaussian_{\dim}(\obs_{i} \mid \hat{\bU} \bmu_{k}, \hat{\bS}_k)^{\hat{\rawclust}_ {ik}}, \\
& \propto \prod_{k=1}^{\K}  \left\{\Gaussian_{ \latentdim}(\bmu_{k} \mid \bnu, \lambda) \times \prod_{i \in \hat{\mathcal{C}}_k}  \Gaussian_{\dim}(\obs_{i} \mid \hat{\bU} \bmu_{k}, \hat{\bS}_k)\right\}.
\end{align*}
Using the same reasoning as in \Cref{bfem:appendix:ProofCAVIupdateMu}, the distributions into bracket happens to be (un-normalized) Gaussians with parameter $\left(\bvarmean_k(\hat{\Clust}), \bvarcovar_k(\hat{\Clust})\right)$. Thus, we have that:
\begin{align}
\p(\bmu \mid \hat{\Clust}, \Obs, \hat{\globalparam}) =  \q\left(\bmu \mid \bvarmean(\hat{\Clust}), \bvarcovar(\hat{\Clust})\right).
\end{align}
Finally, denoting $\entropy(\q) = \Expectation_{\bm{\eta} \sim \q}[\q(\bm{\eta})]$, we may write the expression of the variational bound $\J(\hat{\q}  , \hat{\globalparam})$ with $\hat{\q}$ defined in \Cref{bfem:eq:VariationalDistributionICL}, and get:
\begin{align*}
\J(\hat{\q} , \hat{\globalparam}) &= \Expectation_{(\bmu, \Clust) \sim \hat{\q}}\left[\log\left(\Obs, \Clust, \bmu \mid \hat{\globalparam} \right)\right] + \entropy(\hat{\q}), \\
&= \Expectation_{\bmu \sim 	\q\left(\bmu \mid \bvarmean(\hat{\Clust}), \bvarcovar(\hat{\Clust})\right) }\left[ \Expectation_{\Clust \sim \delta_{\hat{\Clust}} } \left[\log\left(\Obs, \Clust, \bmu \mid \hat{\globalparam} \right)\right]\right] + \entropy\left(\q\left(\bmu \mid \bvarmean(\hat{\Clust}), \bvarcovar(\hat{\Clust})\right) \right), \\
&= \Expectation_{\bmu \sim 	\p(\bmu \mid \hat{\Clust}, \Obs) }\left[ \log\left(\Obs, \hat{\Clust}, \bmu \mid \hat{\globalparam} \right)\right] + \entropy\left(\p(\bmu \mid \hat{\Clust}, \Obs, \hat{\globalparam}) \right), \\
&= \log \p(\Obs, \hat{\Clust} \mid \hat{\globalparam}) - \KL{	\p(\bmu \mid \hat{\Clust}, \Obs, \hat{\globalparam})}{	\p(\bmu \mid \hat{\Clust}, \Obs, \hat{\globalparam})}, \\
& = \log \p(\Obs, \hat{\Clust} \mid \hat{\globalparam}).
\end{align*}
Here, the second line used the fact that $\entropy(\delta_{\hat{\Clust}}) = 0$, and the last equality used the tightness of the variational lower bound when $\q$ is equal to the posterior distribution.

\section{Image denoising}
\label{bfem:appendix:ProofPatchExpectation}

%We first begin by deriving the formula of \Cref{bfem:eq:PatchReconstruction,bfem:eq:ConditionalPatch}. 

\begin{proof}[Proof of \Vref{bfem:prop:PatchExpectation}]
	Denote $\psi_k(\Obs, \bmu) = \Expectation \left[\bpatch_i \mid \obs_i, \bmu_{k}, \rawclust_{ik} = 1 \right]$ and $\bmean_{k} = \bU \bmu_k$. We begin by recalling that \Cref{bfem:eq:ConditionalPatch} is obtained from the fact that $(\bpatch_i, \obs_i) \mid \{\rawclust_{ik}=1\}$ is jointly Gaussian, with covariance $\bS_k - \sigma^2 \bI_\dim$. Thus,
	\begin{align*}
\psi_k(\Obs, \bmu) &= \bU \bmu_k +  \left(\bS_k - \sigma^2 \bI_{\dim}\right) \bS_k^{-1} (\obs_i - \bU \bmu_k), \\
&=  \bU \bmu_k +  \left(\bI_\dim- \sigma^2  \bS_k^{-1}\right) (\obs_i - \bU \bmu_k),
\end{align*}
	This expression involves direct inverse of $\bS_k^{-1}$ which can be avoided thanks to the particular block structure made on the matrix $\bS_k$ in the BDLM model. Keeping the same notations as in \Cref{lemma:TrSkA}, we split $\bS_k^{-1}$ in two parts depending on the discriminative subspace and its orthogonal complement: $\bS_k^{-1} = \tilde{\bD} \bDelta_k^{-1} \tilde{\bD}^\top + \bar{\bD} \bDelta_k^{-1} \bar{\bD}^\top$, with $\tilde{\bD} = [\bU, \bm{0}_{\dim \times (\dim - \latentdim)}]$ and $\bar{\bD} = [\bm{0}_{\dim \times \latentdim}, \bV]$. Then, we have:
	\begin{align*}
	\left\{	 \begin{array}{ll}
	\bar{\bD} \bDelta_k^{-1} \bar{\bD}^\top &= \frac{1}{\sigma^2} \bar{\bD} \bar{\bD}^\top \\
	\bar{\bD}  \bar{\bD}^\top + \tilde{\bD}\tilde{\bD}^\top &= \bI_{\dim}
	\end{array}\right. \implies \bar{\bD} \bDelta_k^{-1} \bar{\bD}^\top =  \frac{1}{\sigma^2} \left(\bI_{\dim} - \tilde{\bD}\tilde{\bD}^\top\right).
	\end{align*}
	Thus, 
	\begin{align*}
		\sigma^2 \bS_k^{-1} &= \sigma^2 \left[ \tilde{\bD} \bDelta_k^{-1} \tilde{\bD}^\top + \bar{\bD} \bDelta_k^{-1} \bar{\bD}^\top \right], \\
		&= \bI_\dim + \tilde{\bD} \left(\sigma^2 \bDelta_k^{-1} - \bI_{\dim}\right) \tilde{\bD}^\top,
	\end{align*}
	and we can thus simplify the expression in \Cref{bfem:eq:ConditionalPatch} into:
	\begin{equation*}
	\psi_k(\Obs, \bmu) =  \bU \bmu_k +  \tilde{\bD} \left( \bI_{\dim} - \sigma^2 \bDelta_k^{-1}\right) \tilde{\bD}^\top(\obs_i - \bU \bmu_k),
	\end{equation*}
	Finally, in BDLM, $\bmu_{k}$ is a random variable and we can further take the expectation of $\psi_k(\Obs, \bmu)$ according to the variational distribution $\q(\bmu_k)$, leaving:
	\begin{equation}
		\psi_k(\Obs) = \Expectation_{\q(\bmu_{k})} \left[ \Expectation \left[\bpatch_i \mid \obs_i, \bmu_{k}, \rawclust_{ik} = 1 \right]\right]=  \bU \bvarmean_k +  \tilde{\bD} \left( \bI_{\dim} - \sigma^2 \bDelta_k^{-1}\right) \tilde{\bD}^\top(\obs_i - \bU \bvarmean_k)
	\end{equation}

	Note that this Proposition closely resembles the Proposition 1 of \textcite{houdard2018high}, this is explained by the similar assumptions made by the HDDC and BDLM models.
\end{proof}

% Biblatex (no instruction from Springer) -> see journal instructions : Name and year parenthesis.
\printbibliography

\end{document}